\documentclass[
    draftcls,
    narroweqnarray,
    inline,
    oneside,
    11pt,
    journal
    ]{IEEEtran}
\onecolumn

\IEEEoverridecommandlockouts

\usepackage{graphicx,cite}
\usepackage{fancybox,subfigure}
\usepackage{color,setspace}
\usepackage{times,amsmath,amssymb,color,multirow,stfloats,extpfeil}
\definecolor{orange}{cmyk}{0,0.5,1,0}
\definecolor{green}{cmyk}{1,0.4,.8,0}
\definecolor{blue}{rgb}{0.2,0.3,0.8}
\definecolor{red}{rgb}{0.8,0.1,0.1}

\newcommand{\thdc}[3]{\;\raisebox{-2ex}{\shortstack[c]{\ensuremath{#1}\footnotesize{=#2} \\  \ensuremath{\gtrless} \\ \ensuremath{#1}\footnotesize{=#3}}} \;}

\def\bone{\mbox{\boldmath $1$}}

\title{Distributed Binary Detection over Fading Channels: Cooperative and Parallel Architectures}
\author{\normalsize Nahal Maleki${\dagger}$, ~\IEEEmembership{IEEE Student Member}, Azadeh Vosoughi${\ddagger}$,~\IEEEmembership{IEEE Senior Member},  \\ Nazanin  Rahnavard${\ddagger}$,~\IEEEmembership{IEEE Member}  \\
\thanks{Copyright (c) 2015 IEEE. Personal use of this material is permitted. However, permission to use this material for any other purposes must be obtained from the IEEE by sending a request to pubs-permissions@ieee.org.}
\thanks{This work is supported by National Science Foundation under Grant CCF-1336123, CCF-1341966, ECCS-1056065 and CCF-0915994. Part of this work is presented in SPAWC 2009 \cite{vosoughi-Ahmadi-spawc}}.
\thanks{${\dagger}$N. Maleki is with the ECE Department, University of Rochester, Rochester,
NY (e-mail: nmalekit@ur.rochester.edu). ${\ddagger}$A. Vosoughi is with the EECS Department, University of Central Florida, Orlando,
FL (e-mail: azadeh@ucf.edu). ${\ddagger}$N. Rahnavard is with the EECS Department, University of Central Florida, Orlando,
FL (e-mail: nazanin@eecs.ucf.edu).}}
\newcommand{\subparagraph}{}
\usepackage[compact]{titlesec}
\titlespacing{\section}{0pt}{*0}{*0}
\titlespacing{\subsection}{0pt}{*0}{*0}
\titlespacing{\subsubsection}{0pt}{*0}{*0}
\begin{document}
\maketitle

\begin{abstract}

This paper considers the problem of binary distributed detection of a known signal in correlated Gaussian sensing noise in a wireless sensor network, where the sensors are restricted to use likelihood ratio test (LRT), and communicate with the fusion center (FC) over bandwidth-constrained channels that are subject to fading and noise. To mitigate the deteriorating effect of fading encountered in the conventional parallel fusion architecture, in which the sensors directly communicate with the FC, %(without cooperation among sensors),
we propose new fusion architectures that enhance the detection performance, via harvesting cooperative gain (so-called ``decision diversity gain''). In particular, we propose: (i) cooperative fusion architecture with Alamouti's space-time coding (STC) scheme at sensors, (ii) cooperative fusion architecture with signal fusion at sensors, and (iii) parallel fusion architecture with local threshold changing at sensors. For these schemes, we derive the LRT and majority fusion rules at the FC, and provide upper bounds on the average error probabilities for homogeneous sensors, subject to uncorrelated Gaussian sensing noise, in terms of signal-to-noise ratio (SNR) of communication and sensing channels. Our simulation results indicate that, when the FC employs the LRT rule, unless for low communication SNR and moderate/high sensing SNR, performance improvement is feasible with the new fusion architectures. When the FC utilizes the majority rule, such improvement is possible, unless for high sensing SNR.
\end{abstract}
\section{Introduction}
The problem of distributed detection with the fusion center (FC) (so-called classical parallel fusion architecture)  has a long and rich history, where each local detector (sensor) processes its observation locally and independently,
%in order to reduce the communication bandwidth required between the sensors and the FC,
and  passes its local binary decision to the FC. The main assumption in the classical works is that the bandwidth-constrained communication channels are {\it error-free} and thus the reliability of the final decision at the FC is determined by the reliability of the local binary decisions.  However, wireless channels are inherently error-prone, due to noise and fading.
%Motivated by the event detection application of WSNs,
An {\it integrated approach} of distributed detection over noisy fading channels was considered in \cite{c7,c8,robust-decision,Magazine-Paper,outage,fusion-DC,ChVVV2004,Chamberland04,Chen-threshold05,Kanchumarthy08,censoring-sensor06,multiple-symbol}, in which the sensors send their modulated local binary decisions to the FC and the FC employs a fusion rule, incorporating
%coherent or non-coherent fusion rules
channel state information
%\footnote{For coherent fusion rules, CSI implies the availablity of complex channels at the FC. For non-coherent fusion rules, full and partial CSI imply the availablity of channel envelopes and statistics of chanenl envelopes at the FC, respectively.}
(CSI), to improve the reliability of the final decision at the FC.
 %the performance of the parallel fusion network
%in the presence of noise and fading.
The performance of these integrated distributed detection is ultimately limited by the communication bounds. %even if all the sensors make perfect local binary decisions, the FC can still make an error due to noisy fading channels.
A common thread in  the schemes discussed in \cite{c7,c8,robust-decision,Magazine-Paper,outage,fusion-DC,ChVVV2004,Chamberland04,Chen-threshold05,Kanchumarthy08,censoring-sensor06,multiple-symbol} is that they are non-cooperative, i.e., there is no information exchange among the sensors. Cooperative wireless communication \cite{c13,Jia-vosoughi-spawc} has been proven to significantly enhance performance
%with respect to the non-cooperative scenario,
in the presence of fading, via
%creating a virtual array of transmission antenna and
invoking spatial diversity, that leads into mitigation of the detrimental fading effects \cite{c13,Jia-vosoughi-spawc}.
%Taking advantage of broadcast nature of the wireless media, cooperation among distributed single antenna nodes invokes spatial diversity, through creating a virtual array of transmission antenna, and mitigates the detrimental effect of fading \cite{c13,Jia-vosoughi-spawc}.
Motivated by the promises of cooperative communication, we propose a new class of integrated distributed detection, which harvests cooperative gain (enabled by at most $1$-bit information exchange among one-hop neighboring nodes) and  improves the performance of the integrated distributed detection \cite{c7,c8} in the presence of fading, via allowing each sensor to send (at most) $2$ information bits to the FC and assuming identical transmit power per node. In particular, we propose three schemes: (i) cooperative fusion architecture with Alamouti's space-time coding (STC) scheme at sensors, in which neighboring sensors exchange $1$ information bit and each sensor sends $2$ information bits to the FC; (ii) cooperative fusion architecture with signal fusion at sensors, in which neighboring sensors exchange $1$ information bit and each sensor sends 1 information bit to the FC; and (iii) parallel fusion architecture with local threshold changing at sensors, in which neighboring sensors do not exchange information and each sensor sends $2$ information bits to the FC. To describe the proposed schemes, suppose ${\cal S}_1$ and ${\cal S}_2$ are two designated cooperative partners. {\it In scheme (i)}, rather than transmitting their local decisions directly to the FC, ${\cal S}_1$ and ${\cal S}_2$ are coordinated to form a transmit cluster, such that they first exchange their local decisions and apply Alamouti's scheme \cite{c9,vosoughi-Ahmadi-spawc} for transmitting the decisions to the FC. Different from most literature on distributed STC,
%for cooperative communication,
which assume a node acts as a relay only for error-free reception \cite{c15}, we consider the fact that the channels between ${\cal S}_1$ and ${\cal S}_2$ are subject to errors, due to noise and fading. {\it In scheme (ii)}, similar to scheme (i), ${\cal S}_1$ and ${\cal S}_2$ exchange their local decisions. Instead of applying Alamouti's scheme, however, each node updates its decision, via optimally fusing its observation with the received signal from its cooperative partner. Updated decisions are transmitted to the FC.  {\it In scheme (iii)}, different from schemes (i) and (ii), there is no explicit information exchange between  ${\cal S}_1$ and ${\cal S}_2$.  Each node forms two decisions, where one is made based on its observation only and the other is obtained based on optimally fusing its observation with its guess of the decision of its cooperative partner. ${\cal S}_1$ and ${\cal S}_2$ apply Alamouti's scheme for transmitting these decisions to the FC. For these three schemes we provide the likelihood ratio test (LRT) and majority fusion rules at the FC. The average (over fading) error probability of decision error at the FC for these schemes depend on signal-to-noise ratio (SNR) of communication (channels between cooperative partners as well as between  the nodes and the FC) and sensing channels (channels between the target and the nodes).
%and the {\it a priori} probabilities of binary hypotheses.
For the proposed schemes we derive upper bounds on the average error probability and investigate how a node should allocate its transmit power for communicating with its cooperative partner and the FC, such that the error is minimized.  These results enable us to quantity the cooperative gain offered by the proposed schemes, with respect to the schemes in \cite{c7,c8}, assuming identical transmit power per node.\\
%(global or local) information exchange
Information exchange among the nodes for {\it consensus-based} distributed detection without the FC has been studied before (examples are  \cite{Swaszek95Aero,Pados95SMS,Hong05SSP,consensus2010}). \cite{Swaszek95Aero,Pados95SMS,Hong05SSP} considered a consensus-based distributed detection system, where the sensors successively update and broadcast their local binary decisions over {\it error-free} links \cite{Swaszek95Aero,Pados95SMS,Hong05SSP}.
%until a common decision is reached.
In \cite{consensus2010} each sensor successively updates its continuous-valued decision variable and passes it to its  neighbors over random links without bandwidth constraint.
%The link model in \cite{consensus2011,consensus2010} is abstract and does not capture noise and fading effects.
%and asymptotic convergence can be reached only in some probabilistic sense and under certain conditions.
Distributed detection in networks {\it with feedback} has been studied before (examples are \cite{c19,c20,c21}) from information theoretic perspective, focusing on the asymptotic regime (i.e., networks with large number of sensors) and quantifying performance in terms of error exponents. These works consider a variety of feedback architectures, including two-message feedback architectures, where each sensor sends its first message to the FC, based on its own observation, and sends its second message, based on the additional information provided by the FC through feedback, where the feedback contains (functions of) the messages generated by (some) other sensors. Different from these works, we consider networks with a finite number of sensors, without feedback from the FC, where (at most) 1-bit information exchange is allowed between two cooperative sensors. Also, we relax the error-free communication constraint, by considering fading effects during the information exchange phase, i.e., we assume a sensor knows the decision of its partner with a limited reliability, that is dictated by the quality of inter-sensor communication channel.
Perhaps the most related work is \cite{c18},
%in which the authors considered a distributed detection system with the FC,
in which each sensor communicates its local binary decision to its neighbors and the sensors communicate their updated binary decisions to the FC. The local decision rules and the fusion rule at the FC are {\it all majority rules} and communication channels are assumed to be {\it error-free} in \cite{c18}.
To the best of our knowledge, for parallel fusion architecture no prior work has studied the impact of local (limited) information exchange on enhancing the performance of the integrated distributed detection systems \cite{c7,c8,robust-decision,Magazine-Paper,outage,fusion-DC,ChVVV2004,Chamberland04,Chen-threshold05,Kanchumarthy08,censoring-sensor06,multiple-symbol} operating in a noisy fading environment.
% global information exchange for non-consensus DD application does not make sense.
Paper organization follows.
%The remainder of the paper is organized as follows.
Section II introduces our sensing model and overviews the integrated distributed detection schemes in \cite{c7,c8} for this work to be self-explanatory. Sections III, IV, V describe schemes (i),(ii),(iii), respectively, and provide local decision and fusion rules  at the sensors and the FC. Section VI provides the performance analysis. Our numerical results are presented in Section VII. Concluding remarks are in Section VIII.
%=================================================================

%{\bf Notations:} A complex Gaussian random variable (RV) $x$ with mean $r$ and variance $s^2$ is denoted as $x \sim {\cal C}{\cal N}(r,s^2)$. The indicator function $\bone_{\{x> a\}}$ is defined as one when $x>a$ and is zero otherwise. $\mathfrak{Re}(.)$ indicates the real part of a complex number. sgn(.) represents the sign function.

%================================================================
\section{Basic Models}
\subsection{Sensing Model}\label{sensing-model}
We consider the binary hypothesis testing problem of detecting a known signal in correlated Gaussian noise based on measurements $x_k$ at $K$ distributed sensors.
The {\it a priori} probabilities of two hypotheses $\mathcal{H}_{0},\mathcal{H}_{1}$ are denoted by $\pi_0,\pi_1$, respectively. The FC is tasked with determining whether the unknown hypothesis is $\mathcal{H}_{0}$ or $\mathcal{H}_{1}$,
based on the information collected from the $K$ sensors.
The measurement $x_k$ of sensor ${\cal S}_k$ under $\mathcal{H}_{0}$ and $\mathcal{H}_{1}$, respectively, are
$x_{k} = w_{k}$ and
$x_{k} = 1+ w_{k}$ for $k=1,...,K$, where sensing noise $w_k$ is zero-mean with variance  $\sigma_{w_k}^2$.
The spatial correlation between noises $w_i, w_j$ are characterized with the correlation coefficient $\rho_{ij}$. We assume sensors are grouped into $S\!=\!K/2$ distinct pairs of cooperative partners $({\cal S}_i,{\cal S}_j)$, where ${\cal S}_i$ knows $\sigma_{w_i}^2$, $\sigma_{w_j}^2$ and $\rho_{ij}$ only and is restricted to use LRT. The FC employs LRT, when all sensing noise variances and pairwise correlation coefficients are available at the FC. In the absence of this knowledge, the FC uses the majority rule.
%============================================================
\subsection{Classical Parallel Fusion Architecture}\label{parallel-classical}
Each sensor ${\cal S}_k$ makes a local binary decision  $u_k \! \in \! \{1,-1\}$ based on its measurement $x_k$. The local decisions $+1$ and $-1$ correspond to the hypotheses  ${\cal H}_1$ and ${\cal H}_0$, respectively.  The local detection performance of ${\cal S}_k$ are characterized with $P_{d_k}=P(u_k=1 | {\cal H}_1)$ and $P_{f_k}=P(u_k=1 | {\cal H}_0)$. To form $u_k$, sensor ${\cal S}_k$ applies the LRT $\frac{f(x_k|{\cal H}_1)}{f(x_k|{\cal H}_0)}\thdc{u_k}{1}{-1}\frac{\pi_0}{\pi_1}$. For the sensing model in Section \ref{sensing-model}, the LRT is reduced to  $x_k \thdc{u_k}{1}{-1}\tau_k$ where $\tau_k=0.5+\sigma_{w_k}^2\ln(\frac{\pi_0}{\pi_1})$. Also,
$P_{d_k}=Q(\frac{\tau_k-1}{\sigma_{w_k}})$ and $P_{f_k}=Q(\frac{\tau_k}{\sigma_{w_k}})$,  where $Q(x)$ denotes the $Q$-function\footnote{Considering the local LRT at sensor $k$, the distributions $f(x_k|{\cal H}_\ell)$ for $\ell=0,1$ and $P_{d_k}, P_{f_k}$ values depend on the distribution of sensing noises. Thus the simplified form of the local rule and $P_{d_k}, P_{f_k}$ expressions would change for non-Gaussian $w_k$. Given the joint pdf $f(w_1,...,w_K)$ at the FC, the expressions $\Lambda$ in  (\ref{Lambda-canonical}) for dependent and (\ref{Lambda-canonical-uncorrelated}) for independent $w_k$'s remain unchanged.}.
The local decisions $u_k$ are transmitted over orthogonal channels subject to noise and fading to the FC. When ${\cal S}_k$ sends $u_k$ the received signal at the FC is
$y_k= u_k h_k + v_k~~\mbox{where}~~h_k \sim {\cal C}{\cal N}(0, \sigma_{h_k}^2),~~v_k \sim {\cal C}{\cal N}(0, \sigma_{v}^2)$.
where $h_k$ represents the fading channel coefficient corresponding to the channel between  ${\cal S}_k$ and the FC. The channel variance is $\sigma^2_{h_k} = {\cal P} \mathcal{G}/d_{k}^{\varepsilon}$, where ${\cal P} $ represents transmit power of ${\cal S}_k$, $d_{k}$ denotes the distance between ${\cal S}_k$  and the FC, $\varepsilon$ is the pathloss exponent, and $\mathcal{G}$ is a constant that depends on the antenna gains and the wavelength. We assume that $\varepsilon$ and $\mathcal{G}$ are identical for all links. The term $v_k$ is the receiver noise at the FC. We assume $v_k$ and $h_k$ are independent. We define $\bar{\gamma}_{h_k}^2=\frac{\sigma_{h_k}^2}{\sigma_v^2}$ as the average received SNR corresponding to node ${\cal S}_k$-FC communication channel. Relying on the received signals $y_k$ and the availability of CSI $h_k$ the FC forms the LRT  $\Lambda=\frac{f(y_1,...,y_K|\mathcal{H}_1)}{f(y_1,...,y_K|\mathcal{H}_0)}$, where $f(y_1,...,y_K|\mathcal{H}_{\ell})$ indicates the joint pdf of $y_1,...,y_K$ given the hypothesis $\mathcal{H}_{\ell}$, to make the final decision. In particular, the FC decides $\mathcal{H}_{1}$ when $\Lambda > \pi_0/\pi_1$ and decides $\mathcal{H}_{0}$ otherwise.  Considering that $\mathcal{H}_{\ell} \rightarrow u_1,...,u_K \rightarrow y_1,..., y_K$ form a Markov chain and also the fact $f(y_1,...,y_K|u_1,...,u_K)=\prod_{k=1}^K f(y_k|u_k)$ we can simplify $\Lambda$ as follows
\begin{eqnarray}\label{Lambda-canonical}
\Lambda= \frac{\sum_{u_1} ...\sum_{u_K} \left(\prod_{k=1}^K f(y_k|u_k)\right)
P(u_1,...,u_K|\mathcal{H}_1)}{ \sum_{u_1} ...\sum_{u_K} \left(\prod_{k=1}^K f(y_k|u_k)\right) P(u_1,...,u_K|\mathcal{H}_0)}.
\end{eqnarray}
Focusing on the terms $f(y_k|u_k)$ in (\ref{Lambda-canonical}) we note that given $h_k, u_k$ we have $y_k \sim {\cal C}{\cal N}(u_k h_k, \sigma_v^2)$. Considering the term $P( u_1,...,u_K |\mathcal{H}_{\ell})$ in (\ref{Lambda-canonical}) we note that it depends on the characteristics of the sensing channel noises described in Section \ref{sensing-model}. When Gaussian sensing noises  are uncorrelated $\rho_{ij}=0$ for all $i \neq j$, this term is simplified to $ \prod_{k=1}^K P(u_k|\mathcal{H}_{\ell})$ and (\ref{Lambda-canonical}) is reduced to \cite{c7,c8}
\begin{equation}\label{Lambda-canonical-uncorrelated}
\Lambda=\prod_{k=1}^{K}\frac{P_{d_k}f(y_{k}|u_{k}=1)+(1-P_{d_k})f(y_{k}|u_{k}=-1)}{P_{f_k}f(y_{k}|u_{k}=1)+(1-P_{f_k})f(y_{k}|u_{k}=-1)}.
\end{equation}
When the parameters of sensing channels are unavailable at the FC, the FC cannot apply the optimal LRT in (\ref{Lambda-canonical}) or (\ref{Lambda-canonical-uncorrelated}). Alternatively, the FC demodulates the channel inputs $u_k$ using  $y_k $ for $k=1,...,K$ and applies the majority rule to the demodulated symbols to reach the final decision, i.e., if the sum of the demodulated symbols is positive the FC decides $\mathcal{H}_{1}$ and otherwise decides $\mathcal{H}_{0}$.
%============================================================
\section{Cooperative Fusion Architecture with STC at Sensors}\label{section-DSTC}
\subsection{Inter-node Communication Channel Model and Local Decision Rules at Sensors} \vspace{-0.2cm}Suppose nodes ${\cal S}_{i}$ and ${\cal S}_{j}$ are within a pair, that is ${\cal S}_{i}$ and ${\cal S}_{j}$  are cooperative partners. Each sensor makes a decision based on its measurement. The nodes exchange their decisions over orthogonal channels subject to noise and fading. Let $u_i$ denote the decision made at ${\cal S}_{i}$ based on $x_i$. Sensor ${\cal S}_{i}$ transmits $\sqrt{1-\alpha}u_i$, where $0< \alpha <1$ is a power normalization factor, to assure that the total transmit power of nodes in this architecture remains the same as that of the classical parallel fusion architecture in Section \ref{parallel-classical}.
When ${\cal S}_{i}$ transmits $\sqrt{1-\alpha}u_i$ the received signal at ${\cal S}_{j}$ is
\begin{equation}\label{dstc-inter-sensor}
r_{ij}=\sqrt{1-\alpha}u_i g_{ij} + \eta_{ij}~~\mbox{where}~~g_{ij} \sim {\cal C}{\cal N}(0,\sigma^2_{{hs}_{ij}}),~~\eta_{ij} \sim {\cal C}{\cal N}(0,\sigma^2_{\eta}).
\end{equation}
%Parameter $1-\alpha$ is the fraction of the transmit power spent for inter-node communication,
$g_{ij}$ represents fading channel coefficient from  ${\cal S}_{i}$ to ${\cal S}_{j}$ and $\eta_{ij}$ is receiver noise at ${\cal S}_{j}$. The channel variance is $\sigma^2_{{hs}_{ij}} = {\cal P} \mathcal{G}/d_{ij}^{\varepsilon}$, where  $d_{ij}$ denotes the distance between ${\cal S}_i$  and ${\cal S}_j$. Noises $\eta_{ij}$, $\eta_{ji}$  and channel coefficients $g_{ij}$, $g_{ji}$ are independent and noises are independent and identically distributed (i.i.d.) across all pairs. Upon receiving $r_{ij}$, ${\cal S}_{j}$ demodulates the channel input $u_i$, using the knowledge of $g_{ij}$
\begin{equation}
\hat{u}_i= sgn (\mathfrak{Re}(r_{ij}/g_{ij})).
\label{demod}
\end{equation}
The pair $({\cal S}_{i}$, ${\cal S}_{j})$ sends the information $ u_i,  \hat{u}_i, u_j,  \hat{u}_j$ to the FC in two consecutive time slots, exploiting Alamouti's STC scheme. In particular, in the $n$th slot, ${\cal S}_{i}$ and ${\cal S}_{j}$ send simultaneously $ \sqrt{\frac{\alpha}{2}} u_i$ and $ \sqrt{\frac{\alpha}{2}} {u}_j$, respectively. In the $(n+1)$th time slot, ${\cal S}_{i}$ and ${\cal S}_{j}$ send simultaneously $- \sqrt{\frac{\alpha}{2}} \hat{u}_j$ and $ \sqrt{\frac{\alpha}{2}} \hat{u}_i$, respectively. Considering the definitions of channel variances, we note that effectively ${\cal S}_{i}$ spends $(1-\alpha){\cal P} $ and $\alpha{\cal P}=\frac{\alpha}{2}{\cal P}+\frac{\alpha}{2}{\cal P}$, respectively, for inter-node and sensor-FC communication.
%=====================================
\subsection{Node-FC Communication Channel Model and Fusion Rule at the FC}\label{DSTC-node-FC}
Let $y_{ij}(n)$ and $y_{ij}(n+1)$ denote the received signals at the FC corresponding to the pair $({\cal S}_i$, ${\cal S}_j)$ during two consecutive time slots. We have
\begin{eqnarray}\label{sig-coop-dstc}
&& y_{ij}(n)=  \sqrt{\frac{\alpha}{2}} (u_i h_i+u_j h_j)+v_{ij}(n),~~
y_{ij}(n+1)=\sqrt{\frac{\alpha}{2}}(-\hat{u}_j h_i+\hat{u}_i h_j)+v_{ij}(n+1)\nonumber\\
&&\mbox{where}~~h_i \sim {\cal C}{\cal N}(0,\sigma^2_{h_i}), ~~ h_j \sim {\cal C}{\cal N}(0,\sigma^2_{h_j}), ~~v_{ij}(n), v_{ij}(n+1)\sim {\cal C}{\cal N}(0,\sigma^2_{v}).
\end{eqnarray}
The term $v_{ij}(n)$ is the receiver noise at the FC during the $n$th time slot. We assume noises $v_{ij}(n),v_{ij}(n+1)$ and  channel coefficients $h_i, h_j$ are independent and noises are i.i.d. across all pairs. Taking a similar processing as the Alamouti decoding \cite{c9}, the FC first forms $z_i,z_j$ using $y_{ij}(n), y_{ij}^*(n+1)$ as follows
\begin{eqnarray*}\label{alamouti-equations}
\left[
  \begin{array}{c}
    z_i \\
    z_j
  \end{array}
\right]&=&\left[
                  \begin{array}{cc}
                    h_i^* & h_j \\
                     h_j^* & -h_i
                  \end{array}
                \right] \left[
  \begin{array}{c}
    y_{ij}(n) \\
    y_{ij}^*(n+1)
  \end{array}
\right] =  \left[
                  \begin{array}{cc}
                    h_i^* & h_j \\
                     h_j^* & -h_i
                  \end{array}
                \right] \left[
  \begin{array}{c}
    v_{ij}(n) \\
    v_{ij}^*(n+1)\end{array}
\right] \nonumber
   \\ &+& \sqrt{\frac{\alpha}{2}}  \left(\left[
                  \begin{array}{cc}
                    |h_i|^2 &  h_j h_i^* \\
                    h_i h_j^* & |h_j|^2
                  \end{array}
                \right] \left[
                         \begin{array}{c}
                           u_i \\
                           u_j
                         \end{array}
                       \right]+\left[
                                 \begin{array}{cc}
                                   |h_j|^2 & -h_jh_i^* \\
                                   -h_ih_j^* & |h_i|^2
                                 \end{array}
                               \right]\left[
                                        \begin{array}{c}
                                          \hat{u}_i \\
                                          \hat{u}_j
                                        \end{array}
                                      \right]\right).
\end{eqnarray*}
Note that if each node allocates equal power for communicating with its cooperative partner and with the FC, i.e., $\alpha=1/2$ and also there is no error during inter-node communication, i.e., $\hat{u}_i=u_i, \hat{u}_j=u_j$, the above equations  reduce to the classical Alamouti's scheme \cite{c9}.
The new noise terms $\delta_{ij}^1\!=\!h_i^* v_{ij}(n)\!+\!h_j v^*_{ij}(n+1)$ and $\delta_{ij}^2\!=\!h_j^* v_{ij}(n)\!-\!h_i v^*_{ij}(n+1)$ are  i.i.d. zero mean complex Gaussian RVs with the variance $\sigma^2=(|h_i|^2+|h_j|^2)\sigma_v^2$. Next, using  the signals $z_i, z_j$ and the CSI $h_i, h_j$ for all pairs the FC forms the LRT $\Lambda= \frac{f(z_i,z_j~\mbox{for~all~pairs}|\mathcal{H}_1)}{f(z_i,z_j~\mbox{for~all~pairs}|\mathcal{H}_0)}$, where  $f(z_i,z_j~\mbox{for~all~pairs}|\mathcal{H}_{\ell})$  indicates the joint pdf of $z_i,z_j$ corresponding to all  pairs given the hypothesis $\mathcal{H}_{\ell}$, to make the final decision. In particular, the FC decides on $\mathcal{H}_{1}$ when $\Lambda >  \pi_0/{\pi_1}$ and decides on $\mathcal{H}_{0}$ otherwise. We note $\mathcal{H}_{\ell} \rightarrow u_i,u_j,\hat{u}_i,\hat{u}_j \rightarrow z_i,z_j$ and $\mathcal{H}_{\ell} \rightarrow u_i,u_j \rightarrow \hat{u}_i,\hat{u}_j$  form Markov chains. Also, $(z_i,z_j)$ are independent across the pairs given $u_i, u_j, \hat{u}_i, \hat{u}_j$ for all pairs.  Furthermore, given $u_i, u_j$ for for all pairs, $(\hat{u}_i, \hat{u}_j)$ are independent across the pairs. Therefore, we write
\begin{eqnarray}
\!\!\!\!\!\!\!\!\!\!\!\!\!\!\!\!&& f(z_i,z_j~\mbox{for~all~pairs}|\mathcal{H}_{\ell})= \sum_{u_i}~\sum_{u_j}~\sum_{\hat{u}_i}~\sum_{\hat{u}_j}  f(z_i,z_j~\mbox{for~all~pairs}|u_i, u_j, \hat{u}_i, \hat{u}_j~\mbox{for~all~pairs}) \nonumber\\  && \times  P(u_i, u_j, \hat{u}_i, \hat{u}_j~\mbox{for~all~pairs} |\mathcal{H}_{\ell}) \nonumber\\
\!\!\!\!\!\!\!\!\!\!\!\!\!\!\!\! && =\sum_{u_i} \sum_{\hat{u}_i}\sum_{u_j} \sum_{\hat{u}_j} \left(\prod_{\mbox{for~all~pairs}} f(z_i, z_j|u_i, u_j, \hat{u}_i, \hat{u}_j ~\mbox{for}~({\cal S}_i,{\cal S}_j)) \nonumber P(\hat{u}_i |u_i ~\mbox{for}~({\cal S}_i,{\cal S}_j))
P(\hat{u}_j |u_j ~\mbox{for}~({\cal S}_i,{\cal S}_j)) \right)\\ \!\!\!\!\!\!\!\!\!\!\!\!\!\!\!\! && \times P( u_i,u_j~\mbox{for~all~pairs} |\mathcal{H}_{\ell}),
\label{dstc_decoupled_pairs}
\end{eqnarray}
where the sums are taken over all values that $u_i, u_j, \hat{u}_i, \hat{u}_j$ can assume.
Focusing on $f(z_i, z_j|u_i, u_j, \hat{u}_i, \hat{u}_j ~\mbox{for}~({\cal S}_i,{\cal S}_j))$ in (\ref{dstc_decoupled_pairs})  we realize that  $z_i,z_j$ are conditionally independent complex Gaussian RVs with the variance $\sigma^2=(|h_i|^2+|h_j|^2)\sigma_v^2$ and the means $\mu_i,\mu_j$ given below
\begin{equation}
\mu_i =\sqrt{\frac{\alpha}{2}} (|h_i|^2u_i+h_j h_i^*u_j+|h_j|^2\hat{u}_i- h_j h_i^*\hat{u}_j),~~
\mu_j =\sqrt{\frac{\alpha}{2}} (h_ih_j^*u_i+|h_j|^2u_j-h_ih_j^*\hat{u}_i+|h_i|^2\hat{u}_j).
\label{dstc-mean}
\end{equation}
Focusing on the term $P(\hat{u}_i |u_i ~\mbox{for}~({\cal S}_i,{\cal S}_j))$ in (\ref{dstc_decoupled_pairs})  and considering (\ref{demod}) one can easily verify the following, assuming that the FC only knows the statistics of inter-node channels $g_{ij}, g_{ji}$ \cite{c19}
\begin{equation}
P(\hat{u}_{i}\neq u_i|u_{i})=1-P(\hat{u}_{i} = u_i|u_{i})=\frac{1}{2}\left(1-\sqrt{\frac{\bar{\gamma}_{{hs}_{ij}}}{1+\bar{\gamma}_{{hs}_{ij}}}}\right)~~\mbox{where}~~\bar{\gamma}_{{hs}_{ij}}=\frac{(1-\alpha)\sigma^2_{{hs}_{ij}}  }{\sigma^2_\eta}
\label{pd-pf-statistics}
\end{equation}
denotes the average received SNR corresponding to $({\cal S}_i,{\cal S}_j)$ inter-node communication.
The term $P( u_i,u_j~\mbox{for~all~pairs} |\mathcal{H}_{\ell})$ in (\ref{dstc_decoupled_pairs}) can be found in terms of the probability of $x_i,x_j$ being in certain intervals for all pairs. For instance, $P( u_i\!=\!1,u_j\!=\!-1~\mbox{for~all~pairs} |\mathcal{H}_{\ell})\!=\!P( x_i \!> \! \tau_i , x_j \!< \! \tau_j~\mbox{for~all~pairs} |\mathcal{H}_{\ell})$. These probabilities can be characterized for the sensing channel model in Section \ref{sensing-model}, where $x_i,x_j$ are jointly correlated Gaussian RVs with known statistics\footnote{For non-Gaussian $w_k$'s, the probability $P( u_i,u_j~\mbox{for~all~pairs} |\mathcal{H}_{\ell})$ should be calculated in terms of the joint pdf $f(w_1,...,w_K)$.}. When Gaussian sensing noises are uncorrelated  we obtain  $P( u_i\!=\!1,u_j\!=\!-1~\mbox{for~all~pairs} |\mathcal{H}_0)\!=\!\prod_{\{i: u_i=1\}} P_{f_{i}} \prod_{\{j: u_j=-1\}} (1-P_{f_{j}} )$ or $P( u_i\!=\!1,u_j\!=\!-1~\mbox{for~all~pairs} |\mathcal{H}_1)\!=\!\prod_{\{i: u_i=1\}} P_{d_{i}} \prod_{\{j: u_j=-1\}} (1-P_{d_{j}})$.
When the parameters of sensing channels are unavailable at the FC, the FC cannot apply the LRT. Alternatively, the FC demodulates the channel inputs for all pairs using the signals $z_i, z_j$ for all pairs and applies the majority rule to the demodulated symbols to reach the final decision.
%============================================================
\section{Cooperative Fusion Architecture with Signal Fusion at Sensors}\label{cooperative-local-fusion-scheme}
\subsection{Inter-node Communication Channel Model and Local Decision Rules at Sensors}\label{cooperative-local-fusion-scheme-A}
Suppose nodes ${\cal S}_{i}$ and ${\cal S}_{j}$ are within a pair. Each sensor makes an initial decision based on its measurement. The nodes exchange their decisions over orthogonal channels subject to noise and fading. Let $u_i$ denote the decision made at ${\cal S}_{i}$ based on $x_i$. When ${\cal S}_{i}$ transmits $\sqrt{1-\alpha}u_i$ the received signal at ${\cal S}_{j}$ is $r_{ij}$ as shown in (\ref{dstc-inter-sensor}). Upon receiving $r_{ij}$, ${\cal S}_{j}$ (rather than demodulating the channel input) updates its initial decision by fusing $r_{ij}$ and its measurement $x_j$. In particular, ${\cal S}_{j}$ forms a local LRT  $\tilde{\lambda}_{j}=\frac{f(r_{ij},x_{j}|\mathcal{H}_1)}{f(r_{ij},x_{j}|\mathcal{H}_0)}$, where $f(r_{ij},x_{j}|\mathcal{H}_{\ell})$ indicates the joint pdf of $r_{ij},x_{j}$ given the hypothesis $\mathcal{H}_{\ell}$, to make a new decision $\tilde{u}_{j}$. Node ${\cal S}_{j}$ lets $\tilde{u}_{j}=1$ when $\tilde{\lambda}_{j} > \pi_0/\pi_1$ and lets $\tilde{u}_{j}=-1$ otherwise. Since  ${\cal H}_{\ell} \rightarrow u_i \rightarrow r_{ij}$ and $x_j \rightarrow {\cal H}_{\ell}, u_i \rightarrow r_{ij}$ form Markov chains, we find
\begin{align}\label{lvd-local}
\tilde{\lambda}_{j}=\frac{\sum_{u_{i}}f(r_{ij}|u_{i})P(u_{i}|x_{j},\mathcal{H}_1)f(x_{j}|\mathcal{H}_1)}{\sum_{u_{i}}f(r_{ij}|u_{i})P(u_{i}|x_{j},\mathcal{H}_0)f(x_{j}|\mathcal{H}_0)}.
\end{align}
Considering $f(r_{ij}|u_{i})$ in (\ref{lvd-local}) we note that given $g_{ij}, u_{i}$ we have $r_{ij}\sim {\cal C}{\cal N}(\sqrt{1-\alpha}u_{i}g_{ij},\sigma_{\eta}^2)$. To find $P(u_i |x_j, \mathcal{H}_{\ell})$ in (\ref{lvd-local}) we note that for the sensing model in Section \ref{sensing-model},  $x_i, x_j$ are jointly Gaussian RVs with the mean ${\ell}$, the variances $\sigma_{w_i}^2, \sigma_{w_j}^2$ and the correlation coefficient $\rho_{i,j}$ under the hypothesis ${\cal H}_{\ell}$.  Using the joint pdf of $x_i, x_j$  and the Bayes' rule $P(u_{i}=1|x_{j},\mathcal{H}_{\ell})=\frac{P(u_{i}=1,x_{j}|\mathcal{H}_{\ell})}{f(x_{j}|\mathcal{H}_{\ell})}$ one can show
\begin{align}\label{prob_u_x}
&P(u_{i}=1|x_{j},\mathcal{H}_{\ell})=1-P(u_{i}=-1|x_{j},\mathcal{H}_{\ell})=Q\left(\frac{\tau_{i}-\rho_{i,j} x_{j} \frac{\sigma_{w_{i}}}{\sigma_{w_{j}}} - {\ell}(1-\rho_{i,j}\frac{\sigma_{w_{i}}}{\sigma_{w_{j}}}) }{\sqrt{(1-\rho_{i,j}^2)}\sigma_{w_{i}}}\right).
\end{align}
At last, we find $f(x_{j}|\mathcal{H}_{\ell})$ in (\ref{lvd-local}) by noting that given $\mathcal{H}_{\ell}$, we have $x_j \sim {\cal C}{\cal N}({\ell},\sigma_{w_j}^2)$. When Gaussian sensing noises are uncorrelated $u_{i}$ and $x_{j}$ are independent\footnote{For non-Gaussian sensing noises, when forming $\tilde{\lambda}_j$, $P(u_i |x_j, \mathcal{H}_{\ell})$ and $f(x_j|\mathcal{H}_{\ell})$ should be calculated, respectively, based on the joint pdf $f(w_i,w_j)$ and the pdf $f(w_j)$. Also, $P(u_{i}|x_{j},\mathcal{H}_{\ell})=P(u_{i}|\mathcal{H}_{\ell})$ when $w_i,w_j$ are mutually independent.} for a  given hypothesis $\mathcal{H}_{\ell}$, leading into  $P(u_{i}|x_{j},\mathcal{H}_{\ell})=P(u_{i}|\mathcal{H}_{\ell})$.
The pair $({\cal S}_{i}, {\cal S}_{j})$ sends $\sqrt{\alpha}\tilde{u}_{i},\sqrt{\alpha}\tilde{u}_{j}$ to the FC over two orthogonal  channels subject to noise and fading. Considering the definitions of channel variances, we note that effectively ${\cal S}_{i}$ spends $(1-\alpha){\cal P} $ and $\alpha{\cal P}$, respectively, for inter-node and sensor-FC communication.
%=======================================
\subsection{Node-FC Communication Channel Model and Fusion Rule at the FC}\label{scheme-local-fusion-comm-part}
Let $y_i$ and $y_j$ denote the received signals at the FC corresponding to the pair $({\cal S}_{i}, {\cal S}_{j})$. We have
\begin{equation*}
y_i=\sqrt{\alpha}\tilde{u}_{i} h_i + v_i,~~y_j=\sqrt{\alpha}\tilde{u}_{j} h_j + v_j, ~~h_i \sim {\cal C}{\cal N}(0, \sigma_{h_i}^2), ~~h_j \sim {\cal C}{\cal N}(0, \sigma_{h_j}^2), ~~ v_i, v_j \sim {\cal C}{\cal N}(0,\sigma_v^2).
\end{equation*}
The terms $v_i$ and $v_j$ are the receiver noises at the FC. We assume that noises and fading coefficients are independent and noises are i.i.d. across the pairs. Next, using  the signals $y_i, y_j$ and the CSI $h_i, h_j$ for all pairs the FC forms the LRT
$\Lambda= \frac{f({y}_i,{y}_j~\text{for all pairs}|\mathcal{H}_1)}{f({y}_i,{y}_j~\text{for all pairs}|\mathcal{H}_0)}$, where  $f({y}_i,{y}_j~\text{for all pairs}|\mathcal{H}_{\ell})$ indicates the joint pdf of $y_i, y_j$ corresponding to all pairs given the hypothesis $\mathcal{H}_{\ell}$, to make the final decision. We note  ${\cal H}_{\ell} \rightarrow \tilde{u}_i,\tilde{u}_j \rightarrow y_i,y_j$ form a Markov chain. Also,  $(y_i,y_j)$ are independent across the pairs given $\tilde{u}_i, \tilde{u}_j$ for all pairs. Hence, we can write
\begin{eqnarray}
&&\!\!f({y}_i,{y}_j~\mbox{for all pairs}|\mathcal{H}_{\ell})=
 \sum_{\tilde{u}_i}~\sum_{\tilde{u}_j}  f({y}_i,{y}_j~\mbox{for all pairs}|\tilde{u}_i,\tilde{u}_j\mbox{ for all pairs})
P(\tilde{u}_i,\tilde{u}_j~\mbox{for all pairs}|\mathcal{H}_{\ell})\nonumber\\
&&\!=\!\sum_{\tilde{u}_i}~\sum_{\tilde{u}_j}\left(\prod_{\mbox{for all pairs}} f({y}_i, y_j |\tilde{u}_i, \tilde{u}_j ~\mbox{for}~({\cal S}_{i}, {\cal S}_j))\right) P(\tilde{u}_i,\tilde{u}_j~\mbox{for all pairs}|\mathcal{H}_{\ell}),
\label{lvd_decoupled_pairs}
\end{eqnarray}
where the sums are taken over all values that $\tilde{u}_i,\tilde{u}_j$ can take.
To find $f(y_i, y_j |\tilde{u}_i, \tilde{u}_j ~\mbox{for}~({\cal S}_{i}, {\cal S}_j))$ in (\ref{lvd_decoupled_pairs}) we realize that in the pair $({\cal S}_{i}, {\cal S}_j)$ given $\tilde{u}_i, \tilde{u}_j$, the variables $y_i$ and $y_j$ are independent complex Gaussian RVs with  the means $\mu_i= \sqrt{\alpha}\tilde{u}_{i}h_{i}, \mu_j= \sqrt{\alpha}\tilde{u}_{j}h_{j}$ and the variance $\sigma_v^2$. To obtain $P(\tilde{u}_i,\tilde{u}_j~\mbox{for all pairs}|\mathcal{H}_{\ell})$ in (\ref{lvd_decoupled_pairs}) we assume that the FC only knows the statistics of inter-node channels $g_{ij}, g_{ji}$. Note that ${\cal H}_{\ell}\rightarrow x_i,x_j,r_{ij},r_{ji} \rightarrow \tilde{u}_i,\tilde{u}_j$ forms a Markov chain. Therefore
\begin{eqnarray}
&&P(\tilde{u}_i,\tilde{u}_j~\mbox{for all pairs}|\mathcal{H}_{\ell})= \nonumber \\
&& \int_{g_{ij}} \int_{g_{ji}} \int_{x_i}\int_{x_j}\int_{r_{ij}}\int_{r_{ji}} P(\tilde{u}_i,\tilde{u}_j~\mbox{for all pairs}| g_{ij}, g_{ji}, x_i, x_j, r_{ij},r_{ji}~\mbox{for all pairs}) \nonumber \\ && \times
f(x_i, x_j, r_{ij}, r_{ji}~\mbox{for all pairs}| \mathcal{H}_{\ell}) f(g_{ij}, g_{ji})  d g_{ij} d g_{ji} d x_i d x_j d r_{ij} d r_{ji}.\label{second-term1-local-fusion}
\end{eqnarray}
Equation (\ref{second-term1-local-fusion}) can be further simplified by noting that given $x_{i}, x_j, r_{ij}, r_{ji}$ for all pairs, the variables $(\tilde{u}_{i},\tilde{u}_{j})$ are independent across the pairs. Furthermore, within a pair, given $x_{i}, x_j, r_{ij}, r_{ji}$, the variables $\tilde{u}_{i}$ and $\tilde{u}_{j}$ are independent. Also, we note that ${\cal H}_{\ell} \rightarrow x_i, x_j \rightarrow r_{ij},r_{ji}$ forms a Markov chain and given $x_i,x_j$ for all pairs, $(r_{ij},r_{ji})$ are independent across the pairs. Besides, within a pair, given $x_i, x_j$, the variables $r_{ij}, r_{ji}$ are independent. Combining all, we can rewrite (\ref{second-term1-local-fusion}) as
\begin{eqnarray}
&&\!\!\!\! \int_{g_{ij}} \int_{g_{ji}} \int_{x_i}\int_{x_j}\int_{r_{ij}}\int_{r_{ji}} \big( \prod_{\mbox{for all pairs}} P(\tilde{u}_{i}| x_{i},r_{ji}~\mbox{for}~({\cal S}_{i},{\cal S}_{j})) P(\tilde{u}_{j}| x_{j},r_{ij}~\mbox{for}~({\cal S}_{i},{\cal S}_{j})) \label{second-term2}\\
&&\!\!\!\!  \times f(r_{ji}| x_{j}~\mbox{for}~({\cal S}_{i},{\cal S}_{j})) f(r_{ij}|x_{i}~\mbox{for}~({\cal S}_{i},{\cal S}_{j})) \big) f({x}_i,{x}_j\mbox{ for all pairs}|\mathcal{H}_{\ell}) f(g_{ij}, g_{ji})  d g_{ij} d g_{ji}  d x_i d x_j d r_{ij} d r_{ji}. \nonumber
\end{eqnarray}
Focusing on the term $P(\tilde{u}_{i}|x_{i},r_{ji}~\mbox{for}~({\cal S}_{i},{\cal S}_{j}))$ in (\ref{second-term2}) we note that $P(\tilde{u}_{i} =1 | x_{i},r_{ji})=\bone_{\{\tilde{\lambda}_{i} > \pi_0/\pi_1\}}$ and $P(\tilde{u}_{i} =-1 | x_{i},r_{ji})=\bone_{\{\tilde{\lambda}_{i} < \pi_0/\pi_1\}}$ where $\tilde{\lambda}_{i}$ depends on the inter-node channels $g_{ji}$, the threshold $\tau_j$, the sensing noise variances $\sigma_{w_i}^2, \sigma_{w_j}^2$ and the correlation coefficient $\rho_{i,j}$.
Similarly, we can find $P(\tilde{u}_{j}|x_{j},r_{ij}~\mbox{for}~({\cal S}_{i},{\cal S}_{j}))$ in (\ref{second-term2}).  Considering the term $f(r_{ji}|x_{j}~\mbox{for}~({\cal S}_{i},{\cal S}_{j}))$ in (\ref{second-term2}) we find
\begin{eqnarray}
f(r_{ji}|x_{j}~\mbox{for}~({\cal S}_{i},{\cal S}_{j}))%&=& f(r_{ji},u_{j}=1|x_{j})+f(r_{ji},u_{j}=-1|x_{j})\nonumber \\
\!&\!=\!&\!f(r_{ji}|u_{j}=1)P(u_{j}=1|x_{j})+f(r_{ji}|u_{j}=-1)P(u_{j}=-1|x_{j})\nonumber \\
\!&\!=\!&\!f(r_{ji}|u_{j}=1) \bone_{\{x_j > \tau_j\}} +f(r_{ji}|u_{j}=-1)\bone_{\{x_j <  \tau_j\}}
\label{coop-fusion-sensor}
\end{eqnarray}
where $f(r_{ji}|u_{j})$ in (\ref{coop-fusion-sensor}) can be found noting that
given $g_{ji},u_j$ we have $r_{ji}\sim {\cal C}{\cal N}(\sqrt{1-\alpha}u_j g_{ji},\sigma_{\eta}^2)$.
Considering the term $f(x_i,x_j\mbox{ for all pairs}|\mathcal{H}_{\ell})$ in (\ref{second-term2}) we note that when Gaussian sensing noises are uncorrelated\footnote{For non-Gaussian sensing noises, $\tilde{\lambda}_i$ would change as explained in the previous footnote. Also, $P(u_{j}|x_{j})$ in (\ref{coop-fusion-sensor}) and $f(x_i,x_j\mbox{ for all pairs}|\mathcal{H}_{\ell})$  in (\ref{second-term2}), respectively, should be calculated based on the pdf $f(w_j)$ and the joint pdf $f(w_1,...,w_K)$.} we obtain  $f(x_i,x_j\mbox{ for all pairs}|\mathcal{H}_{\ell})=\prod_{\mbox{ for all pairs}} f(x_{i}\mbox{ for }({\cal S}_{i},{\cal S}_{j})|\mathcal{H}_{\ell})$ $f(x_{j}\mbox{ for }({\cal S}_{i},{\cal S}_{j})|\mathcal{H}_{\ell})$.  Combining all these one can verify that the probability in (\ref{second-term1-local-fusion}) depends on sensing channels through $\tau_i,\tau_j, \sigma_{w_i}^2, \sigma_{w_j}^2, \rho_{i,j}$ and the average received SNR  $\bar{\gamma}_{{hs}_{ij}}$ corresponding to $({\cal S}_{i}, {\cal S}_j)$ inter-node communication.
When the parameters of sensing channels are unavailable at the FC, the FC demodulates the channel inputs for all pairs using the signals $y_i, y_j$ for all pairs and applies the majority rule to the demodulated symbols to reach the final decision.
%============================================================
\section{Parallel Fusion Architecture with Local Threshold Changing at Sensors}\label{parallel-threshold-chaning-scheme}
\subsection{Local Decision Rules at Sensors}\label{parallel-threshold-chaning-scheme-local-rules}
Suppose nodes ${\cal S}_{i}$ and ${\cal S}_{j}$ are within a pair. Each sensor makes an initial decision based on its measurement. Let $u_i$ denote the decision made at ${\cal S}_{i}$ based on $x_i$. In the absence of inter-node communication, ${\cal S}_{i}$ assumes that the decision $u_j$ (made at ${\cal S}_{j}$ based on $x_j$) is different from $u_i$, i.e., ${\cal S}_{i}$ assumes
$u_j=-u_i$. Next, ${\cal S}_{i}$ forms another decision $\bar{u}_i$ by fusing the assumed decision $u_j$ and its measurement $x_i$. In particular, ${\cal S}_{i}$ forms a local LRT $\bar{\lambda}_i=\frac{f(x_{i},u_{j}=-u_{i}|\mathcal{H}_1)}{f(x_{i},u_{j}=-u_{i}|\mathcal{H}_0)}$, where $f(x_{i},u_{j}=-u_{i}|\mathcal{H}_{\ell})$  indicates the joint pdf of  $x_i$ and the assumed decision  $u_j$ at ${\cal S}_{i}$ given the hypothesis $\mathcal{H}_{\ell}$, to make $\bar{u}_i$. Node ${\cal S}_{i}$  lets $\bar{u}_i=1$ when $\bar{\lambda}_i > \pi_0/ \pi_1$ and lets $\bar{u}_i=-1$ otherwise. We find\footnote{Consider a hypothetical case where ${\cal S}_i$  assumes $u_j\!=\!-1$ and makes a decision $u_i^0$ by optimally fusing the assumed $u_j\!=\!-1$ and  $x_i$.
In particular, ${\cal S}_i$  lets $u_i^0\!=\!1$ when
$\lambda_i^0\!=\!\frac{f(x_i,u_j\!=\!-1|\mathcal{H}_1)}{f(x_i,u_j\!=\!-1|\mathcal{H}_0)} \!>\!\frac{\pi_0}{\pi_1}$ and lets $u_i^0\!=\!-1$ otherwise.
Consider another hypothetical case where ${\cal S}_i$  assumes $u_j\!=\!1$ and makes a decision $u_i^1$ by optimally fusing the assumed $u_j\!=\!1$ and  $x_i$. In particular, ${\cal S}_i$ lets $u_i^1\!=\!1$ when
$\lambda_i^1\!=\!\frac{f(x_i,u_j\!=\!1|\mathcal{H}_1)}{f(x_i,u_j\!=\!1|\mathcal{H}_0)}\!>\!\frac{\pi_0}{\pi_1}$ and lets $u_i^1\!=\!-1$ otherwise.
One can verify
$u_i^0\!=\!1$ when
$x_i\!>\!\tau_{i_1}'$ and $u_i^0\!=\!-1$ otherwise, also $u_i^1\!=\!1$ when $x_i\!>\!\tau_{i_2}'$ and $u_i^1\!=\!-1$ otherwise,
where $\tau_{i_1}'\!=\!0.5\!+\!\sigma^2_{w_i}\ln (\frac{P(u_j\!=\!-1|x_i,\mathcal{H}_0)\pi_0}{P(u_j\!=\!-1|x_i,\mathcal{H}_1)\pi_1})$ and $\tau_{i_2}'\!=\!0.5\!+\!\sigma^2_{w_i}\ln(\frac{P(u_j\!=\!1|x_i,\mathcal{H}_0)\pi_0}{P(u_j=1|x_i,\mathcal{H}_1)\pi_1})$ and $\tau_{i_2}' < \tau_i < \tau_{i_1}'$. For these hypothetical cases, now suppose $x_i\!<\!\tau_i$ and thus $u_i\!=\!-1$. Since $\tau_i\!<\!\tau_{i_1}'$ we have $x_i\!<\!\tau_{i_1}'$ and thus $u_i^0\!=\!-1$, i.e., $u_i^0\!=\!u_i$, whereas $u_i^1$ can $\pm 1$. Therefore, the useful information is embedded in $u_i, u_i^1$, as $u_i^0$ does not convey extra information. Similarly, one can argue that when $x_i\!>\!\tau_i$ the useful information is imbedded in $u_i, u_i^0$, as $u_i^1$ does not convey extra information. In conclusion, node ${\cal S}_i$ should assume $u_j\!=\!-u_i$ to be able to extract more information from $x_i$.}
\begin{equation}\label{threshold-changing}
\bar{\lambda}_i =\frac{P(u_j=-u_i|x_i,\mathcal{H}_1)f(x_i|\mathcal{H}_1)}{P(u_j=-u_i|x_i,\mathcal{H}_0)f(x_i|\mathcal{H}_0)}.
\end{equation}
in which $P(u_j=-u_i|x_i,\mathcal{H}_{\ell})$ is given in (\ref{prob_u_x}) and $f(x_{j}|\mathcal{H}_{\ell})$ is found by noting\footnote{For non-Gaussian sensing noises, similar to $\tilde{\lambda}_i$ in Section \ref{cooperative-local-fusion-scheme-A}, $\bar{\lambda}_i$ would change. In particular, $P(u_j|x_i,\mathcal{H}_\ell)$ and $f(x_i|\mathcal{H}_\ell)$ should be calculated, respectively, based on the joint pdf $f(w_i,w_j)$ and the pdf $f(w_i)$.}
that given $\mathcal{H}_{\ell}$, we have $x_j \sim {\cal C}{\cal N}({\ell},\sigma_{w_j}^2)$. In fact, one can verify that ${\cal S}_i$ finds $u_i, \bar{u}_i$ as the following
\begin{eqnarray}\label{array-threshold-changing}
\left
\{\begin{array}{ll}
 u_i = 1, ~  \bar{u}_i= 1     ~~~\mbox{if}~ x_i > \tau'_{i_1}, &
 u_ i =  -1,~ \bar{u}_i= -1   ~~~\mbox{if}~ x_i < \tau'_{i_2}\\
  u_i = -1, ~  \bar{u}_i= 1    ~~~\mbox{if}~ \tau'_{i_2} < x_i < \tau_i,& ~~~
 u_i=1, ~   \bar{u}_i= -1   ~~~ \mbox{if}~ \tau_{i} < x_i < \tau'_{i_1}
\end{array}\right.
\end{eqnarray}
where the thresholds $\tau'_{i_1}, \tau'_{i_2}$, given in the footnote, depend on $\sigma_{w_i}^2, \rho_{i,j}$ and satisfy $\tau'_{i_2} < \tau_i <\tau'_{i_1}$.
When Gaussian sensing noises are uncorrelated the assumed decision $u_j$ and $x_i$  are independent for a given hypothesis $\mathcal{H}_{\ell}$, leading into $P(u_{j}=-u_{i}|x_i,\mathcal{H}_{\ell})=P(u_{j}=-u_{i}|\mathcal{H}_{\ell})$ in (\ref{threshold-changing}). Consequently, the local LRT $\bar{\lambda}_i$ in (\ref{threshold-changing}) can be further simplified and $\tau'_{i_1}, \tau'_{i_2}$ in (\ref{array-threshold-changing}), respectively, reduce to
%it can be easily verified that node ${\cal S}_{i}$  lets $\bar{u}_i=1$ when $x_i > \tau_{{i}_1}$ or $\tau_{{i}_2} < x_i < \tau_i$ and lets $\bar{u}_i=-1$ when $x_i < \tau_{{i}_2}$ or $\tau_{i} < x_i < \tau_{{i}_1}$,
$\tau_{{i}_1}=0.5+\sigma_{w_{i}}^2\ln(\frac{(1-P_{f_{j}})\pi_0}{(1-P_{d_{j}})\pi_1})$ and $\tau_{{i}_2}=0.5+\sigma_{w_{i}}^2 \ln(\frac{P_{f_{j}}\pi_0}{P_{d_{j}}\pi_1})$ where $\tau_{{i}_2} < \tau_{i} <\tau_{{i}_1}$.
Since in addition to threshold $\tau_i$ employed in scheme (i),  $x_i$ is also compared against two additional thresholds $\tau'_{i_1},\tau'_{i_2}$, we refer to scheme (iii) as  ``local threshold changing''.
The pair $({\cal S}_{i}$, ${\cal S}_{j})$ sends  $ u_i, \bar{u}_i, u_j, \bar{u}_j$ to the FC in two consecutive time slots, exploiting Alamouti's STC scheme. In particular,
in the $n$th slot, ${\cal S}_{i}$ and ${\cal S}_{j}$ can send simultaneously $\frac{u_i}{\sqrt{2}}$ and $\frac{\bar{u}_j }{\sqrt{2}}$, respectively. In the $(n+1)$th time slot, ${\cal S}_{i}$ and ${\cal S}_{j}$ can send simultaneously $- \frac{\bar{u}_i }{\sqrt{2}}$ and $\frac{u_j}{\sqrt{2}}$, respectively. Considering the definitions of channel variances, we note that effectively ${\cal S}_{i}$ spends ${\cal P}= \frac{1}{2}{\cal P}+\frac{1}{2}{\cal P}$ for sensor-FC communication.
%=======================================================
\subsection{Node-FC Communication Channel Model and Fusion Rule at FC}\label{scheme-threshold-chaning-at-sensors}
Let $y_{ij}(n)$ and $y_{ij}(n+1)$ denote the received signals at the FC corresponding to the pair $({\cal S}_{i}, {\cal S}_{j})$ during two consecutive time slots. The signal model in (\ref{sig-coop-dstc}) still holds true, after substituting $\sqrt{\frac{\alpha}{2}}$ with $\frac{1}{\sqrt{2}}$, $u_j$ with  $\bar{u}_j$, $\hat{u}_j$ with $\bar{u}_i$, and $\hat{u}_i$ with $u_j$. We can take similar steps as in Section \ref{DSTC-node-FC} to find the signals $z_i, z_j$ using $y_{ij}(n), y_{ij}^*(n+1)$. Next, using  the signals $z_i, z_j$ and the CSI $h_i, h_j$ for all pairs the FC forms the LRT $\Lambda=\frac{f(z_i,z_j~\mbox{for~all~pairs}|\mathcal{H}_{1})}{f(z_i,z_j~\mbox{for~all~pairs}|\mathcal{H}_{0})}$. We note $\mathcal{H}_{\ell} \rightarrow u_i,u_j,\bar{u}_i,\bar{u}_j \rightarrow z_i,z_j$ forms a Markov chain. Also, $(z_i,z_j)$ are independent across the pairs given $u_i, u_j, \bar{u}_i, \bar{u}_j$ for all pairs.  Therefore
\begin{eqnarray}\label{joint-z-threshold}
\!\!\!\!\!\!\!\!\!\!\!\!\!\!\!\!&& f(z_i,z_j~\mbox{for~all~pairs}|\mathcal{H}_{\ell})= \sum_{u_i~\mbox{for~all~pairs}}~\sum_{u_j~\mbox{for~all~pairs}}~\sum_{\bar{u}_i~\mbox{for~all~pairs}}~\sum_{\bar{u}_j~\mbox{for~all~pairs}} \nonumber \\
\!\!\!\!\!\!\!\!\!\!\!\!\!\!\!\!&& \left(\prod_{\mbox{for~all~pairs}} f(z_i, z_j|u_i, u_j, \bar{u}_i, \bar{u}_j ~\mbox{for}~({\cal S}_i,{\cal S}_j)) \right) P(u_i, u_j, \bar{u}_i, \bar{u}_j~\mbox{for~all~pairs} |\mathcal{H}_{\ell}).
\end{eqnarray}
Focusing on the term  $f(z_i, z_j|u_i, u_j, \bar{u}_i, \bar{u}_j ~\mbox{for}~({\cal S}_i,{\cal S}_j)) $ in (\ref{joint-z-threshold}) we realize that $z_i,z_j$ are conditionally independent complex Gaussian RVs with the variance $\sigma^2=(|h_i|^2+|h_j|^2)\sigma_v^2$ and the means $\mu_i,\mu_j$ given in (\ref{dstc-mean}), after substituting $\sqrt{\frac{\alpha}{2}}$ with $\frac{1}{\sqrt{2}}$, $u_j$ with  $\bar{u}_j$, $\hat{u}_j$ with $\bar{u}_i$, and $\hat{u}_i$ with $u_j$. To find $P(u_i,u_j,\bar{u}_i,\bar{u}_j\mbox{ for all pairs}|\mathcal{H}_{\ell})$ we note this term can be expressed in terms of
the probability of $x_i,x_j$ being in certain intervals for all pairs.
For instance, $P(u_{i}\!=\!\bar{u}_{i}\!=\! 1, u_{j}\!=\!\bar{u}_{j}\!=\!-1\mbox{ for all pairs} |\mathcal{H}_{\ell} )=P(x_{i}\!>\!\tau_{{ij}_1}, x_{j}\!<\!\tau_{{ij}_2} \mbox{ for all pairs}{  |\mathcal{H}_{\ell}} )$.
Now, these probabilities can be easily characterized for the sensing channel model in Section \ref{sensing-model}, where $x_i,x_j$ are jointly correlated Gaussian RVs with known statistics\footnote{For non-Gaussian $w_k$'s, $P(u_i,u_j,\bar{u}_i,\bar{u}_j\mbox{ for all pairs}|\mathcal{H}_{\ell})$ should be calculated in terms of the joint pdf $f(w_1,...,w_K)$.}. When Gaussian sensing noises are uncorrelated the decisions $(u_{i},\bar{u}_{i})$ and $(u_{j},\bar{u}_{j})$ given hypothesis $\mathcal{H}_{\ell}$ are independent across the pairs. Therefore
$P(u_i,u_j,\bar{u}_i,\bar{u}_j\mbox{ for all pairs}|\mathcal{H}_{\ell})=
\prod_{\mbox{for all pairs}} P(u_{i},\bar{u}_{i}\mbox{ for }({\cal S}_{i}, {\cal S}_{j})|\mathcal{H}_{\ell})P(u_{j},\bar{u}_{j}\mbox{ for }({\cal S}_i,{\cal S}_{j})|\mathcal{H}_{\ell})$. When the parameters of sensing channels are unavailable at the FC, the FC demodulates the channel inputs for all pairs using the signals $z_i, z_j$ for all pairs and applies the majority rule to the demodulated symbols to reach the final decision.
%========================================================
\section{Performance Analysis}\label{big-section-analysis-performance}
%This section presents our performance analysis. In particular, in
Section \ref{per-analysis-DSTC} provides an upper bound on the average error probability for scheme (i) of Section \ref{section-DSTC}. Leveraging on this, sections \ref{per-analysis-classical-parallel} and \ref{per-analysis-cooperative-local-fusion}, respectively, provide upper bounds on the average error probability for the classical parallel fusion architecture of Section \ref{parallel-classical} and scheme (ii) of Section \ref{cooperative-local-fusion-scheme}.
For mathematical tractability, we assume  that the Gaussian sensing noises $w_k$ are identically distributed and uncorrelated\footnote{The derivations in this section also hold true for i.i.d sensing noises.}, i.e., we  have  $\sigma_{w_k}^2\!= \!\sigma_{w}^2, \tau_k\!=\!\tau$ and thus $P_{d_k}\!=\!P_d, P_{f_k}\!=\!P_f$. Also, we assume that sensors are positioned equally distant from the FC and thus $\bar{\gamma}_{h}^2=\frac{\sigma_{h}^2}{\sigma_v^2}$. Also, distances between the cooperative partners are assumed equal across the pairs and therefore $\bar{\gamma}_{hs}^2=\frac{(1-\alpha)\sigma_{hs}^2}{\sigma_{\eta}^2}$.
%Unfortunately, the upper bound derivations for scheme (iii) of Section \ref{parallel-threshold-chaning-scheme} are intractable.
%In Appendix B we show that when  $\pi_0, \sigma_{w}^2$ are large, scheme (iii) of Section \ref{parallel-threshold-chaning-scheme} outperforms the classical parallel fusion architecture of Section \ref{parallel-classical}.
In Section \ref{simulation-section} we validate the analytical results of this section via Monte-Carlo simulations.
%=================================================
\subsection{Cooperative Fusion Architecture with STC at Sensors}\label{per-analysis-DSTC}
For performance analysis, suppose each pair of cooperative partners $({\cal S}_i, {\cal S}_j)$ is associated with a unique group index $s$ where $s\!=\!1,...,S$ and $S\!=\! K/2$. We denote the two nodes within the group $s$ as $({\cal S}_{2s-1}, {\cal S}_{2s})$, i.e., we map the indices $i,j$ in Section \ref{section-DSTC} into $2s-1,2s$, respectively.
For LRT fusion rule in Section \ref{DSTC-node-FC}, the conditional error probability is $P_{e|h}\!=\!P_{e_1|h}\!+\!P_{e_2|h}$  where $P_{e_1|h}\!=\!P(\Lambda>\frac{\pi_0}{\pi_1}|{\cal H}_0)\pi_0$ and $P_{e_2|h}\!=\!P(\Lambda<\frac{\pi_0}{\pi_1}|{\cal H}_1)\pi_1$ and
$\Lambda\!=\!\frac{f(z_{2s-1},z_{2s}~\tiny{\mbox{for}}~s=1,...,S|{\cal H}_1)}{f(z_{2s-1},z_{2s}~\tiny{\mbox{for}}~s=1,...,S|{\cal H}_0)}$, conditioned on the channel coefficients $h_{2s-1},h_{2s}$ for $s=1,..., S$ at the FC. The average error probability $\bar{P_e}\!=\!\bar{P}_{e_1}\!+\!\bar{P}_{e_2}$ is obtained by taking the averages of $P_{e_1|h}, P_{e_2|h}$ over the distribution of the channel coefficients. Our goal is to provide upper bounds on $P_{e_1|h}, P_{e_2|h}$ and their corresponding averages $\bar{P}_{e_1}\!=\! \mathbb{E}\{P_{e_1|h}\}, \bar{P}_{e_2}\!=\!\mathbb{E}\{P_{e_2|h}\}$. We use the following notation in our derivations.
To capture all different values that $u_{2s-1}, u_{2s}, \hat{u}_{2s-1},\hat{u}_{2s}$ for $s\!=\!1,...,S$ can take we consider two $K$-length sequences $(a^1_{n_1},a^2_{n_1},...,a^{2s-1}_{n_1},a^{2s}_{n_1}, ..., a^{2S-1}_{n_1}, a^{2S}_{n_1})$ and $(a^1_{m_1},a^2_{m_1},...,a^{2s-1}_{m_1},a^{2s}_{m_1}, ..., a^{2S-1}_{m_1}, a^{2S}_{m_1})$ where $a^{2s-1}_{n_1},a^{2s}_{n_1} \in \{1,-1\}$ are the values assumed by $u_{2s-1}, u_{2s}$ and  $a^{2s-1}_{m_1},a^{2s}_{m_1} \in \{1,-1\}$ are the values assumed by $\hat{u}_{2s-1}, \hat{u}_{2s}$. Also, let $n_1$ and $m_1$, respectively, be the decimal numbers corresponding to the two $K$-length binary sequences, when those $a^k_{n_1}$ and $a^k_{m_1}$ assuming $-1$ value in the sequences are reassigned 0 value.  Let $Q_{n_1}$ denote the number of ones in the sequence $(a^1_{n_1},a^2_{n_1},...,a^{2s-1}_{n_1},a^{2s}_{n_1}, ..., a^{2S-1}_{n_1}, a^{2S}_{n_1})$. Define $F_{n_1,m_1}, F_{n_1}, T_{n_1,m_1}, d_{n_1,m_1}$ as below
\begin{eqnarray}
F_{n_1,m_1}\!&\!=\!&\!\{ u_{2s-1}= a^{2s-1}_{n_1}, u_{2s}=a^{2s}_{n_1}, \hat{u}_{2s-1}=a^{2s-1}_{m_1}, \hat{u}_{2s}=a^{2s}_{m_1}~\mbox{for}~s\!=\!1,...,S\} \\
F_{n_1}\!&\!=\!&\! \{u_{2s-1}=a^{2s-1}_{n_1},u_{2s}=a^{2s}_{n_1}~\mbox{for}~s\!=\!1,...,S\} \label{definition-of-Fn}\\
T_{n_1,m_1}\!&\!=\!&\!  \prod_{s=1}^SP(\hat{u}_{2s-1}=a^{2s-1}_{m_1}|u_{2s-1}=a^{2s-1}_{n_1})P(\hat{u}_{2s}=a^{2s}_{m_1}|u_{2s}=a^{2s}_{n_1}) \\
d_{n_1,m_1}&=&(\pi_1P_d^{Q_{n_1}}(1-P_d)^{K-Q_{n_1}}-\pi_0P_f^{Q_{n_1}}(1-P_f)^{K-Q_{n_1}})T_{n_1,m_1}\nonumber\\
&\times & \prod_{s=1}^S f(z_{2s-1},z_{2s}|u_{2s-1}=a^{2s-1}_{n_1},u_{2s}=a^{2s}_{n_1},\hat{u}_{2s-1}=a^{2s-1}_{m_1},\hat{u}_{2s}=a^{2s}_{m_1}).\label{d-definition}
\end{eqnarray}
Furthermore, let $M$ be an integer that satisfies $\frac{P_d^M(1-P_d)^{K-M}}{P_f^M(1-P_f)^{K-M}}\!>\!\frac{\pi_0}{\pi_1}$ and $\frac{P_d^{M-1}(1-P_d)^{K-M+1}}{P_f^{M-1}(1-P_f)^{K-M+1}}\!<\!\frac{\pi_0}{\pi_1}$, i.e.,
$M$ is the smallest number of nodes that can decide ${\cal H}_1$ while ${\cal H}_1$ is true and the FC decides correctly, had all communication channels were error-free. Define the sets $S_{0}\!=\!\{d_{n_1,m_1}~\mbox{where}~Q_{n_1}\! < \!M\}$ and $S_{1}\!=\!\{d_{n_1,m_1}~\mbox{where}~Q_{n_1}\!\geq \! M\}$. Note that all entries of $S_{0}$ and $S_{1}$ are, respectively, negative and positive. Let $|S_0|$ and $|S_1|$ denote the cardinalities of $S_0$ and $S_1$, respectively. Utilizing the above notations, we can rewrite $f(z_{2s-1},z_{2s}~{\mbox{for}}~s=1,...,S|{\cal H}_{\ell})$ in $\Lambda$ as
\begin{eqnarray}\label{performance-dstc}
\!\!\!\!\!\!\!  f(z_{2s-1},z_{2s}~{\mbox{for}}~s=1,...,S|{\cal H}_{\ell}) \!\!&\!\!=\!\!&\!\!\underset{n_1,m_1}{\sum}P(F_{n_1}|{\cal H}_{\ell})T_{n_1,m_1} \\
 \!\!\!\!\!\!\!    \!\!&\!\! \times \!\!&\!\! {\prod}_{s=1}^S f(z_{2s-1},z_{2s}|u_{2s-1}=a^{2s-1}_{n_1},u_{2s}=a^{2s}_{n_1},\hat{u}_{2s-1}=a^{2s-1}_{m_1},\hat{u}_{2s}=a^{2s}_{m_1}) .\nonumber
\end{eqnarray}
Since sensing noises are identically distributed and uncorrelated we find $P(F_{n_1}|{\cal H}_{\ell})\!=\!{\prod}_{s=1}^S P(u_{2s-1}\!=\!a^{2s-1}_{n_1}|{\cal H}_{\ell})P(u_{2s}\!=\!a^{2s}_{n_1}|{\cal H}_{\ell})$ in (\ref{performance-dstc}), and thus $P(F_{n_1}|{\cal H}_{1}) \!=\! P_d^{Q_{n_1}}(1-P_d)^{K-Q_{n_1}}$ and $P(F_{n_1}|{\cal H}_{0}) \!=\! P_f^{Q_{n_1}}(1-P_f)^{K-Q_{n_1}}$. The term $T_{n_1,m_1}$ in (\ref{performance-dstc}) is calculated using (\ref{pd-pf-statistics}) and depends on the average received SNR $\bar{\gamma}_{hs}$ corresponding to inter-node communication. Combining (\ref{d-definition}), (\ref{performance-dstc}) we have
\begin{eqnarray}
\!\!\!\!\!\!\!\!\!\!\!\!P_{e_1|h}\!\!&\!\!=\!\!&\!\!\pi_0P(\sum_{n_1,m_1}d_{n_1,m_1}>0|{\cal H}_0)\!=\! \pi_0\sum_{n,m}{\cal T}_{e_1|h}
P(F_{n,m}|\mathcal{H}_0) \!=\! \pi_0\sum_{n,m} {\cal T}_{e_1|h}
P_f^{Q_{n}}(1-P_f)^{K-Q_{n}}T_{n,m}\label{pe1-first}\\
\!\!\!\!\!\!\!\!\!\!\!\!P_{e_2|h}\!\!&\!\!=\!\!&\!\! \pi_1P(\sum_{n_1,m_1}d_{n_1,m_1}<0|{\cal H}_1)=\pi_1\sum_{n,m} {\cal T}_{e_2|h}
P(F_{n,m}|{\cal H}_1)=\pi_1\sum_{n,m} {\cal T}_{e_2|h} P_d^{Q_n}(1-P_d)^{K-Q_{n}} T_{n,m}.\label{pe2-upper}
\end{eqnarray}
where ${\cal T}_{e_1|h}\!=\!P(\sum_{n_1,m_1}d_{n_1,m_1}\!>\!0|F_{n,m})$ in (\ref{pe1-first}) and ${\cal T}_{e_2|h}\!=\!1\!-\!{\cal T}_{e_1|h}$ in (\ref{pe2-upper}). Note that $P_{e_1|h}$ in (\ref{pe1-first})  and $P_{e_2|h}$ in (\ref{pe2-upper}) depend on $h_{2s-1},h_{2s}$ for $s\!=\!1,..., S$ only through  ${\cal T}_{e_1|h}$ and ${\cal T}_{e_2|h}$, respectively. These imply that the problem of finding upper bounds on $P_{e_1|h}, P_{e_2|h}$ and their corresponding averages $\bar{P}_{e_1}, \bar{P}_{e_2}$ can be reduced to finding upper bounds on ${\cal T}_{e_1|h}, {\cal T}_{e_2|h}$ and their respective averages $\bar{{\cal T}}_{e_1}\!=\!\mathbb{E}\{{\cal T}_{e_1|h}\}, \bar{{\cal T}}_{e_2}\!=\! \mathbb{E}\{{\cal T}_{e_2|h}\}$. In Appendix A we establish the following
\begin{align}
&\bar{{\cal T}}_{e_1}  < \frac{\bone_{\{Q_n<M\}}}{2\sqrt{|S_1|}} \sum_{d_{n_1,m_1} \in S_1} [\sqrt{G(n,m,n_1,m_1)} \prod_{s=1}^S {\cal D}_1(n,m,n_1,m_1)]+ \bone_{\{Q_n \geq M\}},\\
&\bar{{\cal T}}_{e_2}  < \frac{\bone_{\{Q_n>M\}}}{{|S_0|}} \sum_{d_{n_1,m_1} \in S_0}[\underset{t}{\min} ~(|S_0|G(n,m,n_1,m_1))^t  \prod_{s=1}^{S} {\cal D}_2(n,m,n_1,m_1)] \! + \!\bone_{\{Q_n \leq M\}},\\
&G(n,m,n_1,m_1) =  \frac{(\pi_1P_d^{Q_{n_1}}(1-P_d)^{K-Q_{n_1}}-\pi_0P_f^{Q_{n_1}}(1-P_f)^{K-Q_{n_1}})}{(\pi_0P_f^{Q_{n}}(1-P_f)^{K-Q_{n}}-\pi_1P_d^{Q_{n}}(1-P_d)^{K-Q_{n}})} \times \frac{T_{n_1,m_1}}{T_{n,m}}, \label{G-matrix-definition}\\
&{\cal D}_1(n,m,n_1,m_1) = \left((1+\frac{\alpha \bar{\gamma}_h \bar{a}_1 }{8}) (1+\frac{\alpha \bar{\gamma}_h \bar{a}_2}{8})  -  \frac{\alpha^2 \bar{\gamma}_h^2 \bar{a}_3}{64}\right)^{-1}\label{final-term-pe1},\\
&{\cal D}_2(n,m,n_1,m_1) =  \left((1+\frac{\alpha (t^2-t)\bar{\gamma}_h \bar{a}_1}{2}) (1+\frac{\alpha (t^2-t)\bar{\gamma}_h \bar{a}_2}{2}) - \frac{\alpha^2 (t^2-t)^2\bar{\gamma}_h^2\bar{a}_3}{16}\right)^{-1},\label{final-term-pe2}\\
&\bar{a}_1=  (a^{2s-1}_{n}-a^{2s-1}_{n_1})^2+(a^{2s}_{n}-a^{2s}_{n_1})^2, ~~~~ \bar{a}_2 =    (a^{2s-1}_{m}-a^{2s-1}_{m_1})^2+(a^{2s}_{m}-a^{2s}_{m_1})^2, \nonumber\\
&\bar{a}_3 = (a^{2s-1}_{n}-a^{2s-1}_{n_1})(a^{2s}_{n}-a^{2s}_{n_1})-(a^{2s-1}_{m}-a^{2s-1}_{m_1})(a^{2s}_{m}-a^{2s}_{m_1}). \nonumber
\end{align}
The upper bound on ${\bar {\cal T}}_{e_2}$ depends on $t$. Our simulations indicate that the bound is minimized for $t \approx 0.3$.
Furthermore, the upper bounds depend on the power allocation parameter $\alpha$. In Section \ref{simulation-section} we investigate the optimal $\alpha$ that minimizes these bounds.  In the ideal case when the inter-node communication is error-free we find $a_n^{2s-1}=a_m^{2s-1}$ and $a_n^{2s}=a_m^{2s}$ and thus (\ref{final-term-pe1}) and (\ref{final-term-pe2}), respectively, reduce to
${\cal D}_1(n,m,n_1,m_1)=(1+\frac{\alpha \bar{\gamma}_h \bar{a}_1}{8})^{-2}$,
${\cal D}_2(n,m,n_1,m_1)=(1+\frac{\alpha(t^2-t) \bar{\gamma}_h \bar{a}_1}{2})^{-2}$.
%==================================================================
\subsection{Classical Parallel Fusion Architecture}\label{per-analysis-classical-parallel}
Leveraging on the derivations in Section \ref{per-analysis-DSTC}, we provide upper bounds on ${\bar {\cal T}}_{e_1}, {\bar {\cal T}}_{e_2}$. In fact, the absence of inter-node communication renders the notations in Section \ref{per-analysis-DSTC}  simple, as the indices $m,m_1$ and the decisions $\hat{u}_{2s-1}, \hat{u}_{2s}$ are dropped, $z_{2s-1},z_{2s}$ are substituted with $y_{2s-1},y_{2s}$ and the noises $\delta_s^1, \delta_s^2$ are substituted with $v_{2s-1},v_{2s}$. Consequently the derivations of the upper bounds become rather easy. In particular, instead of $d_{n_1,m_1}$ in (\ref{d-definition}), we define $d_{n_1}$ as the following
\begin{eqnarray}\label{new-d1-for-classical-parallel}
d_{n_1}=(\pi_1P_d^{Q_{n_1}}(1-P_d)^{K-Q_{n_1}}-\pi_0P_f^{Q_{n_1}}(1-P_f)^{K-Q_{n_1}})\prod_{s=1}^S f(y_{2s-1},y_{2s}|u_{2s-1}=a^{2s-1}_{n_1},u_{2s}=a^{2s}_{n_1}).\label{d-definition-parallel}
\end{eqnarray}
Also, the relationship between $P_{e_1|h},{\cal T}_{e_1|h}$ in (\ref{pe1-first}) and $P_{e_2|h},{\cal T}_{e_2|h}$ in (\ref{pe2-upper}) can be revised as the following
\begin{align}\label{Pe-Te-parallel-classical}
&P_{e_1|h}=\pi_0 P(\sum_{n_1} d_{n_1} > 0|{\cal H}_0)=\pi_0 \sum_n {\cal T}_{e_1|h} P( F_n|{\cal H}_0), \nonumber\\
&P_{e_2|h}=\pi_1 P(\sum_{n_1} d_{n_1} < 0|{\cal H}_1)=\pi_1 \sum_n {\cal T}_{e_2|h} P( F_n|{\cal H}_1).
\end{align}
in which $F_n$ is defined in (\ref{definition-of-Fn}).
We redefine $S_{0}\!=\!\{d_{n_1}~\mbox{where}~Q_{n_1}\! < \!M\}$ and $S_{1}\!=\!\{d_{n_1}~\mbox{where}~Q_{n_1}\!\geq \! M\}$, where all entries of $S_{0}$ and $S_{1}$ are, respectively, negative and positive. We have
\begin{eqnarray*}\label{perf-classic-first}
{\cal T}_{e_1|h} \! \!& \!\!< \! \! &  \!\!  \frac{1}{|S_1|} \sum_{d_{n_1} \in S_1}P(|S_1|d_{n_1}>-d_{n}|F_{n})= \frac{1}{|S_1|} \sum_{d_{n_1} \in S_1} P\bigg( \! \zeta(n,n_1) \!>\! -\ln(|S_1|G(n,n_1))\!+\!I(n,n_1) \!\bigg), \nonumber\\
{\cal T}_{e_2|h} \! \!& \!\!< \! \! &  \!\!  \frac{1}{|S_0|} \sum_{d_{n_1} \in S_0}P(d_{n}  < -|S_0|d_{n_1} |F_{n})= \frac{1}{|S_0|} \sum_{d_{n_1} \in S_0} P\bigg( \! \zeta(n,n_1) \!>\! -\ln(|S_0|G(n,n_1))\!+\!I(n,n_1). \!\bigg),\\
\!\!\!\!G(n,n_1)\!\!&\!\!=\!\!&\!\! \frac{\pi_1P_d^{Q_{n_1}}(1-P_d)^{K-Q_{n_1}}-\pi_0P_f^{Q_{n_1}}(1-P_f)^{K-Q_{n_1}}}{\pi_0P_f^{Q_{n}}(1-P_f)^{K-Q_{n}}-\pi_1P_d^{Q_{n}}(1-P_d)^{K-Q_{n}}},\nonumber\\
\!\!\!\!I(n,n_1)\!\!&\!\!=\!\!&\!\! \underset{s=1}{\overset{S}{\sum}}    ~{\cal K}_s(n,n_1) ~\mbox{where}~{\cal K}_s(n,n_1) =\frac{1}{\sigma_v^2}(|h_{2s-1}(a_{n_1}^{2s-1}-a_{n}^{2s-1})|^2+|h_{2s}(a_{n_1}^{2s}-a_{n}^{2s})|^2),\\
\!\!\!\!\zeta(n,n_1)\!\!&\!\!=\!\!&\!\! \sum_{s=1}^{S}\theta_s(n,n_1)~\mbox{where}~\theta_s(n,n_1)=\frac{2}{\sigma_v^2} \mathfrak{Re} \big\{ v_{2s-1}h_{2s-1} (a_{n_1}^{2s-1}-a_{n}^{2s-1})+v_{2s}h_{2s}(a_{n_1}^{2s}-a_{n}^{2s})\big\}.
\label{I-definition}
\end{eqnarray*}
Note that $\zeta(n,n_1)$ is a zero mean Gaussian RV with the variance $2I(n,n_1)$. Using similar techniques in Section \ref{per-analysis-DSTC} we can establish the following
\begin{eqnarray}
{\bar {\cal T}}_{e_1} &< & \frac{\bone_{\{Q_n<M\}}}{2\sqrt{|S_1|}} \sum_{d_{n_1} \in S_1}[\sqrt{G(n,n_1)} \prod_{s=1}^{S} {\cal D}_1(n,n_1)]+ \bone_{\{Q_n \geq M\}}, \label{classical-parallel-cond-bound-1}\\
{\bar {\cal T}}_{e_2}&< &  \frac{\bone_{\{Q_n>M\}}}{{|S_0|}} \sum_{d_{n_1} \in S_0}[\underset{t}{\min} ~(|S_0|G(n,n_1))^t  \prod_{s=1}^{S} {\cal D}_2(n,n_1)]+ \bone_{\{Q_n \leq M\}}, \label{classical-parallel-cond-bound-2}\\
 {\cal D}_1(n,n_1) &=& \left((1+ \frac{\bar{\gamma}_h |a_n^{2s-1}-a_{n_1}^{2s-1}|}{2 }) (1+ \frac{\bar{\gamma}_h |a_n^{2s}-a_{n_1}^{2s}|}{2 })\right)^{-1},\label{final-term-classical-parallel-pe1} \\
 {\cal D}_2(n,n_1) &=& \left((1+ \frac{4(t^2-t)\bar{\gamma}_h |a_n^{2s-1}-a_{n_1}^{2s-1}|}{2 }) (1+ \frac{4(t^2-t)\bar{\gamma}_h |a_n^{2s}-a_{n_1}^{2s}|}{2 })\right)^{-1}.
\end{eqnarray}
This completes our derivations for the upper bounds on ${\bar {\cal T}}_{e_1}, {\bar {\cal T}}_{e_2}$ and thus ${\bar {P}}_{e_1}, {\bar {P}}_{e_2}$. Our simulations indicate that the bound is minimized for $t \approx 0.3$.
%(UNCOMMENT FOR THESIS)
%A close  examination of (\ref{final-term-classical-parallel-pe1}) and (\ref{final-term-classical-parallel-pe2}) reveals that for any $n,n_1$ there exists an $s$ value such that either $(a^{2s-1}_{n}-a^{2s-1}_{n_1})^2$ or $(a^{2s}_{n}-a^{2s}_{n_1})^2$ is zero. This implies that this scheme, as expected, does not provide diversity gain.
%===========================================================
\subsection{Cooperative Fusion Architecture with Signal Fusion at Sensors}\label{per-analysis-cooperative-local-fusion}
Leveraging on the derivations in Sections \ref{per-analysis-DSTC} and \ref{per-analysis-classical-parallel}, we provide upper bounds on ${\bar {\cal T}}_{e_1}, {\bar {\cal T}}_{e_2}$. In fact, the absence of inter-node communication renders the notations in Section \ref{per-analysis-DSTC}  simple, as the indices $m,m_1$ and the decisions $\hat{u}_{2s-1}, \hat{u}_{2s}$ are dropped, $z_{2s-1},z_{2s}$ are substituted with $y_{2s-1},y_{2s}$ and the noises $\delta_s^1, \delta_s^2$ are substituted with $v_{2s-1},v_{2s}$. Thus the derivations of the upper bounds become rather easy. In particular, instead of $d_{n_1,m_1}$ in (\ref{d-definition}) or $d_{n_1}$ in (\ref{new-d1-for-classical-parallel}), we define $d_{n_1}$ as the following
\begin{eqnarray}\label{define-d-n1-local}
d_{n_1}&=&(\pi_1\prod_{s=1}^S P(\tilde{u}_{2s-1}=a^{2s-1}_{n_1},\tilde{u}_{2s}=a^{2s}_{n_1} |{\cal H}_1)-\pi_0 \prod_{s=1}^S P(\tilde{u}_{2s-1}=a^{2s-1}_{n_1},\tilde{u}_{2s}=a^{2s}_{n_1} |{\cal H}_0)) \nonumber \\
& \times & \prod_{s=1}^S f(y_{2s-1},y_{2s}|\tilde{u}_{2s-1}=a^{2s-1}_{n_1},\tilde{u}_{2s}=a^{2s}_{n_1}).
\end{eqnarray}
where $P(\tilde{u}_{2s-1}=a^{2s-1}_{n_1},\tilde{u}_{2s}=a^{2s}_{n_1} |{\cal H}_{\ell})$ in (\ref{define-d-n1-local})
is determined in Section \ref{scheme-local-fusion-comm-part}. In fact, this probability depends on sensing channels through the threshold $\tau$ and the sensing noise variance $\sigma_w^2$ as well as the average received SNR $\bar{\gamma}_{hs}$ corresponding to inter-node communication.  Furthermore, the relationship between $P_{e_1|h},{\cal T}_{e_1|h}$ in (\ref{pe1-first}) and $P_{e_2|h},{\cal T}_{e_2|h}$ in (\ref{pe2-upper}) can be revised as  (\ref{Pe-Te-parallel-classical}), in which $F_{n}= \{\tilde{u}_{2s-1}=a^{2s-1}_{n},\tilde{u}_{2s}=a^{2s}_{n}~\mbox{for}~s\!=\!1,...,S\}$. We also redefine $S_{0}\!=\!\{d_{n_1}~\mbox{where}~d_{n_1}\! < \! 0 \}$ and $S_{1}\!=\!\{d_{n_1}~\mbox{where}~d_{n_1}\!\geq \!  0\}$, where all entries of $S_{0}$ and $S_{1}$ are, respectively, negative and positive. We have
\begin{eqnarray*}\label{perf-LVD-first}
{\cal T}_{e_1|h} \! \!& \!\!< \! \! &  \!\!  \frac{1}{|S_1|} \sum_{d_{n_1} \in S_1}P(|S_1|d_{n_1}>-d_{n}|F_{n}) =  \frac{1}{|S_1|} \sum_{d_{n_1} \in S_1} P\bigg( \! \zeta(n,n_1) \!>\! -\ln(|S_1|G(n,n_1))\!+\!I(n,n_1) \!\bigg), \nonumber\\
{\cal T}_{e_2|h} \! \!& \!\!< \! \! &  \!\!    \frac{1}{|S_0|} \sum_{d_{n_1} \in S_0}P(d_{n}  < -|S_0|d_{n_1} |F_{n}) = \frac{1}{|S_0|} \sum_{d_{n_1} \in S_0} P\bigg( \! \zeta(n,n_1) \!>\! -\ln(|S_0|G(n,n_1))\!+\!I(n,n_1) \!\bigg),
\end{eqnarray*}
in which $G(n,n_1)=\frac{(\pi_1\prod_{s=1}^S P(\tilde{u}_{2s-1}=a^{2s-1}_{n_1},\tilde{u}_{2s}=a^{2s}_{n_1} |{\cal H}_1)-\pi_0 \prod_{s=1}^S P(\tilde{u}_{2s-1}=a^{2s-1}_{n_1},\tilde{u}_{2s}=a^{2s}_{n_1} |{\cal H}_0))}{(\pi_0\prod_{s=1}^S P(\tilde{u}_{2s-1}=a^{2s-1}_{n_1},\tilde{u}_{2s}=a^{2s}_{n_1} |{\cal H}_0)-\pi_1 \prod_{s=1}^S P(\tilde{u}_{2s-1}=a^{2s-1}_{n_1},\tilde{u}_{2s}=a^{2s}_{n_1} |{\cal H}_1))}$,
\begin{eqnarray*}
\!\!\!\!I(n,n_1)\!\!&\!\!=\!\!&\!\! \underset{s=1}{\overset{S}{\sum}}    ~{\cal K}_s(n,n_1) ~\mbox{where}~{\cal K}_s(n,n_1) =\frac{\alpha}{\sigma_v^2}(|h_{2s-1}(a_{n_1}^{2s-1}-a_{n}^{2s-1})|^2+|h_{2s}(a_{n_1}^{2s}-a_{n}^{2s})|^2),\\
\!\!\!\!\zeta(n,n_1)\!\!&\!\!=\!\!&\!\! \sum_{s=1}^{S}\theta_s(n,n_1)~\mbox{where}~\theta_s(n,n_1)=\frac{2 \sqrt{\alpha}}{\sigma_v^2} \mathfrak{Re} \big\{ v_{2s-1}h_{2s-1} (a_{n_1}^{2s-1}-a_{n}^{2s-1})+v_{2s}h_{2s}(a_{n_1}^{2s}-a_{n}^{2s})\big\}.
\label{I-definition}
\end{eqnarray*}
Note that $\zeta(n,n_1)$ is a zero mean Gaussian RV with the variance $2I(n,n_1)$. Using similar techniques in Sections \ref{per-analysis-DSTC} and \ref{per-analysis-classical-parallel} we can establish the following
\begin{eqnarray}
{\bar {\cal T}}_{e_1} &< & \frac{\bone_{\{d_n \in S_0\}}}{2\sqrt{|S_1|}} \sum_{d_{n_1} \in S_1} [\sqrt{G(n,n_1)}\prod_{s=1}^S {\cal D}_1(n,n_1)]+ \bone_{\{d_n \in S_1\}} \label{LVD-cond-bound-1}, \label{Te1-fusion-sensors}\\
{\bar {\cal T}}_{e_2} &< &  \frac{\bone_{\{d_n \in S_1\}}}{{|S_0|}} \sum_{d_{n_1} \in S_1}[\underset{t}{\min} ~(|S_0|G(n,n_1))^t  \prod_{s=1}^{S} {\cal D}_2(n,n_1)]+ \bone_{\{d_n \in S_0\}}, \label{Te2-fusion-sensors}\\
 {\cal D}_1(n,n_1)&=&\left((1+\frac{ \alpha \bar{\gamma}_h (a_n^{2s-1}-a_{n_1}^{2s-1})^2}{4})(1+\frac{\alpha \bar{\gamma}_h (a_n^{2s}-a_{n_1}^{2s})^2}{4})\right)^{-1}, \\
 {\cal D}_2(n,n_1)&=& \left((1+\alpha(t^2-t) \bar{\gamma}_h (a_n^{2s-1}-a_{n_1}^{2s-1})^2)(1+\alpha(t^2-t) \bar{\gamma}_h (a_n^{2s}-a_{n_1}^{2s})^2)\right)^{-1}.
\end{eqnarray}
This completes our derivations for the upper bounds on ${\bar {\cal T}}_{e_1}, {\bar {\cal T}}_{e_2}$ and thus ${\bar {P}}_{e_1}, {\bar {P}}_{e_2}$. Our numerical results show that the bound is minimized for $t \approx 0.3$.
Furthermore, the upper bounds depend on the power allocation parameter $\alpha$. In Section \ref{simulation-section} we investigate the optimal $\alpha$ that minimizes these bounds.
%=====================================================
\subsection{Parallel Fusion Architecture with Local Threshold Changing at Sensors}\label{per-analysis-threshold-changing}
Leveraging on the derivations in sections \ref{per-analysis-DSTC} and \ref{per-analysis-classical-parallel}, we provide upper bounds on ${\bar {\cal T}}_{e_1}, {\bar {\cal T}}_{e_2}$. To capture all different values that $u_{2s-1},u_{2s},\bar{u}_{2s-1},\bar{u}_{2s}$ for $s=1,...,S$ can take we consider a $2K$-length sequence $(a^1_{n_1},a^2_{n_1},a^1_{m_1},a^2_{m_2},..., a^{2S-1}_{n_1}, a^{2S}_{n_1},a^{2S-1}_{m_1}, a^{2S}_{m_1})$ where $a^{2s-1}_{n_1},a^{2s}_{n_1} \! \in \! \{1,-1\}$ and $a^{2s-1}_{m_1},a^{2s}_{m_1} \! \in \! \{1,-1\}$, respectively,
are the values assumed by $u_{2s-1}, u_{2s}$ and  $\bar{u}_{2s-1}, \bar{u}_{2s}$. Let $Q^1_{n_1,m_1},Q^2_{n_1,m_1},Q^3_{n_1,m_1},Q^4_{n_1,m_1}$, respectively, denote the number of cases in the above sequence that $a^{s'}_{n_1}\!=\!a^{s'}_{m_1}\!=\!1$, $a^{s'}_{n_1}\!=\!-a^{s'}_{m_1}\!=\!1$, $a^{s'}_{n_1}\!=\!-a^{s'}_{m_1}\!=\!-1$, and $a^{s'}_{n_1}\!=\!a^{s'}_{m_1}\!=\!-1$ for $s'=2s, 2s-1, s=1,...,S$. Instead of $F_{n_1,m_1}$ in (17) and $d_{n_1,m_1}$ in (18), we redefine them as the following
\begin{eqnarray}\label{d-definition-new}
F_{n_1,m_1}\!&\!=\!&\!\{ u_{2s-1}= a^{2s-1}_{n_1}, u_{2s}=a^{2s}_{n_1}, \bar{u}_{2s-1}=a^{2s-1}_{m_1}, \bar{u}_{2s}=a^{2s}_{m_1}~\mbox{for}~s\!=\!1,...,S\}, \nonumber\\
d_{n_1,m_1}\!&\!=\!&\!(\pi_1 \prod_{j=1}^4 P_{d_j}^{Q^j_{n_1,m_1}} -\pi_0 \prod_{j=1}^4 P_{f_j}^{Q^j_{n_1,m_1}})\\
\!&\!\times\!&\!\prod_{s=1}^S f(z_{2s-1},z_{2s}|u_{2s-1}=a^{2s-1}_{n_1},u_{2s}=a^{2s}_{n_1},\bar{u}_{2s-1}=a^{2s-1}_{m_1},\bar{u}_{2s}=a^{2s}_{m_1}) \nonumber,
\end{eqnarray}
where $P(\bar{u}_i\!=\!1|u_i\!=\!-u_j\!=\!-1,{\cal H}_\ell)\!=\!P(x_i\!>\!\tau_1|{\cal H}_\ell), P(\bar{u}_i\!=\!-1|u_i\!=\!-u_j\!=\!1,{\cal H}_\ell)\!=\!P(\tau\!<\!x_i\!<\! \tau_1|{\cal H}_\ell), P(\bar{u}_i\!=\!1|u_i\!=\!-u_j\!=\!-1,{\cal H}_\ell)\!=\!P(\tau_2\!<\!x_i\!<\!\tau|{\cal H}_\ell)$, $P(\bar{u}_i\!=\!-1|u_i\!=\!-u_j\!=\!-1, {\cal H}_\ell)\!=\!P(x_i\!<\!\tau_2|{\cal H}_\ell)$, respectively, are equal to $P_{d_1},P_{d_2},P_{d_3},P_{d_4}$ under ${\cal H}_1$, and are equal to $P_{f_1},P_{f_2},P_{f_3},P_{f_4}$ under ${\cal H}_0$.
Since sensing noises are identically distributed and uncorrelated we find $P(F_{n_1,m_1}|{\cal H}_{1}) \!=\! \prod_{j=1}^4 P_{d_j}^{Q^j_{n_1,m_1}}$ and $P(F_{n_1,m_1}|{\cal H}_{0}) \!=\! \prod_{j=1}^4 P_{f_j}^{Q^j_{n_1,m_1}}$. Note that the relationship between $P_{e_1|h},{\cal T}_{e_1|h}$ in (\ref{pe1-first}) and $P_{e_2|h},{\cal T}_{e_2|h}$ in (\ref{pe2-upper}) hold true.
Using similar techniques in sections \ref{per-analysis-DSTC} and \ref{per-analysis-classical-parallel} we can establish the following
\begin{eqnarray}\label{final-cond-bound-new}
\bar{{\cal T}}_{e_1}  &<& \frac{\bone_{\{d_{n,m} \in S_0\}}}{2\sqrt{|S_1|}} \sum_{d'_{n_1,m_1} \in S_1} [\sqrt{G(n,m,n_1,m_1)} \prod_{s=1}^S {\cal D}_1(n,m,n_1,m_1)]+ \bone_{\{d_{n,m} \in S_1\}},\\
\bar{{\cal T}}_{e_2}  & < & \frac{\bone_{\{d_{n,m} \in S_1\}}}{{|S_0|}} \sum_{d_{n,m} \in S_0}[\underset{t}{\min} ~(|S_0|G(n,m,n_1,m_1))^t  \prod_{s=1}^{S} {\cal D}_2(n,m,n_1,m_1)]+ \bone_{\{d_{n,m} \in S_0\}},
\end{eqnarray}
in which $S_0\!=\!\{d_{n_1,m_1}\text{where} ~\frac{ \prod_{j=1}^4 P_{d_j}^{Q^j_{n_1,m_1}}}{ \prod_{j=1}^4 P_{f_j}^{Q^j_{n_1,m_1}}}<\frac{\pi_0}{\pi_1}\}$ and $S_1\!=\!\{d_{n_1,m_1}\text{where} ~\frac{ \prod_{j=1}^4 P_{d_j}^{Q^j_{n_1,m_1}}}{ \prod_{j=1}^4 P_{f_j}^{Q^j_{n_1,m_1}}}>\frac{\pi_0}{\pi_1}\}$ and
\begin{eqnarray}
\!\!\!\!\!\!\!  \!\!\!\!\!\!\!   \!\!\!\!\!\!\!\!\!\! && G(n,m,n_1,m_1)= \frac{(\pi_1  \prod_{j=1}^4 P_{d_j}^{Q^j_{n_1,m_1}} -\pi_0  \prod_{j=1}^4 P_{f_j}^{Q^j_{n_1,m_1}} )}{(\pi_1  \prod_{j=1}^4 P_{d_j}^{Q^j_{n,m}} -\pi_0  \prod_{j=1}^4 P_{f_j}^{Q^j_{n,m}})}, \\
  &&{\cal D}_1(n,m,n_1,m_1) =  \left((1+\frac{\alpha \bar{\gamma}_h \bar{a}_1 }{8}) (1+\frac{\alpha \bar{\gamma}_h \bar{a}_2}{8})  -  \frac{\alpha^2 \bar{\gamma}_h^2 \bar{a}_3}{64}\right)^{-1}\label{final-term-pe1-new}\\
\!\!\!\!\!\!\!  \!\!\!\!\!\!\! \!\!\!\!\!\!\!\!\!\! && {\cal D}_2(n,m,n_1,m_1) = \left((1+\frac{\alpha (t^2-t)\bar{\gamma}_h \bar{a}_1}{2}) (1+\frac{\alpha (t^2-t)\bar{\gamma}_h \bar{a}_2}{2}) - \frac{\alpha^2 (t^2-t)^2\bar{\gamma}_h^2 \bar{a}_3}{16}\right)^{-1}\label{final-term-pe2-new}\\
\!\!\!\!\!\!\!  \!\!\!\!\!\!\! \!\!\!\!\!\!\!\!\!\! && \bar{a}_1 =(a^{2s-1}_{n}-a^{2s-1}_{n_1})^2+(a^{2s}_{m}-a^{2s}_{m_1})^2, ~~~~ \bar{a}_2 = (a^{2s-1}_{m}-a^{2s-1}_{m_1})^2+(a^{2s}_{n}-a^{2s}_{n_1})^2 \nonumber\\
\!\!\!\!\!\!\!  \!\!\!\!\!\!\! \!\!\!\!\!\!\!\!\!\! &&  \bar{a}_3= (a^{2s-1}_{n}-a^{2s-1}_{n_1})(a^{2s}_{m}-a^{2s}_{m_1})-(a^{2s-1}_{m}-a^{2s-1}_{m_1})(a^{2s}_{n}-a^{2s}_{n_1}). \nonumber
\end{eqnarray}
This completes our derivations for the upper bounds on ${\bar {\cal T}}_{e_1}, {\bar {\cal T}}_{e_2}$ and thus ${\bar {P}}_{e_1}, {\bar {P}}_{e_2}$.
The upper bound on ${\bar {\cal T}}_{e_2}$ depends on $t$. Our simulations indicate that the bound is minimized for $t \approx 0.3$.
%In the case when $a_n^{2s-1}=a_n^{2s}$ and $a_m^{2s-1}=a_m^{2s}$ {\red what is the physical meaning of this case?} and thus (\ref{final-term-pe1}) and (\ref{final-term-pe2}), respectively, reduce to
%\begin{equation*}
%{\cal D}_1(n,m,n_1,m_1)=(1+\frac{\alpha \bar{\gamma}_h \bar{a}_1}{8})^{-2},~~~~
%{\cal D}_2(n,m,n_1,m_1)=(1+\frac{\alpha(t^2-t) \bar{\gamma}_h \bar{a}_1}{2})^{-2}.
%\end{equation*}
%===================================
\subsection{Comparison of Different Schemes in Asymptotic Regime for Large $S$}
For all four schemes discussed in sections \ref{per-analysis-DSTC},\ref{per-analysis-classical-parallel},\ref{per-analysis-cooperative-local-fusion},\ref{per-analysis-threshold-changing}, from sections $A, B, C, D$ of Appendix B we have established the following, for large $S$
\begin{align*}\label{four-exp}
\!\!\!\!\!\!\!\!\!\!\!\!\!\!\!\!\!\!\!\!\!\! &\bar{P}_e=\bar{P}_{e_{1}}+\bar{P}_{e_{2}}<
\kappa_{l_{11}}e^{S(\mu_{l_{11}}+\frac{1}{2}\sigma^2_{l_{11}})}+\frac{1}{2}e^{-S\frac{\mu^2_{l_{12}}}{2\sigma^2_{l_{12}}}}+\kappa_{l_{21}}e^{S(\mu_{l_{21}}+\frac{1}{2}\sigma^2_{l_{21}})}+\frac{1}{2}e^{-S\frac{\mu^2_{l_{22}}}{2\sigma^2_{l_{22}}}},\\
&\mbox{where}~~\mu_{l_{11}}+\frac{1}{2}\sigma^2_{l_{11}}, \mu_{l_{21}}+\frac{1}{2}\sigma^2_{l_{21}} <0.\nonumber
\end{align*}
Also, $\mu_{l_{11}},\mu_{l_{12}},\mu_{l_{21}},\mu_{l_{22}}$ and $\sigma^2_{l_{11}},\sigma^2_{l_{12}},\sigma^2_{l_{21}},\sigma^2_{l_{22}}$ and $\kappa_{l_{11}}, \kappa_{l_{21}}$ differ for different schemes and do not depend on $S$ (they only depend on SNR$_h$, SNR$_c$ defined in Section \ref{simulation-section} and $\pi_0$). For each scheme we examine these four exponentials and keep the dominant one.
%that has the largest exponent (since we are in asymptotic regime $S \rightarrow \infty$).
Let $\kappa_{l_{x}}e^{-S \gamma_x}$ for $x=a, b, c, d$ be the dominant exponent, respectively, for schemes discussed in sections \ref{per-analysis-DSTC},\ref{per-analysis-classical-parallel},\ref{per-analysis-cooperative-local-fusion},\ref{per-analysis-threshold-changing}, where $\gamma_{x}=\min\{-(\mu_{l_{11}}+\frac{1}{2}\sigma^2_{l_{11}}),\frac{\mu^2_{l_{12}}}{2\sigma_{l_{12}}},-(\mu_{l_{21}}+\frac{1}{2}\sigma^2_{l_{21}}),\frac{\mu^2_{l_{22}}}{2\sigma^2_{l_{22}}}\}$ and $\kappa_{l_{x}}$ be its corresponding multiplicative scalar. When comparing the error exponents of any pair of these four schemes, for instance schemes in  sections \ref{per-analysis-DSTC},\ref{per-analysis-classical-parallel}, we have
%\begin{align}\label{error-exponent-a-b}
$\lim_{S \rightarrow \infty} (\frac{\ln (\kappa_{l_{a}}e^{-S \gamma_a}) }{S}-\frac{\ln (\kappa_{l_{b}}e^{-S \gamma_b})}{S})= \gamma_b -\gamma_a$,
%\end{align}
%
implying that such a difference depends on SNR$_h$, SNR$_c$, $\pi_0$ only and does not change with $S$. This analysis suggests that our numerical findings in Section \ref{simulation-section} on performance comparison between different schemes should not vary much for large $S=\frac{K}{2}$. In fact, our simulation results show that performance comparison between different schemes, given SNR$_h$, SNR$_c$, $\pi_0$, remains the same for $K=10, 14$ and $20$ (due to lack of space we only include the results for $K=20$ in Table \ref{table:incresing-number-sensors}).
%==========================================================
\section{Numerical Results}\label{simulation-section}
In this section, we evaluate and compare performance of the proposed schemes in Sections \ref{section-DSTC}, \ref{cooperative-local-fusion-scheme}, \ref{parallel-threshold-chaning-scheme}, against the conventional scheme in Section \ref{parallel-classical}. For the sakes of presentation, we refer to the schemes in Sections \ref{parallel-classical}, \ref{section-DSTC}, \ref{cooperative-local-fusion-scheme}, and \ref{parallel-threshold-chaning-scheme}, respectively, as ``parallel'', ``STC@sensors'', and ``fusion@sensors'', ``threshold changing@sensors''. We consider  $K\!=\!10$ sensors ($S\!=\!5$ groups of paired sensors). We assume that the sensing noises $w_k$ are identically distributed, i.e., $\sigma_{w_k}^2\!=\!\sigma_w^2$ and $\rho_{ij}\!=\!\rho$ characterizes the correlation. We define SNR$_c\!=\! -20 \log_{10} \sigma_w$ as SNR corresponding to sensing channels. We let the distances between the sensors and the FC $d\!=\!10m$, the distances between the cooperative partners within each group $d_0\!=\!2m$, the variance of receiver noises $\sigma^2_{v}=\sigma^2_{\eta}=-50dBm$,  the pathloss exponent $\varepsilon=2$, and the antenna gain ${\cal G}= -30$dB. To make a fair comparison among different schemes, we enforce the sensors in all schemes to transmit the same power ${\cal P}$. In  ``STC@sensors'' and ``fusion@sensors'' a sensor spends $(1-\alpha) {\cal P}$ and $\alpha {\cal P}$, respectively, for communicating with its cooperative partner and with the FC, where $\alpha$ is different in these two schemes. We define SNR$_h\!=\!10 \log_{10} \bar{\gamma}_h$, in which  $ \bar{\gamma}_h\!=\! \frac{\sigma_h^2}{\sigma_v^2}\!=\!\frac{{\cal P \cal G}}{d^{\varepsilon} \sigma_v^2}$. Our goal is to investigate the average error $\bar{P}_e$ of ``STC@sensors'', ``fusion@sensors'', ``threshold changing@sensors'' against that of ``parallel'', as SNR$_h$ and SNR$_c$ vary and identify different regimes in which these schemes outperform ``parallel''. Note that, in ``STC@sensors'' and  ``fusion@sensors'', given  SNR$_h$ and SNR$_c$, average error $\bar{P}_e$ depends on $\alpha$, i.e., one would expect that there is an optimal power allocation $\alpha^*$ at which $\bar{P}_e$ attains its minimum, given  SNR$_h$ and SNR$_c$. We start with investigating $\alpha^*$.

\underline{{\it Optimal power allocation when the FC employs LRT rule}}:
%Nahal's observations
%DSTC - SNR_h is fixed and SNR_c is changing
%1) alpha^* does not change as pi0 and SNR_c change - OK
%2) Pe for pi0=0.6 is better than pi0=0.7 - I cannot justify
%DSTC - SNR_c is fixed and SNR_h is changing
%1)alpha^* increases as SNR_h increases - I cannot justify
%2)Pe for pi0=0.6 is better than pi0=0.7 - I cannot justify
We start with ``STC@sensors''. Fig. \ref{Fig2} plots $\bar{P}_e$ versus $\alpha$ for SNR$_h\!=\!5$dB, SNR$_c\!=\!2,6,10 $dB and $\pi_0\!=\!0.6$, assuming $\rho\!=\!0$. We observe that $\alpha^* \! \approx \! 0.65$, regardless of the variations in SNR$_c$ values. Fig. \ref{Fig3} plots $\bar{P}_e$ versus $\alpha$ for SNR$_c=6$dB, SNR$_h\!=\!5,10,15 $dB and $\pi_0\!=\!0.6$, assuming $\rho=0$. We observe that $\alpha^*$ increases as SNR$_h$ (or equivalently $\bar{\gamma}_h$) increases, in particular, we obtain $\alpha^* \! \approx \!0.6, 0.7, 0.8$, respectively for SNR$_h=5$dB, 10 dB, 15dB. These observations can be explained considering our analytical results in (\ref{pe1-first})-(\ref{final-term-pe2}) of Section \ref{per-analysis-DSTC}. Recall $T_{n,m},T_{n_1,m_1}$ in (\ref{G-matrix-definition}) capture the errors during inter-node communication and depend on the average received SNR $\bar{\gamma}_{hs}$ corresponding to inter-node communication, which for $\sigma_{\eta}^2=\sigma_v^2$ it reduces to $\bar{\gamma}_{hs}\!=\! (\frac{d}{d_0})^{\varepsilon} (1-\alpha) \bar{\gamma}_{h}$. This implies that $G(n,m,n_1,m_1)$ is decoupled into two fractions, where the first fraction depends on SNR$_c$ (through the local performance indices $P_d, P_f$) and the second one depends on $(\frac{d}{d_0})^{\varepsilon}(1-\alpha)\bar{\gamma}_h$. On the other hand, the inverses of ${\cal D}_1(n,m,n_1,m_1), {\cal D}_2(n,m,n_1,m_1)$ depend on  $\alpha\bar{\gamma}_h$ only, capturing the errors of sensor-FC communication channels. Due to this decoupling of the effective factors in the terms of $\bar{P}_e$, we expect that,  $\alpha^*$ becomes insensitive to variations of SNR$_c$ (for fixed $d,d_0$, SNR$_h$) and varies as SNR$_h$ changes (for fixed $d,d_0$, SNR$_c$). For the scenario where the distance between cooperative partners is shorter than the distance between the nodes and the FC we expect $\alpha^*> 0.5$, i.e., a sensor spends a higher (lower) percentage of its transmit power for communicating with the FC (its cooperative partner). These observations are in agreement with the fact that the local information exchange in ``STC@sensors'' does not affect the local error probability $p_{e_i}\!=\!P(u_i\!=\!-1|{\cal H}_1)\pi_1\!+\!P(u_i\!=\!1|{\cal H}_0)\pi_0$ at ${\cal S}_i$ (which depends on SNR$_c$ through $P_d, P_f$); it rather provides a form of ``decision diversity'', to mitigate the fading effect during sensor-FC communication, i.e., it improves the global performance $\bar{P}_e$ at the FC via reducing the errors during inter-sensor and sensor-FC communication.

%fusion at sensors- SNR_h is fixed and SNR_c is changing
We continue with ``fusion@sensors''. Fig. \ref{Fig6} plots ${\bar P}_e$ versus $\alpha$ for SNR$_h\!=\!5$dB, SNR$_c\!=\!2,6,10 $dB and $\pi_0\!=\!0.6$, assuming $\rho\!=\!0$. We observe that $\alpha^*$ increases as SNR$_c$ increases, in particular, we obtain $\alpha^* \! \approx \! 0.6, 0.7, 0.8$, respectively, for SNR$_c\!=\! 2,6,10 $dB. Comparing this trend with that of ``STC@sensors'' in Fig. \ref{Fig2} we notice that the schemes have different trends.
%fusion at sensors - SNR_c is fixed and SNR_h is changing
Fig. \ref{Fig7} plots $\bar{P}_e$ versus $\alpha$ for SNR$_c=6$dB, SNR$_h\!=\!5,10,15 $dB and $\pi_0\!=\!0.6$, assuming $\rho=0$. We observe that $\alpha^*$ increases as SNR$_h$ (or equivalently $\bar{\gamma}_h$) increases, in particular, we obtain $\alpha^* \! \approx \!0.7, 0.7, 0.85$, respectively for SNR$_h \!=\!5,10,15$ dB.   Comparing this trend with that of ``STC@sensors''  in Fig. \ref{Fig3} we observe that the schemes have similar trends. These observations can be explained considering our analytical results in (\ref{Te1-fusion-sensors}),(\ref{Te2-fusion-sensors}) of Section \ref{per-analysis-cooperative-local-fusion}. Note that similar to ``STC@sensors'', the inverses of ${\cal D}_1(n,n_1), {\cal D}_2(n,n_1)$ depend on $\alpha \bar{\gamma}_h$ only, capturing the errors of sensor-FC communication channels. However, the structure of $G(n,n_1)$ is different from that of $G(n,m,n_1,m_1)$ in ``STC@sensors''. In particular, examining $G(n,n_1)$ reveals that this term depends on $P(\tilde{u}_{2s-1},\tilde{u}_{2s}~\mbox{for~all~pairs} | {\cal H}_{\ell})$ given in (\ref{second-term1-local-fusion}), which as we mentioned in Section \ref{scheme-local-fusion-comm-part}, it depends on SNR$_c$ (through $P_d, P_f$) as well as the average received SNR $\bar{\gamma}_{hs}$ corresponding to inter-node communication. This implies that, different from $G(n,m,n_1,m_1)$ in ``STC@sensors'', the impacts of effective factors SNR$_c$ and $(\frac{d}{d_0})^{\varepsilon} (1-\alpha) \bar{\gamma}_{h}$ in $G(n,n_1)$ cannot be decoupled and hence $\alpha^*$ varies as SNR$_c$ changes (for fixed $d,d_0$, SNR$_h$) or SNR$_h$ changes (for fixed $d,d_0$, SNR$_c$). These observations are in agreement with the fact that the local information exchange in ``fusion@sensors'', different from ``STC@sensors", affects the the local error probability $p_{e_i}\!=\!P(\tilde{u}_i\!=\!-1|{\cal H}_1)\pi_1\!+\!P(\tilde{u}_i\!=\!1|{\cal H}_0)\pi_0$ at ${\cal S}_i$.  Therefore, it improves the global performance $\bar{P}_e$ at the FC via improving the local error probability at the sensors. As SNR$_c$ decreases (for fixed $d,d_0$, SNR$_h$) the reliability of the initial decision $u_i$ at ${\cal S}_i$ (which is based on observation $x_i$) reduces, and hence the local information exchange is more needed to form the new decision $\tilde{u}_i$ with higher reliability, where more local information exchange is translated into a higher (lower) percentage of transmit power for inter-node communication (sensor-FC communication) or equivalently smaller $\alpha^*$.

\underline{{\it Optimal power allocation when the FC employs majority rule}}: Similar observations are made when the FC employs the majority rule. Figs. \ref{Fig4} and  \ref{Fig8}, respectively, plot  ${\bar P}_e$ versus $\alpha$, for ``STC@sensors'' and ``fusion@sensors'', when SNR$_h\!=\!5$dB, SNR$_c\!=\!2,6,10 $dB and $\pi_0\!=\!0.6$. Figs. \ref{Fig5} and \ref{Fig9}, respectively, plot $\bar{P}_e$ versus $\alpha$, for ``STC@sensors'' and ``fusion@sensors'', when SNR$_c=6$dB, SNR$_h\!=\!5,10,15 $dB and $\pi_0\!=\!0.6$. Comparing Fig. \ref{Fig2} with Fig. \ref{Fig4}, Fig. \ref{Fig6} with Fig. \ref{Fig8},
Fig. \ref{Fig3} with Fig. \ref{Fig5},  and Fig. \ref{Fig7} with Fig. \ref{Fig9}, we can make similar observations regarding the variations of $\alpha^*$ as SNR$_c$ or SNR$_h$ change. For each scheme, when we compare the value of $\alpha^*$ for LRT and majority rules, given $d,d_0$, SNR$_c$, SNR$_h$, we find that $\alpha^*$ corresponding to the majority rule is larger than that of the LRT rule, i.e., a sensor spends a higher (lower) percentage of its transmit power for communicating with the FC (its cooperative partner). This is due to the fact that, the majority rule demodulates first the sensor-FC channel outputs to find the channel inputs, rather than using the channel outputs directly for fusion, resembling the concept of ``hard versus soft decoding'' in \cite{c19}. To compensate for the information loss due to demodulation and its negative impact on error, each sensor is required to invest higher percentage of its transmit power for communicating with the FC.
%Nahal's writeup: to compensate for the lack of knowledge of $P_d$ and $P_f$ at the FC.

\underline{\it Performance comparison of different schemes}:
To validate our performance analysis in Section VI, Fig. (\ref{Fig1}) shows $\bar{P}_e$ of ``parallel'', ``STC@sensors'', ``fusion@sensors'', ``threshold changing@sensors'' versus SNR$_h$ for SNR$_c\!=\!6$dB and $\pi_0\!=\!0.6$, to compare the analytical and Monte-Carlo simulation results. We obtain $\bar{P}_e$ of ``STC@sensors'' and ``fusion@sensors'', using $\alpha^*$ corresponding to SNR$_h$ and SNR$_c$ values.  The figure demonstrates a good agreement between theory and simulation. It also shows that, different from conventional communication systems, $\bar{P}_e$ has an error floor at high SNR$_h$. This behavior is due to the fact that $\bar{P}_e$ in our distributed detection system is dependent on SNR$_h$ and SNR$_c$. In fact, had all communication channels were error-free, $\bar{P}_e$ of ``parallel'' would be
\begin{eqnarray}\label{error-floor}
\bar{P}_e \!&\!= \!&\! \pi_1 P\left(\frac{P_d^{n}(1-P_d)^{2S-n}}{P_f^{n}(1-P_f)^{2S-n}}< \frac{\pi_0}{\pi_1}\right) + \pi_0 P\left(\frac{P_d^{n}(1-P_d)^{2S-n}}{P_f^{n}(1-P_f)^{2S-n}}> \frac{\pi_0}{\pi_1}\right) \nonumber \\  \!&\!= \!&\!  \pi_1 \sum_{n=0}^{M-1} \frac{(2S)!}{(2S-n)! n!} P_d^n(1-P_d)^{2S-n} + \pi_0 \sum_{n=M}^{2S}  \frac{(2S)!}{(2S-n)! n!}P_f^n(1-P_f)^{2S-n},
\end{eqnarray}
where $M$
%as previously defined in Section \ref{per-analysis-DSTC},
satisfies $\frac{P_d^{M}(1-P_d)^{2S-M}}{P_f^{M}(1-P_f)^{2S-M}} \! > \! \frac{\pi_0}{\pi_1}$. Equation (\ref{error-floor}) indicates that the error floor depends on SNR$_c$ (through $P_d, P_f$), and
as SNR$_c$ reduces the error floor increases. Fig. (\ref{Fig1}) also shows that ``parallel'' and ``STC@sensors'' have similar error floors, whereas ``fusion@sensors'' has a lower error floor. These observations are in agreement with the fact that the local information exchange in ``STC@sensors'' improves $\bar{P}_e$ via providing ``decision diversity'', without changing the local error probability at ${\cal S}_i$ (which depends on the reliability of $u_i$). On the other hand, the local information exchange in ``fusion@sensors'' improves $\bar{P}_e$ via improving the local error probability at ${\cal S}_i$ (which depends on the reliability of  $\tilde{u}_i$).
For moderate/high SNR$_h$ where the errors during inter-sensor and sensor-FC communication are negligible, $\bar{P}_e$ is governed by the local error probability at ${\cal S}_i$. Since the reliability of local decisions in ``parallel'' and ``STC@sensors'' are identical and the reliability of local decisions in ``fusion@sensors`` exceeds that of ``parallel'' and ``STC@sensors'', we expect that ``parallel'' and ``STC@sensors'' have similar error floors, whereas ``fusion@sensors'' reaches a lower error floor and Fig. (\ref{Fig1})  confirms these.

Table \ref{table:2} tabulates  $\bar{P}_e$ of ``parallel'', ``STC@sensors'', ``fusion@sensors'' and ``threshold changing@sensors'', as  SNR$_h$ and SNR$_c$ vary, for $\pi_0\!=\!0.6$ and $\rho\!=\!0$, when the FC employs the LRT rule. To have a fair comparison among different schemes, we obtain $\bar{P}_e$ of ``STC@sensors'' and ``fusion@sensors'', using $\alpha^*$ corresponding to SNR$_h$ and SNR$_c$ values.
Comparing ``STC@sensors'' and ``parallel'' we note that, for moderate SNR$_h$  and moderate/high SNR$_c$, ``STC@sensors'' outperforms ``parallel'', while the performance gain of ``STC@sensors'' decreases as SNR$_c$ reduces. On the other hand, for low SNR$_h$  ``STC@sensors'' performs worse than ``parallel'', whereas for high SNR$_h$  ``parallel'' and ``STC@sensors''  reach similar error floors. These observations agree with the fact that the local information exchange in ``STC@sensors'' improves $\bar{P}_e$ via providing ``decision diversity'', without changing the local error probability at ${\cal S}_i$ (which depends on SNR$_c$). For moderate/high SNR$_h$ the errors during inter-sensor and sensor-FC communication are small and $\bar{P}_e$ is mainly determined by SNR$_c$. Therefore, lowering SNR$_c$ increases $\bar{P}_e$. On the other hand, for low SNR$_h$ the errors during inter-sensor communication negatively impact the diversity gain of ``STC@sensors''.
Comparing ``fusion@sensors'' and ``parallel'' we note that, for low SNR$_h$  they have similar performance, whereas for moderate/high SNR$_h$ ``fusion@sensors'' outperforms ``parallel'' (regardless of SNR$_c$). In particular, for high SNR$_h$ the error floor of ``fusion@sensors'' is smaller than that of ``parallel''. These observations agree with the facts that for moderate/high SNR$_h$, $\bar{P}_e$ is dominated by the local probability error at ${\cal S}_i$, and the local probability error of ``fusion@sensors'' is smaller than that of ``parallel''.
Comparing ``threshold changing@sensors'' and ``parallel'' we note that  the local decisions in ``threshold changing@sensors'' have an enhanced reliability, due to the fact that $u_i,\bar{u}_i$ at ${\cal S}_i$ are obtained based on comparing the sensor's observation $x_i$ with three thresholds (instead of one). For moderate/high SNR$_h$ ``threshold changing@sensors'' outperforms ``parallel'' (regardless of SNR$_c$). This observation can be explained as follows.
In this SNR$_h$ regime,
 $\bar{P}_e$ is dominated by the local probability error at ${\cal S}_i$. Since the reliability of local decisions in ``threshold changing@sensors'' exceeds that of local decisions in ``parallel'' we expect that ``threshold changing@sensors'' outperforms ``parallel''. Furthermore,  ``threshold changing@sensors''  improves $\bar{P}_e$ over ``parallel'', via providing ``decision diversity'' through Alamouti's STC. For low SNR$_h$ ``threshold changing@sensors'' outperforms ``parallel'' only for low SNR$_c$. This is because in this regime two factors contribute to $\bar{P}_e$: unreliable local decisions and communication channel errors. Since ``threshold changing@sensors'' increases the reliability of local decisions, its performance exceeds that of ``parallel''. On the other hand, for low SNR$_h$ and moderate/high SNR$_c$, where the communication channel errors are the major contributors to $\bar{P}_e$, Alamouti's STC introduces destructive signal interference and degrades performance of ``threshold changing@sensors'' with respect to ``parallel''.

Table \ref{table:2} also tabulates  $\bar{P}_e$ of ``parallel'', ``STC@sensors'', ``fusion@sensors'' and ``threshold changing@sensors'', as  SNR$_h$ and SNR$_c$ vary, for $\pi_0\!=\!0.7$ and $\rho\!=\!0$, when the FC employs the LRT rule. Similar observations can be made as we compare ``STC@sensors" and ``fusion@sensors'' against ``parallel'' for different SNR$_h$ and SNR$_c$ regimes, regardless of $\pi_0$. However, when comparing ``threshold changing@sensors'' against ``parallel'' in low SNR$_h$, we note that the behavior changes as $\pi_0$ increases. In particular, for low SNR$_h$ ``threshold changing@sensors'' outperforms ``parallel'' for low SNR$_c$ when $\pi_0$ is smaller. As $\pi_0$ increases, ``threshold changing@sensors'' outperforms ``parallel'' for low/moderate SNR$_c$, i.e., for low SNR$_h$ the range of SNR$_c$ values over which ``threshold changing@sensors'' outperforms ``parallel'' expands as $\pi_0$ increases. Overall,  Table \ref{table:2} indicates that, for low SNR$_h$ and moderate/high SNR$_c$, the proposed schemes do not have an advantage over ``parallel''. The exception is when $\pi_0$ is large enough ($\pi_0 \geq 0.7$), in which case ``threshold changing@sensors'' outperforms ''parallel''.  On the other hand, for low SNR$_h$ and low SNR$_c$ ``fusion@sensors'' and ``threshold changing@sensors'' outperform ``parallel''. %STC is even worse than parallel in this regime
For moderate/high SNR$_h$, regardless of SNR$_c$,   the schemes ranked from lowest to highest $\bar{P}_e$ are ``threshold changing@sensors'',   ``fusion@sensors'', ``STC@sensors'' and ``parallel''.
%==========majority rule===================
Table \ref{table:4} is similar to Table \ref{table:2}, with the difference that the FC employs the majority rule. Comparing each of the schemes ``STC@sensors'', ''fusion@sensors'' and ``threshold changing@sensors'' against  ``parallel''  for different SNR$_h$ and SNR$_c$ regimes, we observe similar trends for the majority and LRT rules. However, when we compare the schemes to rank them based on their $\bar{P}_e$ we note the differences. In particular, for very high SNR$_c$, ``STC@sensors'' outperforms all the schemes, for high SNR$_c$ none of the proposed schemes has an advantage over ``parallel'', and for moderate/low SNR$_c$ ``fusion@sensors'' outperforms all the schemes.

%==============correlation============================

\underline{\it Impact of correlation on performance comparison}:
Table \ref{table:correlation} tabulates $\bar{P}_e$ of ``parallel'', ``STC@sensors'', ``fusion@sensors'' and ``threshold changing@sensors'', as  SNR$_h$ and SNR$_c$ vary, for $\pi_0\!=\!0.7$ and $\rho\!=\!0.1, 0.2, 0.3, 0.5, 0.8$, when the FC employs the LRT rule. We observe that as $\rho$ increases the performance gap between the proposed schemes  and ``parallel'' reduces. This observation can be explained as follows. As we mentioned before, the performance advantage of ``fusion@sensors'' and ``threshold changing@sensors'' over ``parallel'', when the FC employs the LRT rules, is mainly due to the fact that the local information exchange in ``fusion@sensors'' or three-threshold-based test at the sensors in ``threshold changing@sensors'' would enhance the reliability of the local decisions (compared with ``parallel'') when Gaussian sensing noises are uncorrelated.
As these noises  become correlated and $\rho$ increases, the increase in the reliability of the local decisions in ``fusion@sensors'' and ``threshold changing@sensors'' diminishes and  thus these two schemes start to lose their performance gain over ``parallel''.
For $\rho \! \leq \! 0.2$  the observations made on the performance comparison among these schemes remain the same as $\rho\!=\!0$.  When $\rho$ varies between $0.2\!-\!0.3$, ``threshold changing@sensors'' outperforms others for high SNR$_h$, ``fusion@sensors'' outperforms others for medium SNR$_h$, and ``parallel'' and ``fusion@sensor'' outperform others for low SNR$_h$ (all regardless of SNR$_c$). When $\rho\!=\!0.5$ for high SNR$_h$ and high SNR$_c$ ``threshold changing@sensors'' outperforms others.  For high SNR$_h$ and medium/low SNR$_c$  and for medium SNR$_h$ (regardless of SNR$_c$) ``fusion@sensors'' outperforms others.
For low SNR$_h$ (regardless of SNR$_c$) ``parallel'' and ``fusion@sensor'' outperform others.  When $\rho\!=\!0.8$ ``threshold changing@sensors'' has an inferior performance, regardless of SNR$_h$ and SNR$_c$.
%Although considering correlation among sensors, the reliability of local decision at ``threshold changing@sensors'' is still greater than the other schemes including ``fusion@sensors'', it is not that much high to compensate the effect of destructive interference caused by Alamouti's space-time code, hence causing the ``fusion@sensors'' to have the best performance.
Table \ref{table:correlation} also shows that the performance degradation of ``threshold changing@sensors''  is pronounced, as $\rho$ increases, compared with other schemes.
%This observation can be explained as the following.
Note that at $\rho\!=\!0$ ``threshold changing@sensors'' has the lowest error floor, whereas at $\rho\!=\!1$ all schemes have the same error floor. These imply that the rate of performance degradation of ``threshold changing@sensors'' must be higher than other schemes.

%=================increasing K================================

\underline{\it Impact of increasing $K$}: Table \ref{table:incresing-number-sensors} tabulates $\bar{P}_e$ of ``parallel'', ``STC@sensors'', ``fusion@sensors'' and ``threshold changing@sensors'', as  SNR$_h$ and SNR$_c$ vary, for $\pi_0\!=\!0.6, \rho\!=\!0, K\!=\!20$, when the FC employs the LRT rule. The observations made on the comparison between these scheme for $K\!=\!10$ remain true.

%=================increasing number of partners======================

\underline{\it Discussion on increasing number of cooperative partners in a group}: To investigate how increasing number of partners impacts the performance, we consider a network of $K\!=\!4$ sensors. Suppose sensors are positioned on the circumference of a circle on
$x\!-\!y$ plane, whose center is located at the origin and its diameter is  $2\sqrt{2}m$,  and ${\cal S}_i$ is equally distant from ${\cal S}_j$ and ${\cal S}_k$ such that $d_{ij}\!=\!d_{ik}\!=\!2m, d_{jk}\!=\!2\sqrt{2}m$. Also, the FC is located above the origin (above $x\!-\!y$ plane), such that all sensors are at equal distance of $d\!=\!10m$ from the FC. Let ``STC4@sensors'', ``fusion4@sensors'' and ``threshold changing4@sensors'', respectively, refer to schemes (i), (ii), (iii) with 4 partners in one group and ``STC@sensors'', ``fusion@sensors'' and ``threshold changing@sensors'', respectively, refer to schemes (i), (ii), (iii) with 2 partners in one group (two groups in the network).  Table \ref{table:9} tabulates $\bar{P}_e$ of all schemes as  SNR$_h$ and SNR$_c$ vary, for $\pi_0\!=\!0.6, 0.7$ and $\rho\!=\!0$, when the FC employs LRT rule. Comparing all these schemes, we observe that for low SNR$_h$ either ``parallel'' or ''threshold changing@sensors'' outperforms others (depending on SNR$_c$), whereas for moderate/high SNR$_h$ ``threshold changing@sensors'' outperforms others, including ``fusion4@sensors''. These observations suggest that  no performance gain is achieved as the number of cooperative partners increases beyond $2$. In the following we provide our intuitive reasoning on how we expect schemes (i),(ii),(iii) perform, as the number of cooperative partners in a group increases, assuming $K$ and sensor placements are fixed. Going from ``STC@sensors'' to ``STC4@sensors'', for a fixed transmit power per sensor ${\cal P}$, we expect the power consumption for inter-sensor communication $(1-\alpha){\cal P}$ increases, as the average distances between sensors within a group increase. This leaves a sensor with a smaller power, i.e., smaller $\alpha{\cal P}$, for its communication with the FC. For moderate/high SNR$_c$, the relative performance of ``STC@sensors'' and ``STC4@sensors'' depends on ${\cal P}$ value. Our simulations show that for ${\cal P}\!> \!32$mW,
$\alpha{\cal P}$ is large enough that ``STC4@sensors'' provides a larger ``decision diversity gain'' than that of ``STC@sensors'' during sensor-FC communication, and thus the former outperforms the latter.  For low SNR$_c$, however, these schemes have similar performance, since performance in this regime is limited by the reliability of local decisions at sensors (which are the same for both schemes). Overall, these imply that for wireless sensor networks that typically operate within $ 0.12 \! \leq \! {\cal P} \! \leq \! 36$mW \cite{power-sensors}, increasing the number of cooperative partners in a group beyond 2 does not have much practical incentive. Similarly, going from ``fusion@sensors'' to ``fusion4@sensors'',  we expect $(1-\alpha){\cal P}$ increases, while $\alpha{\cal P}$ decreases. However, different from scheme (i), in scheme (ii) local information exchange affects the reliability of local decisions. Hence, the increase of the reliability of the local decisions in ``fusion4@sensors'' due to the increase of $(1-\alpha){\cal P}$ still compensates for a less reliable sensor-FC communication due to the decrease of $\alpha{\cal P}$, which leads to the observation that  ``fusion4@sensors'' outperforms ``fusion@sensors'' for ${\cal P} \! \geq \! 10$mW. Going to ``fusion6@sensors'' for $K\!=\!6$, however, $\alpha$ decreases further, such that even for a large ${\cal P} \! \approx \! 40$mW, the unreliability of the sensor-FC communication due to the decrease of $\alpha{\cal P}$ leads to the observation that ``fusion6@sensors'' performs worse than ``fusion@sensors'' (see Table \ref{table:fusion6}). We expect similar performance degradation as we increase the number of partners in a group beyond 6.
Going from ``threshold changing@sensors'' to ``threshold changing4@sensors'', the former outperforms the latter for all SNR$_c$ and SNR$_h$. This is due to the fact that, as the number of cooperative partners in a group increases, the chance that the corresponding information matrix transmitted by this group to the FC deviates from the conventional orthogonal STC matrix increases, leading to destructive signal interference at the FC and diminishing the ``decision diversity gain'' of STC. We conjecture similar performance degradation as we increase the number of partners in a group beyond 4.

%===================inhomogeneous sensor's displacement================================

\underline{\it Homogeneous versus inhomogeneous sensor placement}: We consider a network of $K\!=\!4$ sensors, consisting of two groups, where sensors are positioned on the circumference of a circle on $x\!-\!y$ plane, whose center is located at the origin and its diameter is  $20$m. The distance between two sensors in each group is $d_0=2$m. For homogeneous placement, we assume that the FC is located at the origin and for inhomogeneous placement, we move the FC toward one of the groups, such that the distance between the FC and the two groups are $4$m and $16$m. Table \ref{table:heterogenous} tabulates $\bar{P}_e$ of ``parallel'', ``STC@sensors'', ``fusion@sensors'' and ``threshold changing@sensors'', as  SNR$_h$ and SNR$_c$ vary, for $\pi_0\!=\!0.6, \rho\!=\!0$, when the FC employs the LRT rule, for homogeneous and inhomogeneous placements. This table shows that our findings on
comparison between different schemes is exactly the same as Table \ref{table:2}, which was another example of a homogenous placement.
On the other hand, for inhomogeneous placement, ``threshold changing@sensors'' outperforms other schemes, regardless of ${\cal P}$ and SNR$_c$ values. Since ``threshold changing@sensors'' has the lowest error floor among all schemes, when one group of sensors becomes closer to the FC, the enhanced reliability of the information delivered  to the FC by this group leads to the superior performance of this scheme. Comparing ``STC@sensors'' and ``parallel'', we note the former performs worse than the latter in inhomogeneous placement. Since ``STC@sensors'' and ``parallel'' have similar error floor, placing one group of sensors closer to the FC does not change the reliability of the information provided by this group to the FC in either schemes. However, since the other group of sensors becomes farther away from the FC, the quality of the information delivered to the FC by this group decreases (due to destructive signal interference of STC), leading to the inferior performance of ``STC@sensors''.

%=====================Conclusions========================
\section{Conclusions}
For the problem of binary distributed detection in a wireless sensor network, we have proposed novel cooperative and parallel fusion architectures, to combat fading effects encountered in the conventional parallel fusion architecture. In particular, we have proposed: (i) cooperative fusion architecture with Alamouti's STC scheme at sensors, (ii) cooperative fusion architecture with signal fusion at sensors, and (iii) parallel fusion architecture with local threshold changing at sensors. While there is a limited local information exchange among the sensors (1-bit message) in schemes (i) and (ii), there is no explicit information exchange in scheme (iii). For these schemes, we derived the optimal LRT and the suboptimal majority fusion rules and analyzed their performance, in terms of communication and sensing SNRs. Our numerical results show that, when the FC employs the LRT rule, unless for low communication SNR and moderate/high sensing SNR, performance improvement is feasible with the new cooperative and parallel fusion architectures, while scheme (iii) outperforms others. When the FC utilizes the majority rule, such improvement is possible, unless for high sensing SNR. In particular, for very high sensing SNR scheme (i) outperforms, whereas for moderate/low sensing SNR scheme (ii) outperforms others.

%================================================
\section*{Appendix A}
{\underline{\it Upper Bounds on ${\cal T}_{e_1|h}$ in (\ref{pe1-first}) and its average $\bar{\cal T}_{e_1}$}}: For $Q_{n} <  M$
%(i.e., the sensing channels are reliable) we can find an upper bound on ${\cal T}_{e_1|h}$ smaller than one. Assuming  $Q_n <M$
we have
\begin{eqnarray}
{\cal T}_{e_1|h}=P(\sum_{d_{n_1,m_1} \in S_1}d_{n_1,m_1}> - \sum_{d_{n_1,m_1} \in S_0}d_{n_1,m_1} |F_{n,m})< P(\sum_{d_{n_1,m_1} \in S_1}d_{n_1,m_1}> - d_{n,m} |F_{n,m}).\label{p-ineq}
\end{eqnarray}
where the bound
%upper bound on ${\cal T}_{e_1|h}$
in (\ref{p-ineq}) is obtained noting that $d_{n,m} \! \in \! S_0$ and $\sum_{d_{n_1,m_1} \in S_0}d_{n_1,m_1}  \! < \!d_{n,m} $.
To further bound (\ref{p-ineq}) we define the interval $x$ and the function $\varphi$ as the following
\begin{align}
x=\mathbb{E}\{\sum_{d_{n_1,m_1} \in S_1} C_{n_1,m_1} d_{n_1,m_1} |F_{n,m}\},~\varphi(x)=P(\sum_{d_{n_1,m_1} \in S_1} C_{n_1,m_1} d_{n_1,m_1}>- d_{n,m}|F_{n,m}).
\end{align}
where
%continuous-valued
constants $C_{n_1,m_1}$ take values in the interval $[0,|S_1|]$. Our numerical results suggest that for small $|S_1|$, $\varphi$ is convex over $x$.
%(see Fig. \ref{fig1}).
Invoking the inequality $\varphi(\frac{\sum_{i=1}^n x_i}{n})\! \leq \! \frac{\sum_{i=1}^n \varphi(x_i)}{n}$, where the points $x_1,...,x_n$ belong to $x$ \cite{convex-book}, and letting $n\!=\!|S_1|$ and $x_i\!=\!\mathbb{E}\{|S_1| d_{n_1,m_1} |F_{n,m}\} $ for $i=1,...,|S_1|$ we establish below
\begin{eqnarray}
\!\!\!\!\!\!\varphi(\sum_{d_{n_1,m_1} \in S_1}\mathbb{E}\{d_{n_1,m_1}|F_{n,m}\})&\!=\!&\varphi(\frac{1}{|S_1|} \sum_{d_{n_1,m_1} \in S_1}\mathbb{E}\{|S_1| d_{n_1,m_1} |F_{n,m}\})\nonumber\\
 &\! \leq \!& \frac{1}{|S_1|} \sum_{d_{n_1,m_1}\in S_1}\varphi(\mathbb{E}\{|S_1|d_{n_1,m_1}|F_{n,m}\}). \label{phi-inq}
\end{eqnarray}
The inequality in (\ref{phi-inq}) implies that the upper bound on ${\cal T}_{e_1|h}$ in (\ref{p-ineq}) can be further bounded as
\begin{equation}
P(\sum_{d_{n_1,m_1} \in S_1}d_{n_1,m_1} >- d_{n,m}|F_{n,m}) \leq \frac{1}{|S_1|} \sum_{d_{n_1,m_1} \in S_1}P(|S_1|d_{n_1,m_1}>-d_{n,m}|F_{n,m}).\label{last-upper-bound-pe1}
\end{equation}
The new bound on ${\cal T}_{e_1|h}$ in (\ref{last-upper-bound-pe1}) can be presented  in closed-form, considering  the definitions of $d_{n,m}$ and $d_{n_1,m_1}$ in (\ref{d-definition}) and noting that, conditioned on $u_{2s-1}, u_{2s}, \hat{u}_{2s-1}, \hat{u}_{2s}$, the terms $z_{2s-1},z_{2s}$ are independent complex Gaussian RVs with the variance $\sigma^2=(|h_{2s-1}|^2+|h_{2s}|^2)\sigma_v^2$ and the means $\mu^{n_1,m_1}_{2s-1}, \mu^{n_1,m_1}_{2s}$ for $d_{n_1,m_1}$ and $\mu^{n,m}_{2s-1}, \mu^{n,m}_{2s}$ for $d_{n,m}$.
%(these means are calculated using (\ref{dstc-mean})).
%To find this closed-form expression, we first
Mapping the noises $\delta_{ij}^1, \delta_{ij}^2$ in Section \ref{DSTC-node-FC} into $\delta_s^1, \delta_s^2$,
%After some straightforward calculations,
we find
\begin{equation}\label{convex-one}
\!\!\!\!\!\!\!\!\!P(|S_1|d_{n_1,m_1}\!>\!-d_{n,m}|F_{n,m})\!=\!P\big( \! \zeta(n,m,n_1,m_1) > -\ln(|S_1|G(n,m,n_1,m_1))\!+\!I(n,m,n_1,m_1) \!\big),
\end{equation}
where $G(n,m,n_1,m_1)$ is defined in (\ref{G-matrix-definition}) and
$I(n,m,n_1,m_1)\!=\! \underset{s=1}{\overset{S}{\sum}}    ~{\cal K}_s(n,m,n_1,m_1)$ in which
\begin{eqnarray*}
\!\!\!\!{\cal K}_s(n,m,n_1,m_1)  \!\!&\!\!=\!\!&\!\! \frac{1}{\sigma^2}(|\mu^{n_1,m_1}_{2s-1}-\mu^{n,m}_{2s-1}|^2 + |\mu^{n_1,m_1}_{2s}-\mu^{n,m}_{2s}|^2) \nonumber \\
\!\!&\!\!=\!\!&\!\! \frac{\alpha}{2\sigma_v^2}(|h_{2s-1}(a_{n_1}^{2s-1}-a_{n}^{2s-1})+h_{2s}(a_{n_1}^{2s}-a_{n}^{2s})|^2\nonumber\\
\!\!&\!\!+\!\!&\!\!|h_{2s-1}(a_{m_1}^{2s-1}-a_{m}^{2s-1})+h_{2s}(a_{m_1}^{2s}-a_{m}^{2s})|^2),\\
\!\!\!\!\zeta(n,m,n_1,m_1)\!\!&\!\!=\!\!&\!\! \sum_{s=1}^{S}\theta_s(n,m,n_1,m_1)\nonumber\\
\mbox{where}~\theta_s(n,m,n_1,m_1)&=&\frac{2}{\sigma^2} \mathfrak{Re} \big\{ \delta_s^1(\mu^{n_1,m_1}_{2s-1}-\mu^{n,m}_{2s-1})^*+\delta_s^2(\mu^{n_1,m_1}_{2s}-\mu^{n,m}_{2s})^*\big\}.
\label{I-definition}
\end{eqnarray*}
Recall
%from Section \ref{DSTC-node-FC}
%that
% the noises
$\delta_s^1$ and $\delta_s^2$ are i.i.d zero mean complex Gaussian RVs with the variance $\sigma^2$. Hence,   $\zeta(n,m,n_1,m_1)$ is a zero mean Gaussian RV with the variance $2 I(n,m,n_1,m_1)$. Thus we can express (\ref{convex-one}) as
\begin{equation}\label{P-in-terms-of-Q}
P(|S_1|d_{n_1,m_1}>-d_{n,m}|F_{n,m})=Q\left(\frac{- \ln (|S_1|G(n,m,n_1,m_1))+I(n,m,n_1,m_1)}{\sqrt{2 I(n,m,n_1,m_1)}}\right).
\end{equation}
Note that $I(n,m,n_1,m_1)$ depends on the coefficients $h_{2s-1},h_{2s}$ %through the means $\mu^{n_1,m_1}_{2s-1}, \mu^{n_1,m_1}_{2s},\mu^{n,m}_{2s-1}, \mu^{n,m}_{2s}$ and the variance $\sigma^2$;
, whereas $G(n,m,n_1,m_1)$ is independent of these coefficients. In fact, $G(n,m,n_1,m_1)$ depends on sensing channels through $P_d, P_f$ and the average received SNR $\bar{\gamma}_{hs}$ corresponding to inter-node communication through $T_{n,m},T_{n_1,m_1}$.
One can verify that when $\pi_0 > \pi_1$,  we have
%the term $G(n,m,n_1,m_1)$ takes very small values and thus
$- \ln (|S_1|G(n,m,n_1,m_1))\!+\!I(n,m,n_1,m_1)\!>\!0$.
%This allows us to use the Chernoff bound of $Q$-function $Q(x)\!<\!\frac{1}{2}e^{-\frac{x^2}{2}}$ for $x\!>\!0$, in order to find an upper bound on the probability in (\ref{P-in-terms-of-Q}).
Combining (\ref{p-ineq}), (\ref{last-upper-bound-pe1}), (\ref{P-in-terms-of-Q}), using the Chernoff bound of $Q$-function $Q(x)\!<\!\frac{1}{2}e^{-\frac{x^2}{2}}$ for $x\!>\!0$, and also noting that  $ 0\!< \! e^{\frac{-(\ln(|S_1|G(n,m,n_1,m_1)))^2}{4 I(n,m,n_1,m_1)}}\!<\! 1$ and thus can be dropped without decreasing the upper bound, we find %the following bound
\begin{align}\label{final-cond-bound}
{\cal T}_{e_1|h} < \frac{\bone_{\{Q_n<M\}}}{2\sqrt{|S_1|}} \sum_{d_{n_1,m_1} \in S_1}\sqrt{G(n,m,n_1,m_1)}e^{-\frac{I(n,m,n_1,m_1)}{4}}+ \bone_{\{Q_n \geq M\}}.\end{align}
Finally, to find an upper bound on ${\bar {\cal T}}_{e_1}$ we need to take average of $e^{-\frac{I(n,m,n_1,m_1)}{4}}$  in (\ref{final-cond-bound}) over $h_{2s-1},h_{2s}$ for $s=1,...,S$.  Since $h_{2s-1},h_{2s} \sim {\cal C}{\cal N}(0,\sigma_h^2)$ are i.i.d across the pairs we have
\begin{eqnarray}\label{expectation-s}
\mathbb{E}\{e^{-\frac{I(n,m,n_1,m_1)}{4}}\}\!\!&\!\! = \!\!& \!\!\prod_{s=1}^S \int_{h_{2s-1}} \int_{h_{2s}}    e^{\frac{-{\cal K}_s(n,m,n_1,m_1)}{4}} e^{-\frac{(|h_{2s-1}|^2+|h_{2s}|^2)}{\sigma_h^2}}d_{h_{2s-1}} d_{h_{2s}}.
\end{eqnarray}
After some
%tedious but straightforward
calculations (\ref{expectation-s}) is reduced to $\prod_{s=1}^S {\cal D}_1(n,m,n_1,m_1)$ where ${\cal D}_1(n,m,n_1,m_1)$ is given in (\ref{final-term-pe1}).
The upper bound on ${\bar {\cal T}}_{e_1}$ is obtained by substituting $e^{-\frac{I(n,m,n_1,m_1)}{4}}$  in (\ref{final-cond-bound}) with
$\prod_{s=1}^S {\cal D}_1(n,m,n_1,m_1)$.
This completes our derivations for the upper bound on ${\bar {\cal T}}_{e_1}$. %=================================================

{\underline{\it Upper Bounds on ${\cal T}_{e_2|h}$ in (\ref{pe2-upper}) and and its average $\bar{\cal T}_{e_2}$}}: For $Q_{n}> M$
%(i.e., the sensing channels are not reliable) we can obtain an upper bound on ${\cal T}_{e_2|h}$ smaller than one. Assuming $Q_n > M$ and following similar steps as we have taken in deriving the upper bound of ${\cal T}_{e_1|h}$
we have
\begin{equation}
{\cal T}_{e_2|h}<  P(d_{n,m}<- \sum_{d_{n_1,m_1} \in S_0} d_{n_1,m_1}|F_{n,m}) < \frac{1}{|S_0|}\sum_{d_{n_1,m_1} \in S_0}P(d_{n,m}<-|S_0|d_{n_1,m_1}|F_{n,m}).\label{last-upper-bound-pe2}.
\end{equation}
noting that $d_{n,m} \! \in \! S_1$ and $\sum_{d_{n_1,m_1} \in S_1}d_{n_1,m_1} \! >\! d_{n,m} $.
The new bound on ${\cal T}_{e_2|h}$ in (\ref{last-upper-bound-pe2}) can be found in closed-form, via examining  the definitions of $d_{n,m}$ and $d_{n_1,m_1}$ in (\ref{d-definition}) and noting that, conditioned on $u_{2s-1}, u_{2s}, \hat{u}_{2s-1}, \hat{u}_{2s}$, the terms $z_{2s-1},z_{2s}$ are independent complex Gaussian RVs with the variance $\sigma^2=(|h_{2s-1}|^2+|h_{2s}|^2)\sigma_v^2$ and the means $\mu^{n_1,m_1}_{2s-1}, \mu^{n_1,m_1}_{2s}$ for $d_{n_1,m_1}$ and $\mu^{n,m}_{2s-1}, \mu^{n,m}_{2s}$ for $d_{n,m}$.
%(the means are calculated using (\ref{dstc-mean})).
%After some straightforward calculations, we find
Therefore
\begin{eqnarray}\label{cher-bound}
\!\!\!\!\!\!\!\!\!P( d_{n,m}<-|S_0|d_{n_1,m_1}|F_{n,m})\!=\!P\big(\zeta(n,m,n_1,m_1) \!>\!  -\ln(|S_0|G(n,m,n_1,m_1))\!+\!I(n,m,n_1,m_1) \big).
\end{eqnarray}
Comparing (\ref{cher-bound}), (\ref{convex-one}), it seems natural to write (\ref{cher-bound}) in terms of $Q$-function and apply Chernoff bound  to reach a bound.
%upper bound on (\ref{cher-bound}).
However, different from (\ref{convex-one}),
%one can verify that
when $\pi_0 \! > \! \pi_1$
%the term $G(n,m,n_1,m_1)$ in (\ref{cher-bound}) takes large values and thus
we no longer have $- \! \ln (|S_0|G(n,m,n_1,m_1)) \! + \! I(n,m,n_1,m_1)\!>\!0$. We use an alternative
%Alternatively, to find an upper bound on (\ref{cher-bound}) we recall the Chernoff
bound, which states $P(\sum_{s=1}^{S} x_s\!<\!a)\!<\!\underset{t, t>0}{\min}~e^{ta}\prod_{s=1}^{S} \mathbb{E}\{e^{-tx_s}\}$ when $x_1,...,x_S$ are independent RVs \cite{convex-book}. Combining (\ref{last-upper-bound-pe2}), (\ref{cher-bound}), letting $x_s\!=\!-\theta_s(n,m,n_1,m_1)$, $a\!=\!\ln(|S_0|G(n,m,n_1,m_1))\!-\!I(n,m,n_1,m_1)$ in (\ref{cher-bound}) and using the alternative bound, we find
%can establish the following bound
\begin{eqnarray}\label{final-cond-bound-pe2}
\!\!\!\!\!\!\!\!\!\!\!\!\!\!\!\!\!\!\!\!\!\!{\cal T}_{e_2|h} &\!<\!& \frac{\bone_{\{Q_n>M\}}}{{|S_0|}}\nonumber\\
 & \times &\sum_{d_{n_1,m_1} \in S_0}[\underset{t}{\min} ~e^{ -t I(n,m,n_1,m_1)}(|S_0|G(n,m,n_1,m_1))^t  \prod_{s=1}^{S}\mathbb{E}\{e^{t \theta_s(n,m,n_1,m_1)}\}]+ \bone_{\{Q_n \leq M\}}.
\end{eqnarray}
Noting that $-\theta_s(n,m,n_1,m_1) \sim {\cal C}{\cal N}(0, 2{\cal K}_s(n,m,n_1,m_1) )$ and using the moment generating function results we find $\mathbb{E}\{e^{t \theta_s(n,m,n_1,m_1)}\}\!=\! e^{ t^2 {\cal K}_s(n,m,n_1,m_1)}$. This implies we can rewrite (\ref{final-cond-bound-pe2}) as the following
\begin{equation}\label{final-cond-bound-last}
{\cal T}_{e_2|h} \!<\! \frac{\bone_{\{Q_n>M\}}}{{|S_0|}} \sum_{d_{n_1,m_1} \in S_0}[\underset{t}{\min} ~(|S_0|G(n,m,n_1,m_1))^t  \prod_{s=1}^{S} e^{ (t^2-t) {\cal K}_s(n,m,n_1,m_1)}]+ \bone_{\{Q_n \leq M\}}.
\end{equation}
To find a bound on ${\bar {\cal T}}_{e_2}$ we need to take average of $e^{(t^2-t) {\cal K}_s(n,m,n_1,m_1)}$  in (\ref{final-cond-bound-last}) over $h_{2s-1},h_{2s}$.
%Since $h_{2s-1},h_{2s} \sim {\cal C}{\cal N}(0,\sigma_h^2)$ after some tedious but straightforward calculations
One can verify that this term is equal to ${\cal D}_2(n,m,n_1,m_1)$ in (\ref{final-term-pe2}).
The upper bound on ${\bar {\cal T}}_{e_2}$ is obtained by substituting $e^{(t^2-t) {\cal K}_s(n,m,n_1,m_1)}$  in (\ref{final-cond-bound-last}) with
(\ref{final-term-pe2}). This completes our derivations for the bound on ${\bar {\cal T}}_{e_2}$.
\section*{Appendix B}
We analyze in details the behavior of our upper bounds on the average error probability for large $S$. In short, our analysis shows that, the difference between the error exponents of any two schemes (of the four schemes) depends on SNR$_h$, SNR$_c$ and $\pi_0$ only, and does not change with $S$ (recall $S$ is the number of two-sensor groups and $K\!=\!2S$ is the total number of sensors in the network). Therefore, our findings on performance comparison between different schemes for $\rho=0$ remain the same when $S$ increases.\\
Our detailed analysis follows. First for the ease of notation we define several new random vectors as the following:
let $d_s^{(P)}$ and $d_s^{'(P)}$ be two independent and identically distributed random vectors that have the same distribution as the random vector  $[u_{2s-1},u_{2s}]$; Let $d_s^{(i)}$ and $d_s^{'(i)}$ be two independent and identically distributed random vectors that have the same distribution as the random vector   $[u_{2s-1},u_{2s},\hat{u}_{2s-1},\hat{u}_{2s}]$; Let $d_s^{(ii)}$ and $d_s^{'(ii)}$ be two independent and identically distributed random vectors that have the same distribution as the random vector   $[\tilde{u}_{2s-1},\tilde{u}_{2s}]$; And let $d_s^{(iii)}$ and $d_s^{'(iii)}$ be two independent and identically distributed random vectors that have the same distribution as the random vector   $[u_{2s-1},u_{2s},\bar{u}_{2s-1},\bar{u}_{2s}]$.

%======================================================================
\subsection{Classical Parallel Fusion Architecture}
Using the definitions of ${P}_{e_1|h}$ and ${P}_{e_2|h}$ in (28), $\bar{P}_{e_1}$ and $\bar{P}_{e_2}$ can be written as
\begin{equation}\label{p-e1-response}
\bar{P}_{e_1}=\pi_0\sum_n \bar{{\cal T}}_{e_1}P(F_n|{\cal H}_0), ~~~\bar{P}_{e_2}=\pi_1\sum_n \bar{{\cal T}}_{e_2}P(F_n|{\cal H}_0)
\end{equation}
where upper bounds on $\bar{{\cal T}}_{e_1}$ and $\bar{{\cal T}}_{e_2}$ are derived and  $P(F_{n}|{\cal H}_{1}) \!=\! P_d^{Q_{n}}(1-P_d)^{K-Q_{n}}$ and $P(F_{n}|{\cal H}_{0}) \!=\! P_f^{Q_{n}}(1-P_f)^{K-Q_{n}}$ are obtained in Section VI.B. Substituting these into (\ref{p-e1-response}) we can write the following
\begin{equation}
\bar{P}_{e_1}<\bar{P}_{e_{11}}+\bar{P}_{e_{12}}, ~~~
\bar{P}_{e_2}<\bar{P}_{e_{21}}+\bar{P}_{e_{22}}\nonumber
\end{equation}
where
\begin{eqnarray}
\bar{P}_{e_{11}}&=&\sum_n \frac{\bone_{\{Q_n<M\}}}{2\sqrt{|S_1|}} \sum_{d_{n_1} \in S_1}[\sqrt{G(n,n_1)} \prod_{s=1}^{S} {\cal D}_1(n,n_1)] \pi_0P_f^{Q_{n}}(1-P_f)^{K-Q_{n}}\nonumber\\
\bar{P}_{e_{12}}&=&\sum_n(\bone_{\{Q_n>M\}})\pi_0P_f^{Q_{n}}(1-P_f)^{K-Q_{n}}\nonumber\\
\bar{P}_{e_{21}}&=&\sum_n \frac{\bone_{\{Q_n>M\}}}{|S_0|} \sum_{d_{n_0} \in S_0}[\underset{t}{\min} ~(|S_0|G(n,n_1))^t \prod_{s=1}^{S} {\cal D}_1(n,n_1)] \pi_1P_d^{Q_{n}}(1-P_1)^{K-Q_{n}}\label{pe-21-first}\\
&=&\sum_n \frac{\bone_{\{Q_n>M\}}}{|S_0|^{1-t_0}} \sum_{d_{n_0} \in S_0}[G(n,n_1)^{t_0} \prod_{s=1}^{S} {\cal D}_1(n,n_1)] \pi_1P_d^{Q_{n}}(1-P_1)^{K-Q_{n}} \nonumber\\
\bar{P}_{e_{22}}&=&\sum_n(\bone_{\{Q_n<M\}})\pi_1P_d^{Q_{n}}(1-P_d)^{K-Q_{n}}\nonumber
\end{eqnarray}
where $t_0$ is the value that minimizes $\bar{P}_{e_{21}}$ in (\ref{pe-21-first}). Recall that
$\bar{P}_{e_{12}}$ and $\bar{P}_{e_{22}}$ are the error floors (due to sensing noises), when communication channel is error-free. In the following, we discuss $\bar{P}_{e_{12}}$, $\bar{P}_{e_{11}}$, $\bar{P}_{e_{22}}$, $\bar{P}_{e_{21}}$ in asymptotic regime, as $S \rightarrow \infty$.
$\bullet$ $\bar{P}_{e_{12}}$:  we have $\bar{P}_{e_{12}}=P(L_{12}>\frac{\pi_0}{\pi_1}|{\cal H}_0)$ where the continuous random variable $L_{12}=\prod_{s=1}^{S} \frac{f(d_s^{(P)}|{\cal H}_1)}{f(d_s^{(P)}|{\cal H}_0)}$.
Therefore $\ln L_{12} =\sum_{s=1}^S L_{s_{12}}$, where the continuous random variable $L_{s_{12}}=\ln (\frac{f(d_s^{(P)}|{\cal H}_1)}{f(d_s^{(P)}|{\cal H}_0)})$. Since $L_{s_{12}}$'s are i.i.d random variables with mean $\mu_{l_{12}}=\mathbb{E}_{d_s^{(P)}|{\cal H}_0}\{L_{s_{12}}\}$ and variance $\sigma^2_{l_{12}}=VAR_{d_s^{(P)}|{\cal H}_0}(L_{s_{12}})$ that do not depend on $s$, we invoke the central limit theorem for large $S$ to obtain
\begin{align}
\bar{P}_{e_{12}}=Q(\frac{S\mu_{l_{12}}-\ln(\frac{\pi_0}{\pi_1})}{\sigma_{l_{12}}\sqrt{S}})\approx Q(\frac{\sqrt{S}\mu_{l_{12}}}{\sigma_{l_{12}}})<\underbrace{\kappa_{l_{12}}}_{=1/2}e^{-S\frac{\mu^2_{l_{12}}}{\sigma^2_{l_{12}}}}\nonumber
\end{align}

%====================================================================

$\bullet$ $\bar{P}_{e_{11}}$: After some tedious but straightforward mathematical manipulations, we reach
\begin{align}\label{Pe11}
\bar{P}_{e_{11}} &\overset{(a)}{=} \sum_n \frac{\bone_{\{Q_n<M\}}}{2\sqrt{|S_1|}}\nonumber\\
&\times \sum_{d_{n_1} \in S_1}[\sqrt{\frac{1-\frac{\pi_0P_f^{Q_{n_1}}(1-P_f)^{K-Q_{n_1}}}{\pi_1P_d^{Q_{n_1}}(1-P_d)^{K-Q_{n_1}}}}{1-\frac{\pi_1P_d^{Q_{n}}(1-P_d)^{K-Q_{n}}}{\pi_1P_f^{Q_{n}}(1-P_f)^{K-Q_{n}}}}} \sqrt{\pi_0P_f^{Q_{n}}(1-P_f)^{K-Q_{n}}}\sqrt{\pi_1P_d^{Q_{n_1}}(1-P_d)^{K-Q_{n_1}}}\prod_{s=1}^{S} {\cal D}_1(n,n_1)]\nonumber\\
&\overset{(b)}{<}
\frac{1}{2\sqrt{|S_1|}}\frac{1}{\sqrt{1-\frac{\pi_1P_d^{M-1}(1-P_d)^{K-M+1}}{\pi_1P_f^{M-1}(1-P_f)^{K-M+1}}}}
\sum_n\sum_{n_1}\sqrt{\pi_0P_f^{Q_{n}}(1-P_f)^{K-Q_{n}}}\sqrt{\pi_1P_d^{Q_{n_1}}(1-P_d)^{K-Q_{n_1}}}\prod_{s=1}^{S} {\cal D}_1(n,n_1)\nonumber\\
&\overset{(c)}{=}
\underbrace{\frac{\sqrt{\pi_0\pi_1}}{2\sqrt{|S_1|}}\frac{1}{\sqrt{1-\frac{\pi_1P_d^{M-1}(1-P_d)^{K-M+1}}{\pi_0P_f^{M-1}(1-P_f)^{K-M+1}}}}}_{=\kappa_{l_{11}}}\nonumber\\
&\sum_{a_n}\sum_{a_{n_1}}\Big(\prod_{s=1}^{S} P(u_{2s-1}=a_n^{2s-1},u_{2s}=a_n^{2s}|{\cal H}_0) P(u_{2s-1}=a_{n_1}^{2s-1},u_{2s}=a_{n_1}^{2s}|{\cal H}_1) \nonumber\\
&\times \prod_{s=1}^{S} \frac{{\cal D}_1(n,n_1)}{\sqrt{P(u_{2s-1}=a_n^{2s-1},u_{2s}=a_n^{2s}|{\cal H}_0)}\sqrt{ P(u_{2s-1}=a_{n_1}^{2s-1},u_{2s}=a_{n_1}^{2s}|{\cal H}_1)}}\Big)
\end{align}
where $(a)$ follows by substituting $G(n,n_1)$ from Section VI.B  in $\bar{P}_{e_{11}}$, $(b)$ follows from the fact that the added terms in the righthand side of the inequality are all positive, $\frac{\pi_1P_d^{Q_n}(1-P_d)^{K-Q_n}}{\pi_1P_f^{Q_n}(1-P_f)^{K-Q_n}}<\frac{\pi_1P_d^{M-1}(1-P_d)^{K-M+1}}{\pi_0P_f^{M-1}(1-P_f)^{K-M+1}}<1,~Q_n<M$ and also $0<\frac{\pi_0P_f^{Q_{n_1}}(1-P_f)^{K-Q_{n_1}}}{\pi_1P_d^{Q_{n_1}}(1-P_d)^{K-Q_{n_1}}}<1$ when $d_{n_1} \in S_1$, and (c) follows from the substitution $P_f^{Q_{n}}(1-P_f)^{K-Q_{n}}=\prod_{s=1}^{S} (P(u_{2s-1}=a_n^{2s-1},u_{2s}=a_n^{2s}|{\cal H}_0))$ and $P_d^{Q_{n_1}}(1-P_d)^{K-Q_{n_1}}=\prod_{s=1}^{S} (P(u_{2s-1}=a_{n_1}^{2s-1},u_{2s}=a_{n_1}^{2s}|{\cal H}_1))$.

We define the discrete random variable $L_{11}$ such that $\ln L_{11}=\sum _{s=1}^{S} L_{s_{11}}$, where the discrete random variable $L_{s_{11}}=\ln (\frac{D_1(d_s^{(P)},d_s^{'(P)})}{\sqrt{P(d_s^{(P)}|{\cal H}_1)}\sqrt{P(d_s^{'(P)}|{\cal H}_0)}})$. Our intuition behind defining the discrete random variable $L_{s_{11}}$ was that, $D_1(n,n_1)$ defined in Section VI.B, can be viewed as a realization of a discrete random variable. To find the corresponding random variable, we substitute $[a^{2s-1}_n,a^{2s}_n]$ and $[a^{2s-1}_{n_1},a^{2s}_{n_1}]$, respectively, with $d_s^{(P)}$ and $d_s^{'(P)}$ in $D_1(n,n_1)$.

Note that the probability of the discrete random variable $L_{11}$ assuming the particular value of \\$\prod_{s=1}^{S} \frac{{\cal D}_1(n,n_1)}{\sqrt{P(u_{2s-1}=a_n^{2s-1},u_{2s}=a_n^{2s}|{\cal H}_0)}\sqrt{P(u_{2s-1}=a_{n_1}^{2s-1},u_{2s}=a_{n_1}^{2s}|{\cal H}_1)}}$ is equal to $\prod_{s=1}^{S} P(u_{2s-1}=a_n^{2s-1},u_{2s}=a_n^{2s}|{\cal H}_0) P(u_{2s-1}=a_{n_1}^{2s-1},u_{2s}=a_{n_1}^{2s}|{\cal H}_1)$. Hence $(c)$ in (\ref{Pe11}) can be interpreted as calculating the expectation of a random variable, that is
\begin{align}
\bar{P}_{e_{11}} <
\kappa_{l_{11}} \mathbb{E}_{d_1^{(P)}|{\cal H}_0,...,d_S^{(P)}|{\cal H}_0,d_1^{'(P)}|{\cal H}_1,...,d_S^{'(P)}|{\cal H}_1}\{L_{11}\}\nonumber
\end{align}
Note that $L_{s_{11}}$'s are i.i.d random variable with mean $\mu_{l_{11}}=\mathbb{E}_{d_s^{(P)}|{\cal H}_0,d_s^{'(P)}|{\cal H}_1}\{L_{s_{11}}\}$     and variance $\sigma^2_{l_{11}}=VAR_{d_s^{(P)}|{\cal H}_0,d_s^{'(P)}|{\cal H}_1}(L_{s_{11}})$  that do not depend on $s$. Also, we have verified that $\mu_{l_{11}}+\frac{1}{2}\sigma^2_{l_{11}}<0$. We invoke the central limit theorem for large $S$ to say that that $\ln L_{11}$ is a Gaussian random variable with mean $S\mu_{l_{11}}$ and variance $S\sigma^2_{l_{11}}$. This implies that $L_{11}$ has log-normal distribution with mean $e^{S(\mu_{l_{11}}+\frac{1}{2}\sigma^2_{l_{11}})}$, that is
\begin{align}
\bar{P}_{e_{11}}< \kappa_{l_{11}}e^{S(\mu_{l_{11}}+\frac{1}{2}\sigma^2_{l_{11}})}\nonumber
\end{align}

%========================================================================

$\bullet$ $\bar{P}_{e_{22}}$: Following similar steps as we took for calculating  $\bar{P}_{e_{12}}$ for large $S$, we find
\begin{align}
\bar{P}_{e_{22}}=\Phi(\frac{\sqrt{S}\mu_{l_{22}}}{\sigma_{l_{22}}})= Q(-\frac{\sqrt{S}\mu_{l_{22}}}{\sigma_{l_{22}}}) <\underbrace{\kappa_{l_{22}}}_{=1/2}e^{-S\frac{\mu^2_{l_{22}}}{\sigma^2_{l_{22}}}}\nonumber
\end{align}
where $\mu_{l_{22}}=\mathbb{E}_{d_s^{(P)}|{\cal H}_1}\{L_{s_{22}}\}$, $\sigma^2_{l_{22}}=VAR_{d_s^{(P)}|{\cal H}_1}(L_{s_{22}})$ and the continuous random variable  $L_{s_{22}}=\ln (\frac{f(d_s^{(P)}|{\cal H}_1)}{f(d_s^{(P)}|{\cal H}_0)})$.\\

%========================================================================

$\bullet$ $\bar{P}_{e_{21}}$: Following similar steps as we took for calculating  $\bar{P}_{e_{11}}$ for large $S$, we find
\begin{align}
\bar{P}_{e_{21}}< \underbrace{\frac{1}{|S_0|^{1-t_0}}\frac{\pi_0^{t_0}\pi_1^{1-t_0}}{(1-\frac{\pi_0P_f^{M}(1-P_f)^{K-M}}{\pi_1P_d^{M}(1-P_d)^{K-M+1}})^{t_0}}}_{=\kappa_{l_{21}}}e^{S(\mu_{l_{21}}+\frac{1}{2}\sigma^2_{l_{21}})} \nonumber
\end{align}
where $\mu_{l_{21}}=\mathbb{E}_{d_s^{(P)}|{\cal H}_1,d_s^{'(P)}|{\cal H}_0}\{L_{s_{21}}\}$ and $\sigma^2_{l_{21}}=VAR_{d_s^{(P)}|{\cal H}_1,d_s^{'(P)}|{\cal H}_0}(L_{s_{21}})$ and the discrete random variable
%\begin{align}
$L_{s_{21}}=\ln (\frac{D_2(d_s^{(P)},d_s^{'(P)})}{(P(d_s^{'(P)}|{\cal H}_0))^{1-{t_0}}(P(d_s^{(P)})|{\cal H}_1))^{t_0}})$.
%\end{align}
Also, we have verified that $\mu_{l_{21}}+\frac{1}{2}\sigma^2_{l_{21}}<0$.
To find the discrete random variable $D_2(d_s^{(P)},d_s^{'(P)})$ we substitute $[a^{2s-1}_n,a^{2s}_n]$ and $[a^{2s-1}_{n_1},a^{2s}_{n_1}]$, respectively, with $d_s^{(P)}$ and $d_s^{'(P)}$ in $D_2(n,n_1)$ of Section VI.B.

%============================================================================

\subsection{Cooperative Fusion Architecture with STC at Sensors}
Following similar steps as Section A above, we redefine $\bar{P}_{e_{11}}$, $\bar{P}_{e_{12}}$, $\bar{P}_{e_{21}}$ and $\bar{P}_{e_{22}}$ and write the following
\begin{equation}
\bar{P}_{e_1}<\bar{P}_{e_{11}}+\bar{P}_{e_{12}}, ~~~
\bar{P}_{e_2}<\bar{P}_{e_{21}}+\bar{P}_{e_{22}}\nonumber
\end{equation}
where
\begin{eqnarray}
\bar{P}_{e_{11}}&=&\sum_n\frac{\bone_{\{Q_n<M\}}}{2\sqrt{|S_1|}} \sum_{d_{n_1,m_1} \in S_1} [\sqrt{G(n,m,n_1,m_1)} \prod_{s=1}^S {\cal D}_1(n,m,n_1,m_1)]\pi_0P_f^{Q_{n}}(1-P_f)^{K-Q_{n}}\nonumber\\
\bar{P}_{e_{12}}&=&\sum_n \bone_{\{Q_n>M\}}\pi_0P_f^{Q_{n}}(1-P_f)^{K-Q_{n}}\nonumber\\
\bar{P}_{e_{21}}&=&\sum_n\frac{\bone_{\{Q_n>M\}}}{{|S_0|}} \sum_{d_{n_1,m_1} \in S_0}[\underset{t}{\min} ~ (|S_0|G(n,m,n_1,m_1))^{t}  \prod_{s=1}^{S} {\cal D}_2(n,m,n_1,m_1)]\pi_0P_f^{Q_{n}}(1-P_f)^{K-Q_{n}}\label{pe-21-second}\\
&=&\sum_n\frac{\bone_{\{Q_n>M\}}}{{|S_0|}^{1-t_0}} \sum_{d_{n_1,m_1} \in S_0}[G(n,m,n_1,m_1)^{t_0}  \prod_{s=1}^{S} {\cal D}_2(n,m,n_1,m_1)]\pi_0P_f^{Q_{n}}(1-P_f)^{K-Q_{n}}\nonumber\\
\bar{P}_{e_{22}}&=&\sum_n \bone_{\{Q_n<M\}}\pi_1P_d^{Q_{n}}(1-P_d)^{K-Q_{n}}\nonumber
\end{eqnarray}
where $t_0$ is the value that minimizes $\bar{P}_{e_{21}}$ in (\ref{pe-21-second}). Recall that
$\bar{P}_{e_{12}}$ and $\bar{P}_{e_{22}}$ are the error floors, when communication channel is error-free. In the following, we discuss $\bar{P}_{e_{12}}$, $\bar{P}_{e_{11}}$, $\bar{P}_{e_{22}}$, $\bar{P}_{e_{21}}$ in asymptotic regime, as $S \rightarrow \infty$.

%============================================================================

$\bullet$ $\bar{P}_{e_{12}}$: Since this scheme, i.e., scheme (i), has the same error floor as the scheme discussed in Section A above, $\bar{P}_{e_{12}}$ in this section is equal to $\bar{P}_{e_{12}}$ in Section A above.

%============================================================================

$\bullet$ $\bar{P}_{e_{11}}$: Following the same steps taken in Section A above for calculating $\bar{P}_{e_{11}}$ for large $S$, we find
\begin{align}
\bar{P}_{e_{11}}<\underbrace{\frac{\sqrt{\pi_0\pi_1}}{2\sqrt{|S_1|}}\frac{1}{\sqrt{1-\frac{\pi_1P_d^{M-1}(1-P_d)^{K-M+1}}{\pi_0P_f^{M-1}(1-P_f)^{K-M+1}}}}}_{=\kappa_{l_{11}}} e^{S(\mu_{l_{11}}+\frac{1}{2}\sigma^2_{l_{11}})}\nonumber
\end{align}
where $\mu_{l_{11}}=\mathbb{E}_{d_s^{(i)}|{\cal H}_0,d_s^{'(i)}|{\cal H}_1}\{L_{s_{11}}\}$, $\sigma^2_{l_{11}}=VAR_{d_s^{(i)}|{\cal H}_0,d_s^{'(i)}|{\cal H}_1}(L_{s_{11}})$ and the discrete random variable
\begin{align*}
L_{s_{11}}=\ln \big(\frac{D_1(d_s^{(i)},d_s^{'(i)})}{\sqrt{P(d_s^{'(i)}|{\cal H}_1)}\sqrt{P(d_s^{(i)}|{\cal H}_0)}}\big)
\frac{P(\hat{u'}_{2s-1}|u'_{2s-1})P(\hat{u'}_{2s}|u'_{2s})}{P(\hat{u}_{2s-1}|u_{2s-1})P(\hat{u}_{2s}|u_{2s})}
\end{align*}
Also, we have verified that $\mu_{l_{11}}+\frac{1}{2}\sigma^2_{l_{11}}<0$.
To find the discrete random variable $D_1(d_s^{(i)},d_s^{'(i)})$ we substitute $[a^{2s-1}_n,a^{2s}_n,a^{2s-1}_m,a^{2s}_m]$ and $[a^{2s-1}_{n_1},a^{2s}_{n_1},a^{2s-1}_{m_1},a^{2s}_{m_1}]$ respectively with $d_s^{(i)}$ and $d_s^{'(i)}$ in $D_1(n,m,n_1,m_1)$ in Section VI.A.

%============================================================================

$\bullet$ $\bar{P}_{e_{22}}$: Since this scheme, i.e., scheme (i), has the same error floor as the scheme discussed in Section A, $\bar{P}_{e_{22}}$ in this section is equal to $\bar{P}_{e_{22}}$ in Section A above.

%============================================================================

$\bullet$ $\bar{P}_{e_{21}}$: Following the same steps taken in Section A for calculating $\bar{P}_{e_{21}}$ for large $S$, we find
\begin{align}
\bar{P}_{e_{21}}<\underbrace{\frac{1}{|S_0|^{1-{t_0}}}\frac{\pi_0^{t_0}\pi_1^{1-{t_0}}}{(1-\frac{\pi_0P_f^{M}(1-P_f)^{K-M}}{\pi_1P_d^{M}(1-P_d)^{K-M}})^{t_0}}}_{=\kappa_{l_{21}}}e^{S(\mu_{l_{21}}+\frac{1}{2}\sigma^2_{l_{21}})}\nonumber
\end{align}
where $\mu_{l_{21}}=\mathbb{E}_{d_s^{(i)}|{\cal H}_1,d_s^{'(i)}|{\cal H}_0}\{L_{s_{21}}\}$, $\sigma^2_{l_{21}}=VAR_{d_s^{(i)}|{\cal H}_1,d_s^{'(i)}|{\cal H}_0}(L_{s_{21}})$ and the discrete random variable
\begin{align*}
L_{s_{21}}=\ln (\frac{D_2(d_s^{(i)},d_s^{'(i)})}{(P(d_s^{'(i)}|{\cal H}_0))^{1-{t_0}}(P(d_s^{(i)}|{\cal H}_1))^{t_0}})\frac{(P(\hat{u'}_{2s-1}|u'_{2s-1})P(\hat{u'}_{2s}|u_{2s}))^{t_0}}{(P(\hat{u}_{2s-1}|u_{2s-1})P(\hat{u}_{2s}|u_{2s}))^{t_0}}
\end{align*}
Also, we verified that $\mu_{l_{21}}+\frac{1}{2}\sigma^2_{l_{21}}<0$.
To find the discrete random variable $D_2(d_s^{(i)},d_s^{'(i)})$ we substitute $[a^{2s-1}_n,a^{2s}_n,a^{2s-1}_m,a^{2s}_m]$ and $[a^{2s-1}_{n_1},a^{2s}_{n_1},a^{2s-1}_{m_1},a^{2s}_{m_1}]$ respectively with $d_s^{(i)}$ and $d_s^{'(i)}$ in $D_2(n,m,n_1,m_1)$ of Section VI.A.

%============================================================================

\subsection{Cooperative Fusion Architecture with Signal Fusion at Sensors}
Similar to previous sections, we redefine $\bar{P}_{e_{11}}$, $\bar{P}_{e_{12}}$, $\bar{P}_{e_{21}}$ and $\bar{P}_{e_{22}}$ and write the following
\begin{equation}
\bar{P}_{e_1}<\bar{P}_{e_{11}}+\bar{P}_{e_{12}}, ~~~
\bar{P}_{e_2}<\bar{P}_{e_{21}}+\bar{P}_{e_{22}}\nonumber
\end{equation}
where
\begin{eqnarray}
\bar{P}_{e_{11}}&=&\sum_n\frac{1-\bone_{\{Q^1_n,Q^2_n,Q^3_n\}}}{2\sqrt{|S_1|}} \sum_{d_{n_1} \in S_1} [\sqrt{G(n,n_1)} \prod_{s=1}^S ({\cal D}_1(n,n_1)]\pi_0P(\tilde{u}_{2s-1}=a_n^{2s-1},\tilde{u}_{2s}=a_n^{2s}|{\cal H}_0))\nonumber\\
\bar{P}_{e_{12}}&=&\sum_n(\bone_{\{Q^1_n,Q^2_n,Q^3_n\}})\pi_0\prod_{s=1}^S P(\tilde{u}_{2s-1}=a_n^{2s-1},\tilde{u}_{2s}=a_n^{2s}|{\cal H}_0)\nonumber\\
\bar{P}_{e_{21}}&=&\sum_n\frac{\bone_{\{Q^1_n,Q^2_n,Q^3_n\}}}{(|S_0|)} \nonumber\\
&\times&\sum_{d_{n_1} \in S_1} [\underset{t}{\min} ~ (|S_0|G(n,n_1))^{t} \prod_{s=1}^S ({\cal D}_1(n,n_1)]\pi_0P(\tilde{u}_{2s-1}=a_n^{2s-1},\tilde{u}_{2s}=a_n^{2s}|{\cal H}_1)]\label{pe-21-third-C}\\
&=&\sum_n\frac{\bone_{\{Q^1_n,Q^2_n,Q^3_n\}}}{(|S_0|)^{1-{t_0}}} \sum_{d_{n_1} \in S_1} [G(n,n_1)^{t_0} \prod_{s=1}^S ({\cal D}_1(n,n_1)]\pi_0P(\tilde{u}_{2s-1}=a_n^{2s-1},\tilde{u}_{2s}=a_n^{2s}|{\cal H}_1))\nonumber\\
\bar{P}_{e_{22}}&=&\sum_n(1-\bone_{\{Q^1_n,Q^2_n,Q^3_n\}})\pi_1\prod_{s=1}^S P(\tilde{u}_{2s-1}=a_n^{2s-1},\tilde{u}_{2s}=a_n^{2s}|{\cal H}_1)\nonumber
\end{eqnarray}
where $t_0$ is the value that minimizes $\bar{P}_{e_{21}}$ in (\ref{pe-21-third-C}). Recall that
$\bar{P}_{e_{12}}$ and $\bar{P}_{e_{22}}$ are the error floors, when communication channel is error-free. In the following, we discuss $\bar{P}_{e_{12}}$, $\bar{P}_{e_{11}}$, $\bar{P}_{e_{22}}$, $\bar{P}_{e_{21}}$ in asymptotic regime, as $S \rightarrow \infty$.

%============================================================================

$\bullet$ $\bar{P}_{e_{12}}$: Following the same steps taken in previous sections for calculating $\bar{P}_{e_{12}}$ for large $S$, we find
\begin{align}
\bar{P}_{e_{12}}\approx Q(\frac{\sqrt{S}\mu_{l_{12}}}{\sigma_{l_{12}}})<\underbrace{\kappa_{l_{12}}}_{=1/2}e^{-S\frac{\mu^2_{l_{12}}}{\sigma^2_{l_{12}}}}\nonumber
\end{align}

where $\mu_{l_{12}}=\mathbb{E}_{d_s^{(ii)}|{\cal H}_0}\{L_{s_{12}}\}$, $\sigma^2_{l_{12}}=VAR_{d_s^{(ii)}|{\cal H}_0}(L_{s_{12}})$ and the continuous random variable $L_{s_{12}}=\ln (\frac{f(d_s^{(ii)}|{\cal H}_1)}{f(d_s^{(ii)}|{\cal H}_0)})$.
%============================================================================

$\bullet$ $\bar{P}_{e_{11}}$: Following the same steps taken in previous sections for calculating $\bar{P}_{e_{11}}$ for large $S$, we find
\begin{align}\label{p-e-11-c}
\bar{P}_{e_{11}}<\underbrace{\frac{1}{2\sqrt{|S_1|}}\frac{\sqrt{\pi_0\pi_1}}{\sqrt{1-\text{LRT}_{\max}}}}_{=\kappa_{l_{11}}}e^{S(\mu_{l_{11}}+\frac{1}{2}\sigma^2_{l_{11}})}
\end{align}
where $\mu_{l_{11}}=\mathbb{E}_{d_s^{(ii)}|{\cal H}_0,d_s^{'(ii)}|{\cal H}_1}\{L_{s_{11}}\}$, $\sigma^2_{l_{11}}=VAR_{d_s^{(ii)}|{\cal H}_0,d_s^{'(ii)}|{\cal H}_1}(L_{s_{11}})$ and the discrete random variable
%\begin{align}
$L_{s_{11}}=\ln (\frac{D_1(d_s^{(ii)},d_s^{'(ii)})}{\sqrt{P(d_s^{'(ii)}|{\cal H}_1)}\sqrt{P(d_s^{(ii)}|{\cal H}_0)}})$.
%\end{align}
Also, we have verified that $\mu_{l_{11}}+\frac{1}{2}\sigma^2_{l_{11}}<0$.
To find the discrete random variable $D_1(d_s^{(ii)},d_s^{'(ii)})$ we substitute $[a^{2s-1}_n,a^{2s}_n]$ and $[a^{2s-1}_{n_1},a^{2s}_{n_1}]$ respectively with $d_s^{(ii)}$ and $d_s^{'(ii)}$ in $D_1(n,n_1)$ of Section VI.C.
Also, $\text{LRT}_{\max}$ in (\ref{p-e-11-c}) is
\begin{align*}
&\text{LRT}_{\max}=\max_{n}\big(\frac{\pi_1\prod_{s=1}^SP(\tilde{u}_{2s-1}=a_{n}^{2s-1},\tilde{u}_{2s}=a_{n}^{2s}|{\cal H}_1)}{\pi_0\prod_{s=1}^SP(\tilde{u}_{2s-1}=a_{n}^{2s-1},\tilde{u}_{2s}=a_{n}^{2s}|{\cal H}_0)}, \\ &\mbox{given}~\frac{\pi_1\prod_{s=1}^SP(\tilde{u}_{2s-1}=a_{n}^{2s-1},\tilde{u}_{2s}=a_{n}^{2s}|{\cal H}_1)}{\pi_0\prod_{s=1}^SP(\tilde{u}_{2s-1}=a_{n}^{2s-1},\tilde{u}_{2s}=a_{n}^{2s}|{\cal H}_0)}<1\big)
\end{align*}

%============================================================================

$\bullet$ $\bar{P}_{e_{22}}$: Following the same steps taken in previous sections for calculating $\bar{P}_{e_{22}}$ for large $S$, we find
\begin{align}
\bar{P}_{e_{22}}=\Phi(\frac{\sqrt{S}\mu_{l_{22}}}{\sigma_{l_{22}}})<\underbrace{\kappa_{l_{22}}}_{=1/2}e^{-S\frac{\mu^2_{l_{22}}}{\sigma^2_{l_{22}}}}\nonumber
\end{align}
where $\mu_{l_{22}}=\mathbb{E}_{d_s^{(ii)}|{\cal H}_1}\{L_{s_{22}}\}$, $\sigma^2_{l_{22}}=VAR_{d_s^{(ii)}|{\cal H}_1}(L_{s_{22}})$ and the continuous random variable $L_{s_{22}}=\ln (\frac{f(d_s^{(ii)}|{\cal H}_1)}{f(d_s^{(ii)}|{\cal H}_0)})$.

%============================================================================

$\bullet$ $\bar{P}_{e_{21}}$: Following the same steps taken in previous sections for calculating $\bar{P}_{e_{21}}$ for large $S$, we find
\begin{align}
\bar{P}_{e_{21}}<\underbrace{\frac{1}{(|S_0|)^{1-{t_0}}}\frac{\pi_0^{1-{t_0}}\pi_1^{t_0}}{(1-\text{LRT}_{\max})^{t_0}}}_{=\kappa_{l_{21}}}e^{S(\mu_{l_{21}}+\frac{1}{2}\sigma^2_{l_{21}})}\nonumber
\end{align}
where $\mu_{l_{21}}=\mathbb{E}_{d_s^{(ii)}|{\cal H}_1,d_s^{(ii)}|{\cal H}_0}\{L_{s_{21}}\}$, $\sigma^2_{l_{21}}=VAR_{d_s^{(ii)}|{\cal H}_1,d_s^{(ii)}|{\cal H}_0}(L_{s_{21}})$ and the discrete random variable
$L_{s_{21}}=\ln (\frac{D_2(d_s^{(ii)},d_s^{'(ii)})}{(P(d_s^{'(ii)}|{\cal H}_0))^{1-{t_0}}(P(d_s^{(ii)}|{\cal H}_1))^{t_0}})$. Also, we have verified that $\mu_{l_{21}}+\frac{1}{2}\sigma^2_{l_{21}}<0$.
To find the discrete random variable $D_2(d_s^{(ii)},d_s^{'(ii)})$ we substitute $[a^{2s-1}_n,a^{2s}_n,a^{2s-1}_m,a^{2s}_m]$ and $[a^{2s-1}_{n_1},a^{2s}_{n_1},a^{2s-1}_{m_1},a^{2s}_{m_1}]$ respectively with $d_s^{(i)}$ and $d_s^{'(i)}$ in $D_2(n,n_1)$ of Section VI.C.

%============================================================================

\subsection{Parallel Fusion Architecture with Local Threshold Changing at Sensors}
Similar to previous sections, we redefine $\bar{P}_{e_{11}}$, $\bar{P}_{e_{12}}$, $\bar{P}_{e_{21}}$ and $\bar{P}_{e_{22}}$ and write the following
\begin{equation}
\bar{P}_{e_1}<\bar{P}_{e_{11}}+\bar{P}_{e_{12}}, ~~~
\bar{P}_{e_2}<\bar{P}_{e_{21}}+\bar{P}_{e_{22}}\nonumber
\end{equation}
where
\begin{eqnarray}
\bar{P}_{e_{11}}&=&\sum_{n,m}\frac{1-\bone_{\{Q^1_{n,m},Q^2_{n,m},Q^3_{n,m}\}}}{2\sqrt{|S_1|}} \sum_{d_{n_1,m_1} \in S_1} [\sqrt{G(n,m,n_1,m_1)} \prod_{s=1}^S ({\cal D}_1(n,m,n_1,m_1)]\nonumber\\
&\times& \pi_0 P(u_{2s-1}=a_n^{2s-1},u_{2s}=a_n^{2s},\bar{u}_{2s-1}=a_m^{2s-1},\bar{u}_{2s}=a_m^{2s}|{\cal H}_0))\nonumber\\
\bar{P}_{e_{12}}&=&\sum_{n,m}(\bone_{\{Q^1_{n,m},Q^2_{n,m},Q^3_{n,m},Q^4_{n,m}\}})\pi_0\prod_{s=1}^S P(u_{2s-1}=a_n^{2s-1},u_{2s}=a_n^{2s},\bar{u}_{2s-1}=a_m^{2s-1},\bar{u}_{2s}=a_m^{2s}|{\cal H}_0))\nonumber\\
\bar{P}_{e_{21}}&=&\sum_{n,m}\frac{\bone_{\{Q^1_{n,m},Q^2_{n,m},Q^3_{n,m}\}}}{(|S_0|)} \sum_{d_{n_1,m_1} \in S_1} [\underset{t}{\min} ~(|S_0|G(n,m,n_1,m_1))^{t} \prod_{s=1}^S ({\cal D}_2(n,m,n_1,m_1)]\nonumber\\
&\times&\pi_1P(u_{2s-1}=a_n^{2s-1},u_{2s}=a_n^{2s},\bar{u}_{2s-1}=a_m^{2s-1},\bar{u}_{2s}=a_m^{2s}|{\cal H}_1))\label{pe-21-third}\\
&=&\sum_{n,m}\frac{\bone_{\{Q^1_{n,m},Q^2_{n,m},Q^3_{n,m}\}}}{(|S_0|)^{1-{t_0}}} \sum_{d_{n_1,m_1} \in S_1} [G(n,m,n_1,m_1)^{t_0} \prod_{s=1}^S ({\cal D}_2(n,m,n_1,m_1)]\nonumber\\
&\times& \pi_1P(u_{2s-1}=a_n^{2s-1},u_{2s}=a_n^{2s},\bar{u}_{2s-1}=a_m^{2s-1},\bar{u}_{2s}=a_m^{2s}|{\cal H}_1))\nonumber\\
\bar{P}_{e_{22}}&=&\sum_n (1-\bone_{\{Q^1_{n,m},Q^2_{n,m},Q^3_{n,m},Q^4_{n,m}\}})\pi_1\prod_{s=1}^S P(u_{2s-1}=a_n^{2s-1},u_{2s}=a_n^{2s},\bar{u}_{2s-1}=a_m^{2s-1},\bar{u}_{2s}=a_m^{2s}|{\cal H}_1)\nonumber
\end{eqnarray}
where $t_0$ is the value that minimizes $\bar{P}_{e_{21}}$ in (\ref{pe-21-third}). Recall that
$\bar{P}_{e_{12}}$ and $\bar{P}_{e_{22}}$ are the error floors, when communication channel is error-free. In the following, we discuss $\bar{P}_{e_{12}}$, $\bar{P}_{e_{11}}$, $\bar{P}_{e_{22}}$, $\bar{P}_{e_{21}}$ in asymptotic regime, as $S \rightarrow \infty$.

%============================================================================

$\bullet$ $\bar{P}_{e_{12}}$: Following the same steps taken in previous sections for calculating $\bar{P}_{e_{12}}$ for large $S$, we find
\begin{align}
\bar{P}_{e_{12}}=Q(\frac{\sqrt{S}\mu_{l_{12}}}{\sigma_{l_{12}}})<\underbrace{\kappa_{l_{12}}}_{=1/2}e^{-S\frac{\mu^2_{l_{12}}}{\sigma^2_{l_{12}}}}\nonumber
\end{align}
where $\mu_{l_{12}}=\mathbb{E}_{d_s^{(iii)}|{\cal H}_0}\{L_{s_{12}}\}$, $\sigma^2_{l_{12}}=VAR_{d_s^{(iii)}|{\cal H}_0}(L_{s_{12}})$ and the continuous random variable $L_{s_{12}}=\ln (\frac{f(d_s^{(iii)}|{\cal H}_1)}{f(d_s^{(iii)}|{\cal H}_0)})$.

%============================================================================

$\bullet$ $\bar{P}_{e_{11}}$: Following the same steps taken in previous sections for calculating $\bar{P}_{e_{11}}$ for large $S$, we find
\begin{align}\label{p-e-11-D}
\bar{P}_{e_{11}}<\underbrace{\frac{1}{2\sqrt{|S_1|}}\frac{\sqrt{\pi_0\pi_1}}{\sqrt{1-\text{LRT}_{\max}}}}_{=\kappa_{l_{11}}}e^{S(\mu_{l_{11}}+\frac{1}{2}\sigma^2_{l_{11}})}
\end{align}
where $\mu_{l_{11}}=\mathbb{E}_{d_s^{(iii)}|{\cal H}_0,d_s^{'(iii)}|{\cal H}_1}\{L_{s_{11}}\}$, $\sigma^2_{l_{11}}=VAR_{d_s^{(iii)}|{\cal H}_0,d_s^{'(iii)}|{\cal H}_1}(L_{s_{11}})$ and discrete random variable
%\begin{align}
$L_{s_{11}}=\ln (\frac{D_1(d_s^{(iii)},d_s^{'(iii)})}{\sqrt{P(d_s^{'(iii)}|{\cal H}_1)}\sqrt{P(d_s^{(iii)}|{\cal H}_0)}})$.
%\end{align}
Also, we have verified that $\mu_{l_{11}}+\frac{1}{2}\sigma^2_{l_{11}}<0$.
To find the discrete random variable $D_1(d_s^{(iii)},d_s^{'(iii)})$ we substitute $[a^{2s-1}_n,a^{2s}_n,a^{2s-1}_m,a^{2s}_m]$ and $[a^{2s-1}_{n_1},a^{2s}_{n_1},a^{2s-1}_{m_1},a^{2s}_{m_1}]$ respectively with $d_s^{(i)}$ and $d_s^{'(i)}$ in $D_1(n,m,n_1,m_1)$ of Section VI.D. Also, $\text{LRT}_{\max}$ in (\ref{p-e-11-D}) is
\begin{align}
&\text{LRT}_{\max}=\max_{n,m}\big(\frac{\pi_1\prod_{s=1}^SP(u_{2s-1}=a_n^{2s-1},u_{2s}=a_n^{2s},\bar{u}_{2s-1}=a_m^{2s-1},\bar{u}_{2s}=a_m^{2s}|{\cal H}_1)}{\pi_0\prod_{s=1}^SP(u_{2s-1}=a_n^{2s-1},u_{2s}=a_n^{2s},\bar{u}_{2s-1}=a_m^{2s-1},\bar{u}_{2s}=a_m^{2s}|{\cal H}_0)},~\mbox{given}\nonumber\\
&\frac{\pi_1P(u_{2s-1}=a_n^{2s-1},u_{2s}=a_n^{2s},\bar{u}_{2s-1}=a_m^{2s-1},\bar{u}_{2s}=a_m^{2s}|{\cal H}_1)}{\pi_0P(u_{2s-1}=a_n^{2s-1},u_{2s}=a_n^{2s},\bar{u}_{2s-1}=a_m^{2s-1},\bar{u}_{2s}=a_m^{2s}|{\cal H}_0)}<1\big)\nonumber
\end{align}

%============================================================================

$\bullet$ $\bar{P}_{e_{22}}$: Following the same steps taken in previous sections for calculating $\bar{P}_{e_{22}}$ for large $S$, we find
\begin{align}
\bar{P}_{e_{22}}=\Phi(\frac{\sqrt{S}\mu_{l_{22}}}{\sigma_{l_{22}}})<\underbrace{\kappa_{l_{22}}}_{=1/2}e^{-S\frac{\mu^2_{l_{22}}}{\sigma^2_{l_{22}}}}\nonumber
\end{align}
where $\mu_{l_{22}}=\mathbb{E}_{d_s^{(iii)}|{\cal H}_1}\{L_{s_{22}}\}$, $\sigma^2_{l_{22}}=VAR_{d_s^{(iii)}|{\cal H}_1}(L_{s_{22}})$ and the continuous random variable $L_{s_{22}}=\ln (\frac{f(d_s^{(iii)}|{\cal H}_1)}{f(d_s^{(iii)}|{\cal H}_0)})$.

%============================================================================

$\bullet$ $\bar{P}_{e_{21}}$: Following the same steps taken in previous sections for calculating $\bar{P}_{e_{21}}$ for large $S$, we find
\begin{align}
\bar{P}_{e_{21}}<\underbrace{\frac{1}{(|S_0|)^{1-{t_0}}}\frac{\pi_0^{1-{t_0}}\pi_1^{t_0}}{(1-\text{LRT}_{\max})^{t_0}}}_{=\kappa_{l_{21}}}e^{S(\mu_{l_{21}}+\frac{1}{2}\sigma^2_{l_{21}})}\nonumber
\end{align}
where $\mu_{l_{21}}=\mathbb{E}_{d_s^{(iii)}|{\cal H}_1,d_s^{'(iii)}|{\cal H}_0}\{L_{s_{21}}\}$, $\sigma^2_{l_{21}}=VAR_{d_s^{(iii)}|{\cal H}_1,d_s^{'(iii)}|{\cal H}_0}(L_{s_{21}})$, and the discrete random variable
%\begin{align}
$L_{s_{21}}=\ln (\frac{D_2(d_s^{(iii)},d_s^{'(iii)})}{(P(d_s^{'(iii)}|{\cal H}_0))^{1-{t_0}}(P(d_s^{(iii)}|{\cal H}_1))^{t_0}})$.
%\end{align}
Also, we have verified that $\mu_{l_{21}}+\frac{1}{2}\sigma^2_{l_{21}}<0$.
To find the discrete random variable $D_2(d_s^{(iii)},d_s^{'(iii)})$ we substituting $[a^{2s-1}_n,a^{2s}_n,a^{2s-1}_m,a^{2s}_m]$ and $[a^{2s-1}_{n_1},a^{2s}_{n_1},a^{2s-1}_{m_1},a^{2s}_{m_1}]$ respectively with $d_s^{(i)}$ and $d_s^{'(i)}$ in $D_2(n,m,n_1,m_1)$ of Section VI.D.

\begin{figure}
  \centering
  \includegraphics[scale=0.2]{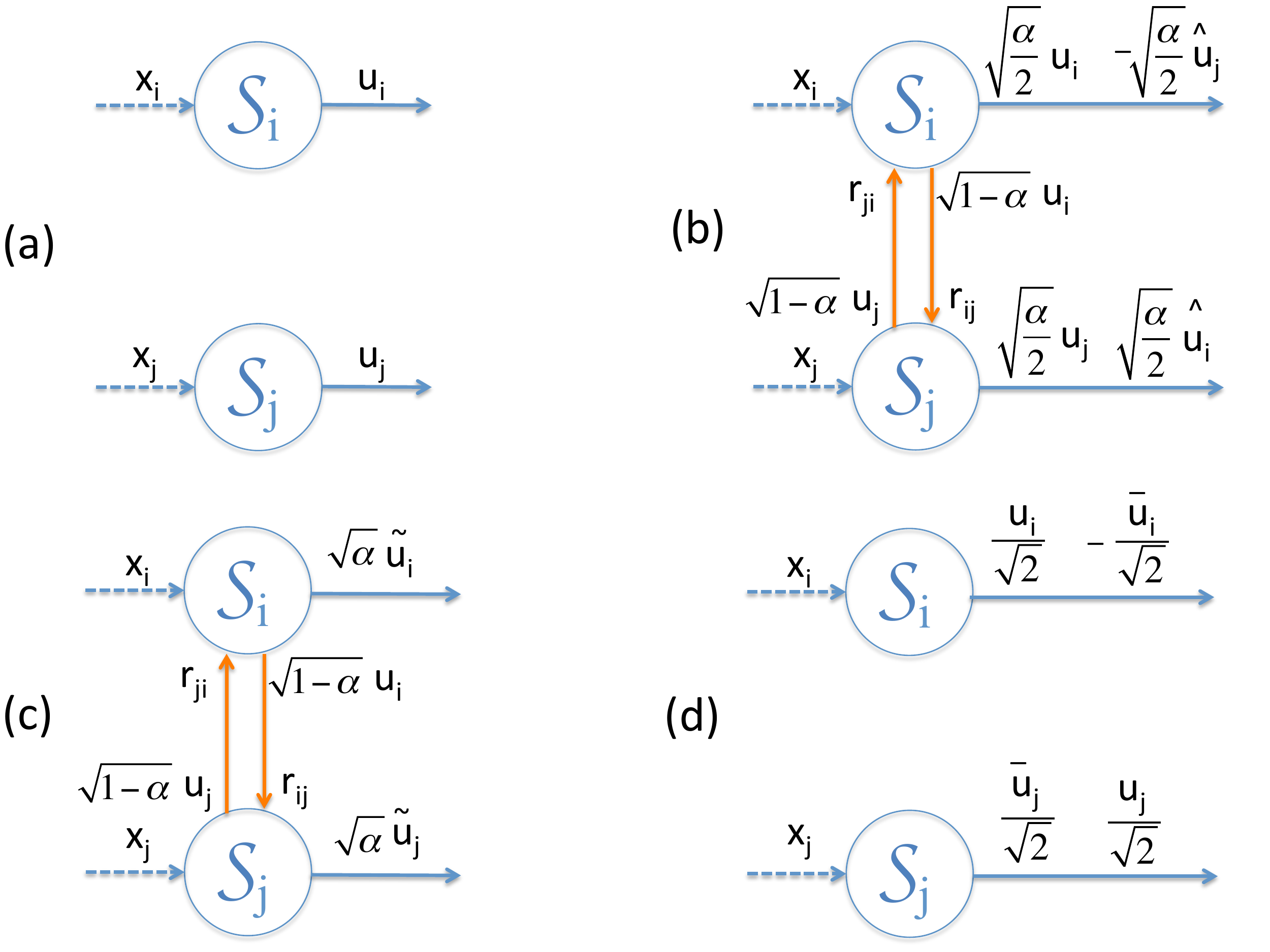}
  \caption {(a) Conventional parallel fusion architecture, (b) cooperative fusion architecture with STC at sensors, (c) cooperative fusion architecture with signal fusion at sensors, (d) parallel fusion architecture with local threshold changing at sensors.}\label{fig-conexity}
\end{figure}

\begin{figure}[b]
\centering
\subfigure[]{
\includegraphics[width=2.8in]{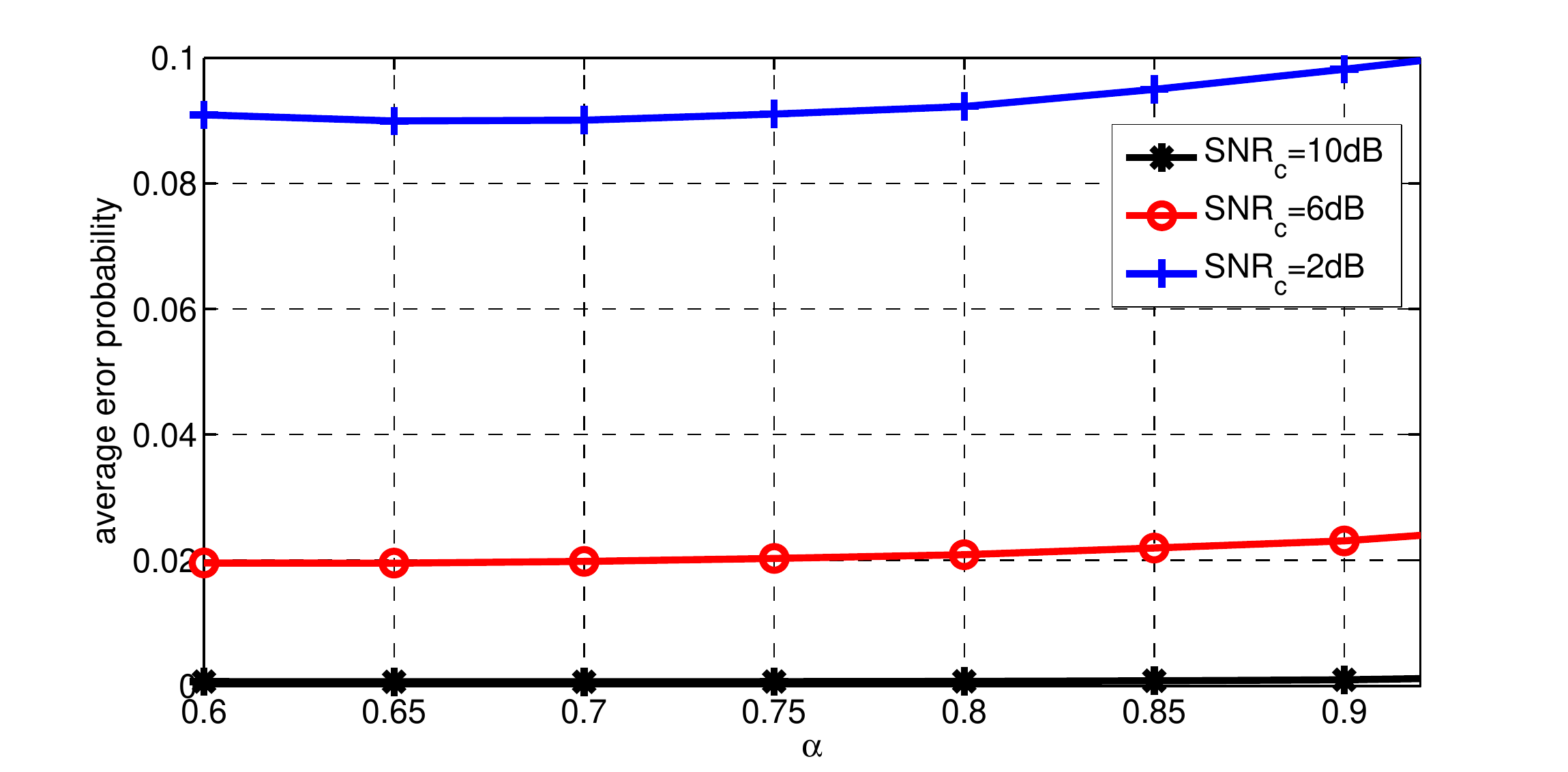}
\label{Fig2}
}
\subfigure[]{
\includegraphics[width=2.8in]{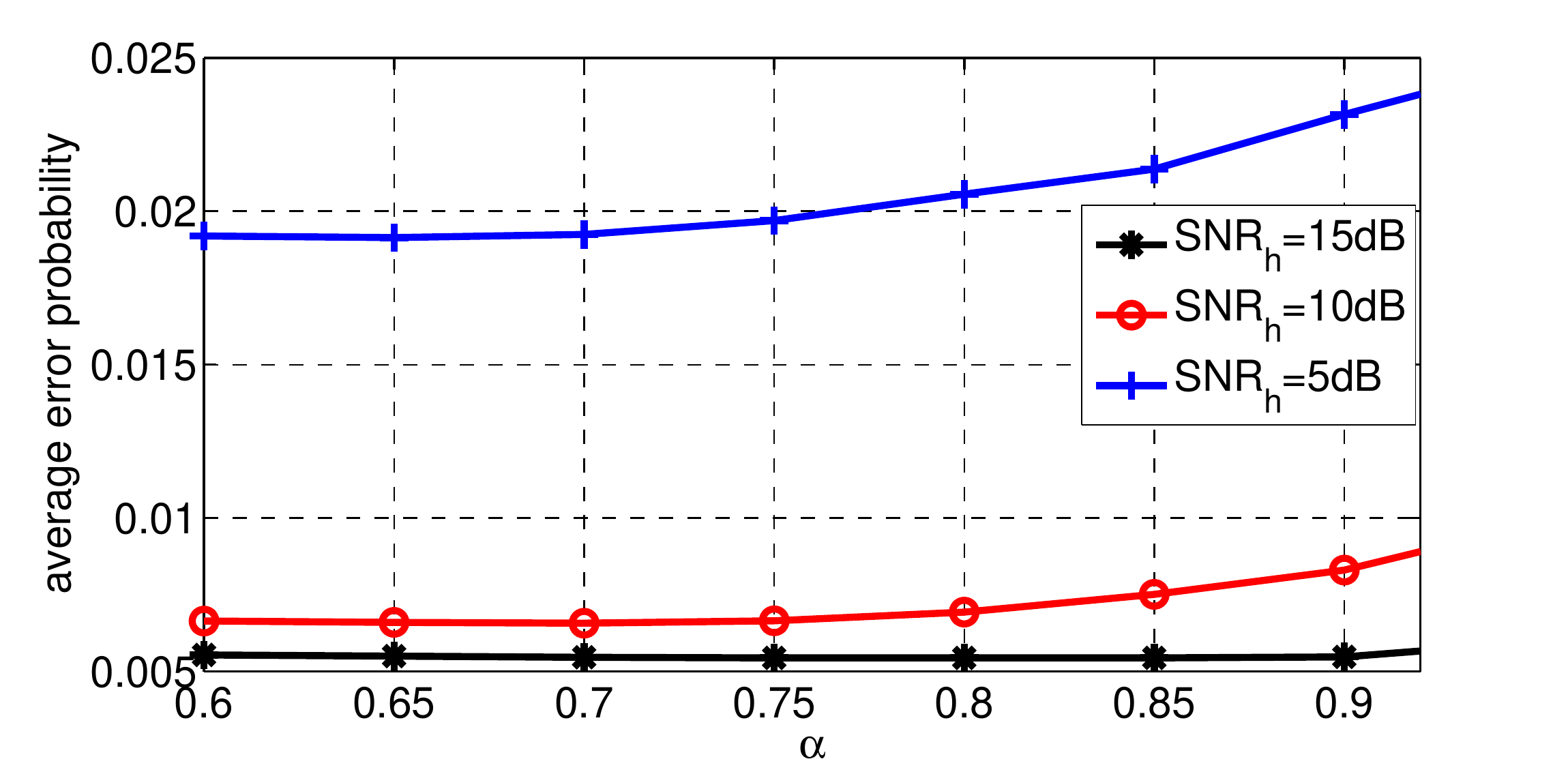}
\label{Fig3}
}
\caption{``STC@sensors'': LRT rule (a) SNR$_h$=5dB with SNR$_c$=2,6,10dB (b) SNR$_c$=6dB with SNR$_h$=5,10,15dB}
\label{}
\end{figure}
\begin{figure}[b]
\centering
\subfigure[]{
\includegraphics[width=2.8in]{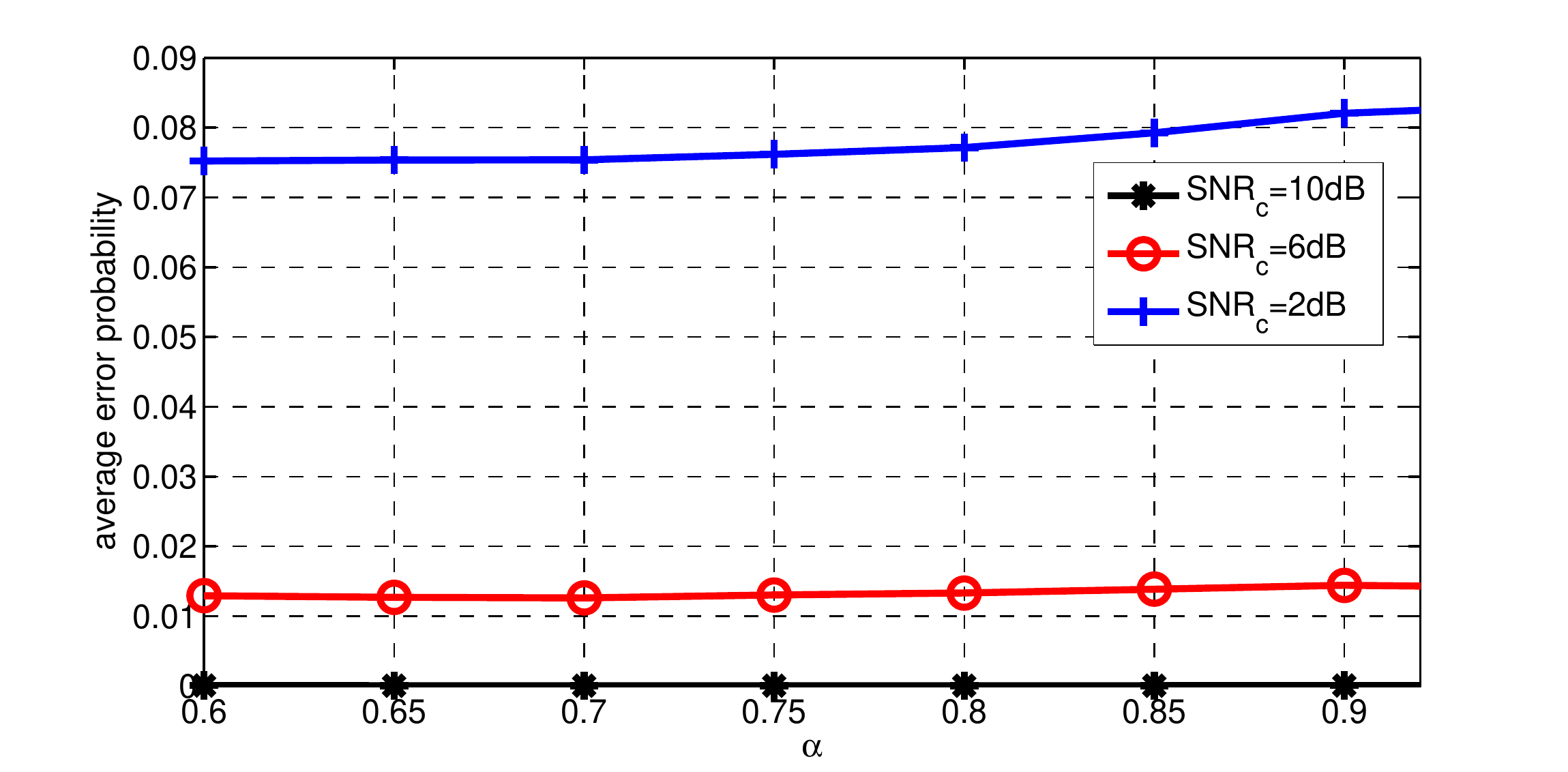}
\label{Fig6}
}
\subfigure[]{
\includegraphics[width=2.8in]{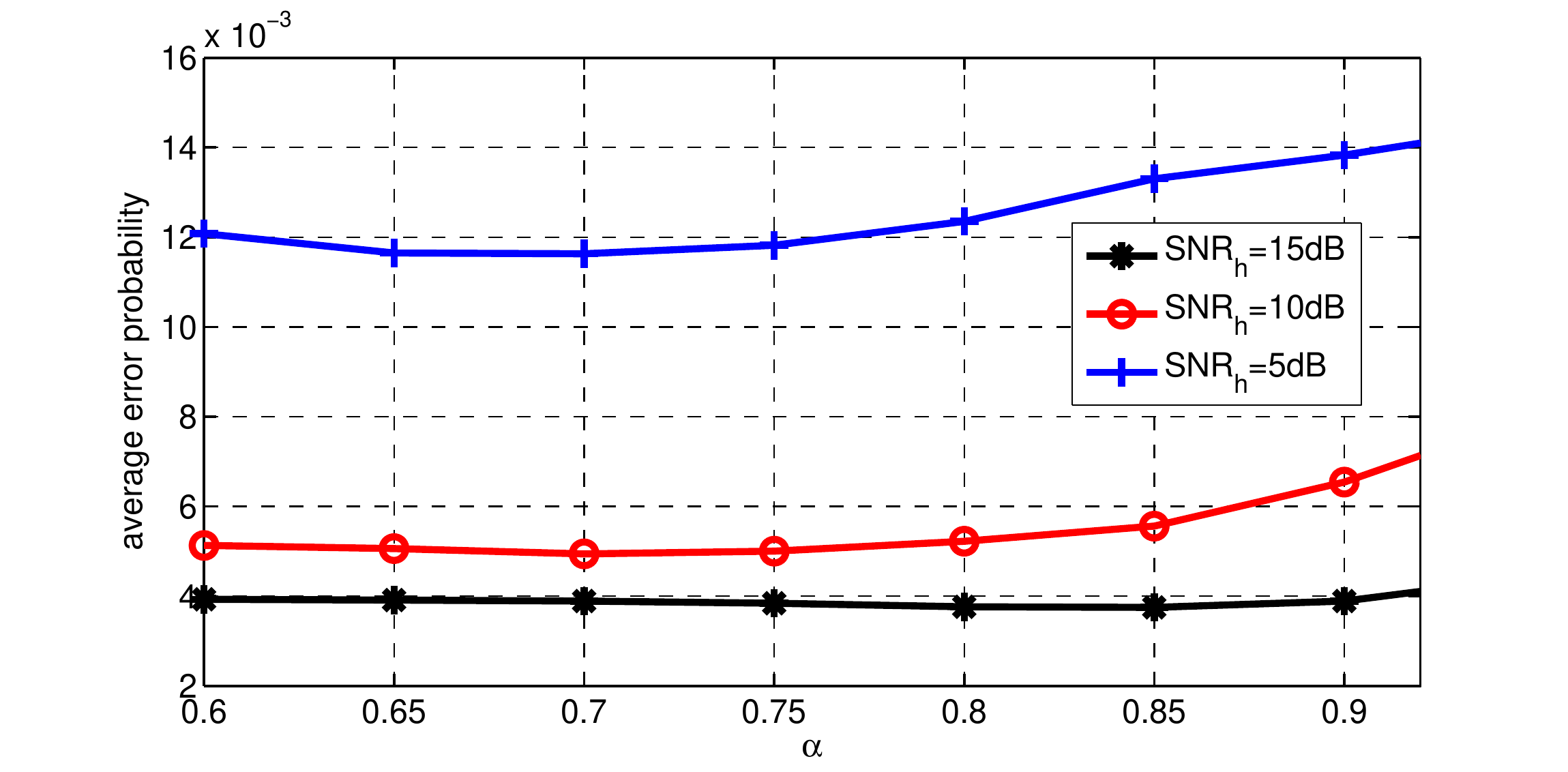}
\label{Fig7}
}
\caption{``fusion@sensors'': LRT rule (a) SNR$_h$=5dB with SNR$_c$=2,6,10dB (b) SNR$_c$=6dB with SNR$_h$=5,10,15dB}
\label{}
\end{figure}
\begin{figure}[b]
\centering
\subfigure[]{
\includegraphics[width=2.8in]{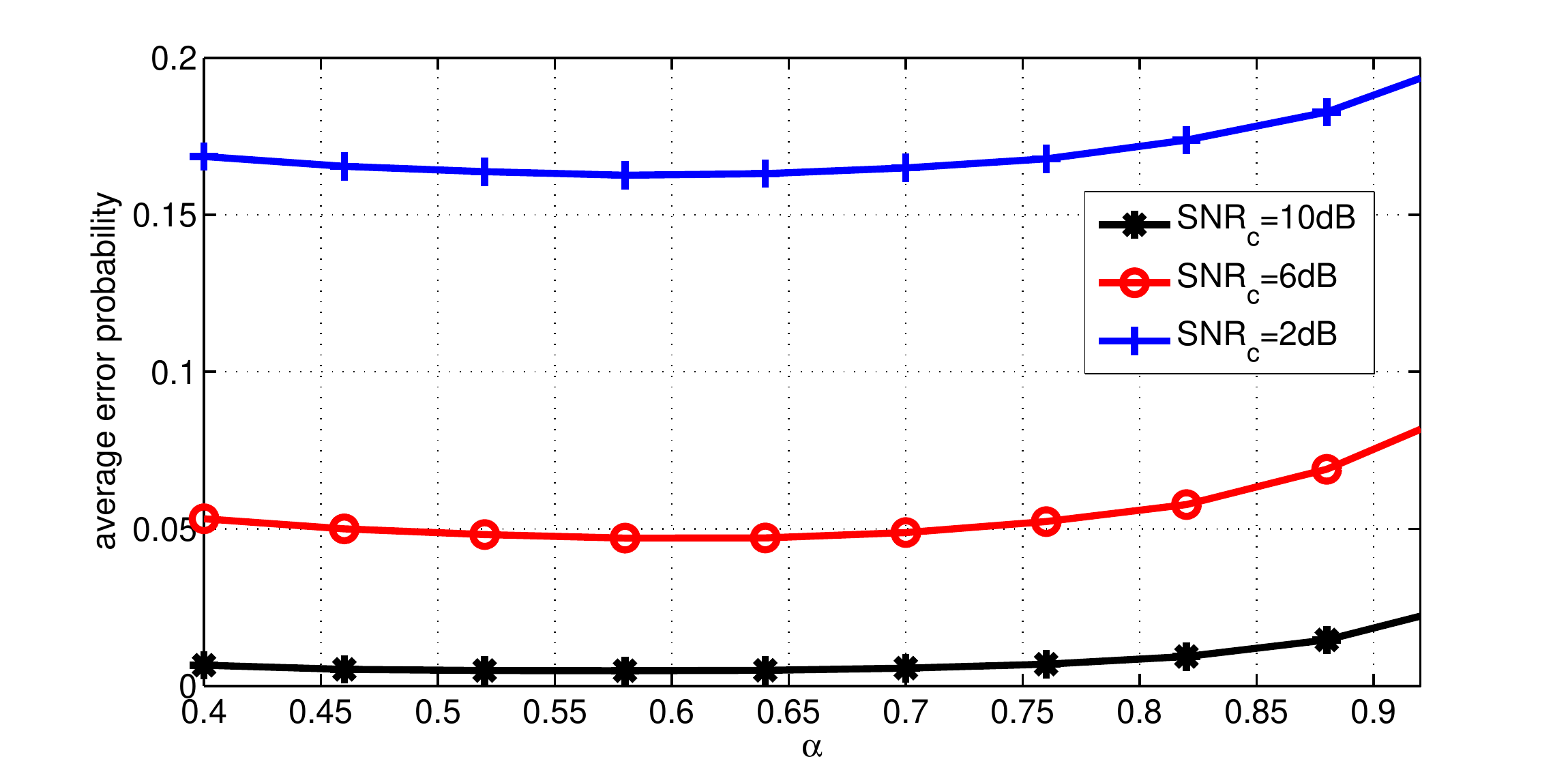}
\label{Fig4}
}
\subfigure[]{
\includegraphics[width=2.8in]{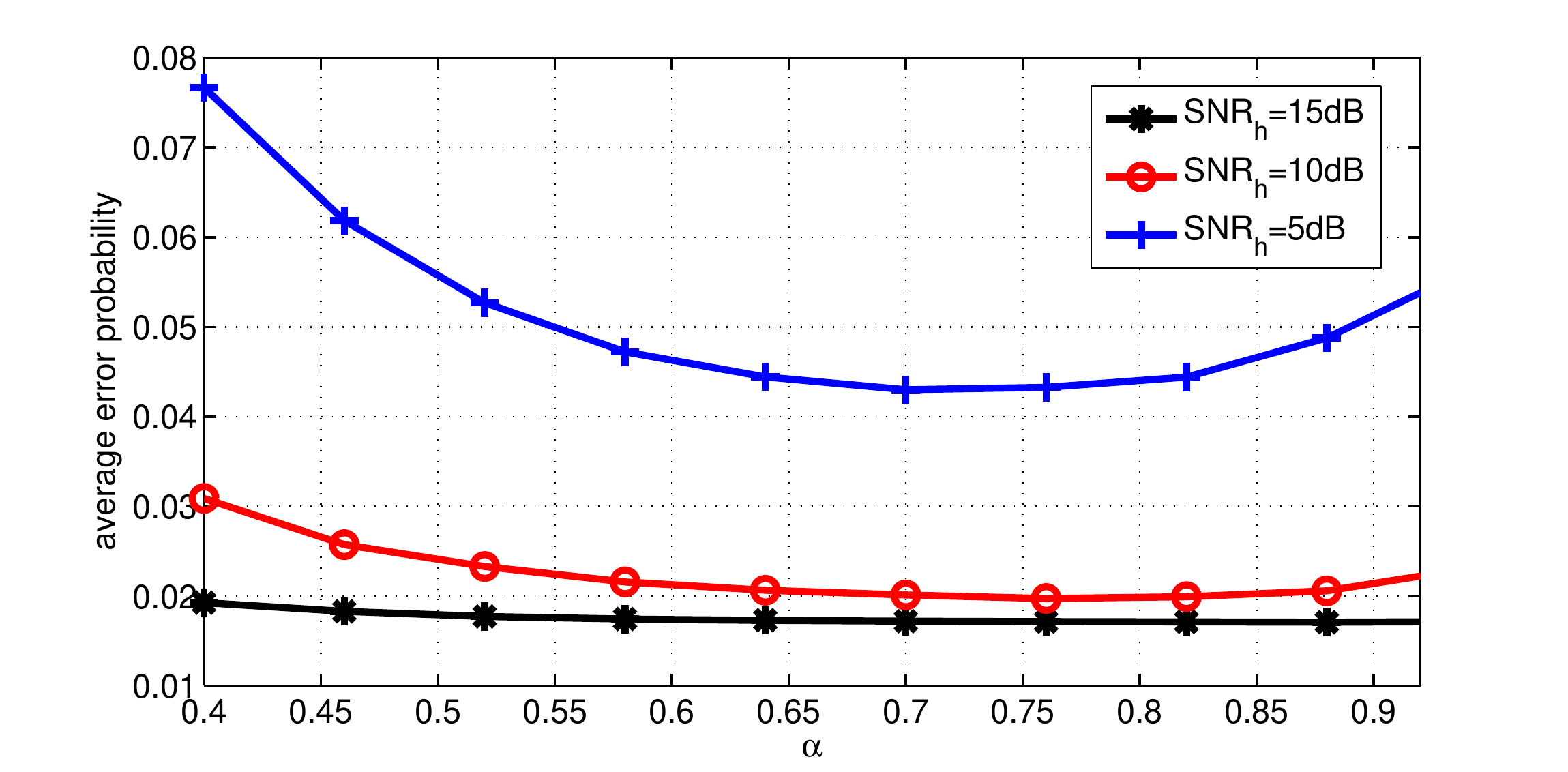}
\label{Fig5}
}
\caption{``STC@sensors'': majority rule (a) SNR$_h$=5dB with SNR$_c$=2,6,10dB (b) SNR$_c$=6dB with SNR$_h$=5,10,15dB}
\label{}
\end{figure}
\begin{figure}[b]
\centering
\subfigure[]{
\includegraphics[width=2.8in]{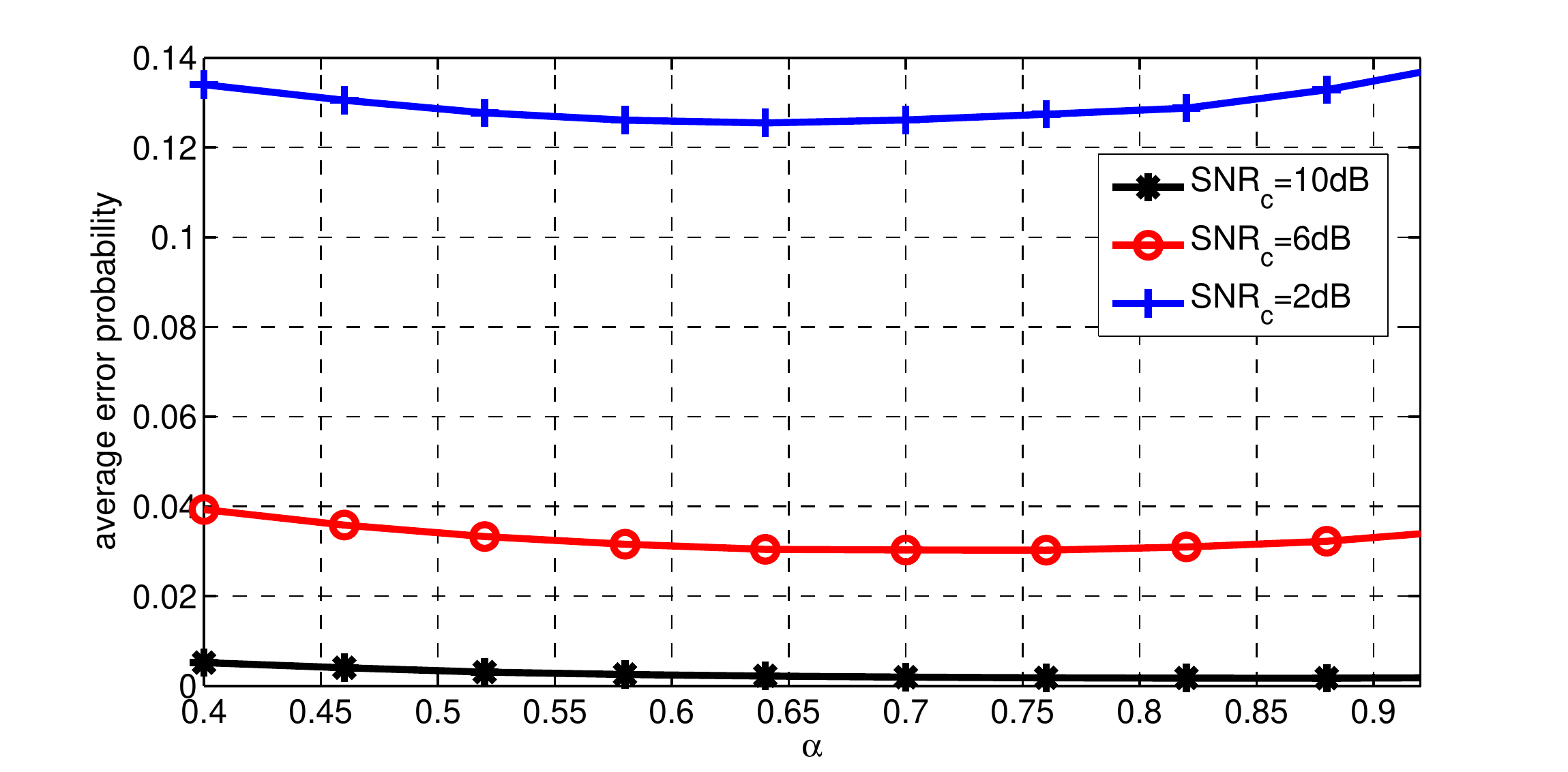}
\label{Fig8}
}
\subfigure[]{
\includegraphics[width=2.8in]{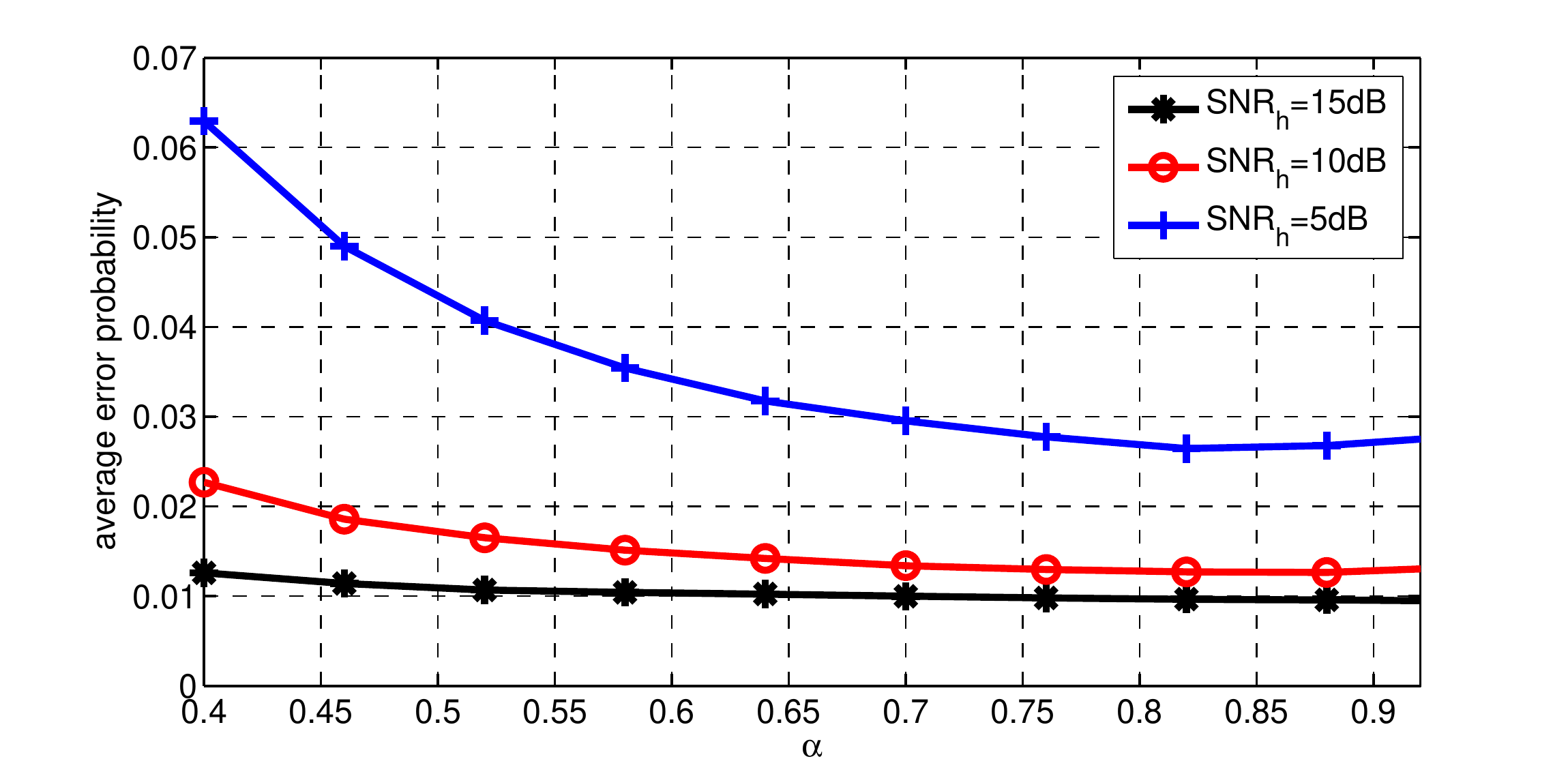}
\label{Fig9}
}
\caption{``fusion@sensors'': majority rule (a) SNR$_h$=5dB with SNR$_c$=2,6,10dB (b) SNR$_c$=6dB with SNR$_h$=5,10,15dB}
\label{}
\end{figure}
%
%%========theory versus simulation=====================
\begin{figure}
\centering
\includegraphics[scale=0.4]{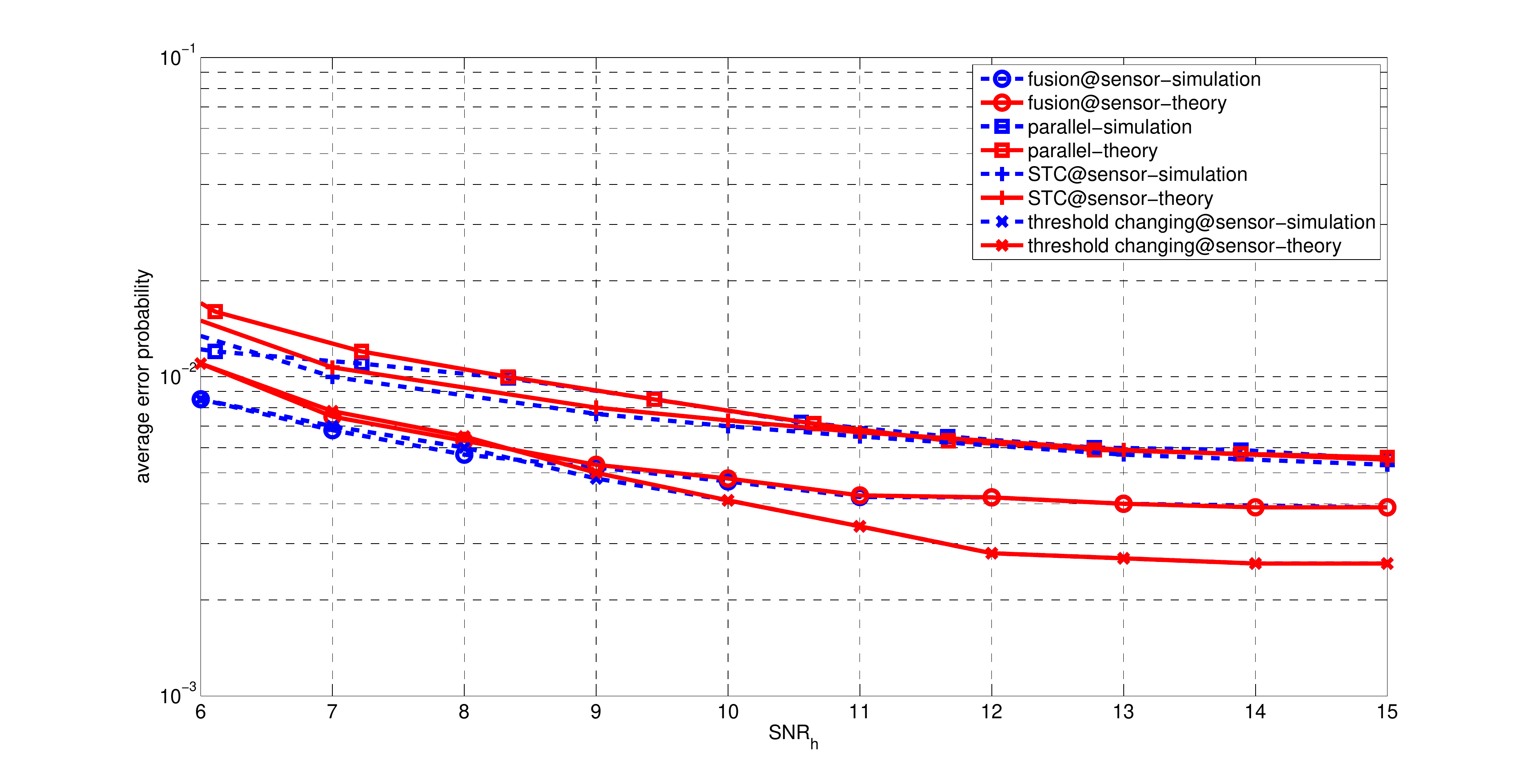}
\caption {Monte-Carlo simulation versus analytical results}\label{Fig1}
\end{figure}

%====================all schemes, optimal rule=================
\begin{table}
\footnotesize
%\hrulefill
% The spacer can be tweaked to stop underfull vboxes.
\vspace*{4pt}
\caption{all schemes, LRT rule, $\rho=0$, $K=10$} % title of Table
%\centering % used for centering table
\hspace{-1cm}
\begin{tabular}{||l|c|c|c|c|c|c|c|c|c|l||}
\hline

 & \multicolumn{3}{|c|}{SNR$_c=10$dB}&\multicolumn{3}{|c|}{SNR$_c=6$dB}&\multicolumn{3}{|c|}{SNR$_c=2$dB}&\multicolumn{1}{|l|}{}\\
\hline

  SNR$_h$&$5$dB & $10$dB & $15$dB & $5$dB & $10$dB & $15$dB & $5$dB & $10$dB & $15$dB&\\
  parallel&$1.9e{-4}$ & $3.8e{-5}$ & $2.0e{-5}$&$1.2e{-2}$ & $7.3e{-3}$ & $5.6e{-3}$&$7.6e{-2}$ & $6.6e{-2}$ & $6.4e{-2}$&\multirow {4}{*}{$\pi_0=0.6$}\\
  STC&$8.7e{-4}$ &$3.1e{-5}$ &$1.4e{-5}$&$1.8e{-2}$ &$6.9e{-3}$ &$5.4e{-3}$&$8.9e{-2}$ &$6.5e{-2}$ &$6.3e{-2}$&\\
  fusion&$1.9e{-4}$ &$1.7e{-5}$ &$1.4e{-5}$&$1.2e{-2}$ &$4.8e{-3}$ &$3.9e{-3}$&$7.4e{-2}$ &$5.5e{-2}$ &$5.2e{-2}$&\\
  threshold&$3.4e{-4}$ &$1.7e{-5}$ &$3.0e{-6}$&$1.2e{-2}$ &$4.4e{-3}$ &$2.6e{-3}$&$7.5e{-2}$ &$4.7e{-2}$ &$3.7e{-2}$&\\
\hline

  parallel &$3.0e{-4}$ & $4.3e{-5}$ & $2.3e{-5}$&$1.3e{-2}$ & $7.4e{-3}$ & $6.7e{-3}$&$8.1e{-2}$ & $7.0e{-2}$ & $6.6e{-2}$&\multirow {4}{*}{$\pi_0=0.7$}\\
  STC&$8.9e{-4}$ &$2.9e{-5}$ &$1.9e{-5}$&$1.9e{-2}$ &$7.1e{-3}$ &$6.2e{-3}$&$9.4e{-2}$ &$6.9e{-2}$ &$6.5e{-2}$&\\
  fusion&$2.3e{-4}$ &$2.0e{-5}$ &$7e{-6}$&$1.3e{-2}$ &$5.9e{-3}$ &$5.4e{-3}$&$7.9e{-2}$ &$5.9e{-2}$ &$5.5e{-2}$&\\
  threshold&$3.9e{-4}$ &$1.5e{-5}$ &$0.0e{-6}$&$1.2e{-2}$ &$3.9e{-3}$ &$2.4e{-3}$&$7.1e{-2}$ &$4.6e{-2}$ &$3.5e{-2}$&\\

\hline
  \end{tabular}
 \label{table:2} % is used to refer this table in the text
\end{table}
%====================all schemes, majority rule=================
\begin{table}
\footnotesize
%\hrulefill
% The spacer can be tweaked to stop underfull vboxes.
\vspace*{1pt}
\caption{all schemes, majority rule, $\rho=0$, $K=10$} % title of Table
%\centering % used for centering table
\hspace{-1cm}
\begin{tabular}{||l|c|c|c|c|c|c|c|c|c|c|c|l||}
\hline
&\multicolumn{2}{|c|}{SNR$_c = 13$dB}& \multicolumn{3}{|c|}{SNR$_c=10$dB}&\multicolumn{3}{|c|}{SNR$_c=6$dB}&\multicolumn{3}{|c|}{SNR$_c=2$dB}&\multicolumn{1}{|l|}{}\\
\hline

  SNR$_h$&$5$dB & $10$dB &$5$dB & $10$dB & $15$dB & $5$dB & $10$dB & $15$dB & $5$dB & $10$dB & $15$dB &\\
  parallel &$2.4e{-4}$&$4e{-6}$&$1.3e{-3}$ & $4.0e{-5}$ & $9.0e{-5}$&$3.2e{-2}$ & $2.2e{-2}$ & $1.8e{-2}$&$1.4e{-1}$ & $1.3e{-1}$ & $1.3e{-1}$ & \multirow {4}{*}{$\pi_0=0.6$}\\
  STC &$1.4e{-3}$&$2e{-6}$&$4.2e{-3}$ &$2.3e{-4}$ &$6.0e{-5}$&$4.4e{-2}$ &$2.0e{-2}$ &$1.7e{-2}$&$1.5e{-1}$ &$1.3e{-1}$ &$1.3e{-1}$ &\\
  fusion &$1.7e{-4}$&$3e{-6}$&$1.3e{-3}$ &$4.0e{-5}$ &$3e{-5}$&$2.7e{-2}$ &$1.3e{-2}$ &$1.0e{-2}$&$1.2e{-1}$ &$9.7e{-2}$ &$9.0e{-2}$ &\\
  threshold&&&$2.4e{-2}$ &$1.3e{-2}$ &$1.1e{-3}$&$1.2e{-1}$ &$1.0e{-1}$ &$9.7e{-2}$&$2.3e{-1}$ &$2.2e{-1}$ &$2.1e{-1}$&\\

    \hline

    \hline

  parallel &&&$2.0e{-3}$ & $5.3e{-5}$ & $2.0e{-4}$&$6.1e{-2}$ & $4.2e{-2}$ & $3.7e{-2}$&$2.0e{-1}$ & $2.0e{-1}$ & $2.0e{-1}$&\multirow {4}{*}{$\pi_0=0.7$}\\
  STC &&&$4.9e{-3}$ &$4.7e{-4}$ &$1.6e{-5}$&$6.8e{-2}$ &$3.9e{-2}$ &$3.6e{-2}$&$1.9e{-1}$ &$2.0e{-1}$ &$2.0e{-1}$&\\
  fusion &&&$1.2e{-3}$ &$1.8e{-4}$ &$6e{-5}$&$2.3e{-2}$ &$1.4e{-2}$ &$1.2e{-2}$&$9.2e{-2}$ &$1.0e{-1}$ &$9.8e{-2}$&\\
  threshold&&&$2.2e{-2}$ &$1.3e{-2}$ &$1.0e{-3}$&$1.0e{-1}$ &$9.6e{-2}$ &$9.0e{-2}$&$2.0e{-1}$ &$1.9e{-1}$ &$1.9e{-1}$&\\

\hline

\end{tabular}
  \label{table:4} % is used to refer this table in the text
\end{table}

%====================all schemes, correlation=================
%\begin{table}
%\footnotesize
%%\hrulefill
%% The spacer can be tweaked to stop underfull vboxes.
%\vspace*{4pt}
%\caption{all schemes, LRT rule, $\rho$=0.1, $\pi_0=0.6$} % title of Table
%\centering % used for centering table
%\begin{tabular}{||l|c|c|l||}
%\hline
% & \multicolumn{3}{|c|}{SNR$_c=10$dB}\\
%\hline
%
%  SNR$_h$&$5$dB & $10$dB & $15$dB \\
%  parallel &$8.0e{-4}$ & $5.0e{-4}$ & $3.2e{-4}$\\
%  STC &$1.7e{-3}$ &$4.9e{-4}$ &$3.1e{-4}$\\
%  fusion &$8.1e{-4}$ &$4.8e{-4}$ &$3.2e{-4}$\\
%  threshold&$9.7e{-4}$ &$4.9e{-4}$ &$2.8e{-4}$\\
%
%
%\hline
%\end{tabular}
%  \label{table:6} % is used to refer this table in the text
%\end{table}

\begin{table}
\footnotesize
\vspace*{4pt}
\caption{all schemes, LRT rule, $\pi_0=0.7$, $K=10$} % title of Table

\hspace{-1cm}
\begin{tabular}{||l|c|c|c|c|c|c|c|c|c|l||}

\hline
 & \multicolumn{3}{|c|}{SNR$_c=10$dB}&\multicolumn{3}{|c|}{SNR$_c=6$dB}&\multicolumn{3}{|c|}{SNR$_c=2$dB}&\multicolumn{1}{|l|}{}\\
 \hline
  SNR$_h$&$5$dB & $10$dB & $15$dB & $5$dB & $10$dB & $15$dB & $5$dB & $10$dB & $15$dB & \\
  \hline
  parallel  &$7.4e{-4}$ &$5.3e{-4}$ &$3.8e{-4}$  &$2.8e{-2}$&$2.4e{-2}$ &$2.2e{-2}$&$1.10e{-1}$&$1.02e{-1}$ & $9.95e{-2}$ & \multirow{4}{*}{$\rho=0.1$} \\
  STC       &$7.7e{-4}$ &$4.3e{-4}$ &$3.8e{-4}$  &$3.2e{-2}$&$1.8e{-2}$ &$2.0e{-2}$&$1.18e{-1}$&$1.02e{-1}$ & $9.89e{-2}$ & \\
  fusion    &$6.9e{-4}$ &$3.0e{-4}$ &$2.2e{-4}$  &$2.5e{-2}$&$1.7e{-2}$ &$1.6e{-2}$&$1.05e{-1}$&$9.5e{-2}$  & $9.06e{-2}$ & \\
  threshold &$9.3e{-4}$ &$2.2e{-4}$ &$1.5e{-4}$  &$2.6e{-2}$&$1.5e{-2}$ &$1.3e{-2}$&$1.09e{-1}$&$8.7e{-2}$  & $7.81e{-2}$ & \\
\hline
  parallel  &$2.7e{-3}$ &$1.9e{-3}$ &$1.7e{-3}$  &$4.0e{-2}$&$3.6e{-2}$ &$3.5e{-2}$&$1.38e{-1}$&$1.32e{-1}$  & $1.31e{-1}$ & \multirow{4}{*}{$\rho=0.2$}  \\
  STC       &$3.4e{-3}$ &$1.8e{-3}$ &$1.6e{-3}$  &$4.4e{-2}$&$3.5e{-2}$ &$3.5e{-3}$&$1.43e{-1}$&$1.32e{-1}$  & $1.32e{-1}$ &\\
  fusion    &$2.5e{-3}$ &$1.3e{-3}$ &$1.1e{-3}$  &$3.9e{-2}$&$3.3e{-2}$ &$3.3e{-2}$&$1.37e{-1}$&$1.29e{-1}$  & $1.25e{-1}$ &\\
  threshold &$3.1e{-2}$ &$1.4e{-3}$ &$1.0e{-3}$  &$4.3e{-2}$&$3.4e{-2}$ &$3.0e{-2}$&$1.42e{-1}$&$1.29e{-1}$  & $1.24e{-1}$ &\\
\hline
  parallel  &$4.8e{-3}$ &$3.8e{-3}$ &$3.5e{-3}$  &$5.7e{-2}$&$5.2e{-2}$ &$5.2e{-2}$&$1.60e{-1}$&$1.54e{-1}$  & $1.54e{-1}$ & \multirow{4}{*}{$\rho=0.3$}  \\
  STC       &$5.8e{-3}$ &$3.5e{-3}$ &$3.5e{-3}$  &$6.0e{-2}$&$5.2e{-2}$ &$5.1e{-3}$&$1.64e{-1}$&$1.56e{-1}$  & $1.56e{-1}$ & \\
  fusion    &$4.8e{-3}$ &$3.4e{-3}$ &$3.3e{-3}$  &$5.6e{-2}$&$5.0e{-2}$ &$4.9e{-2}$&$1.60e{-1}$&$1.52e{-1}$  & $1.52e{-1}$ & \\
  threshold &$3.1e{-2}$ &$3.5e{-3}$ &$2.7e{-3}$  &$6.1e{-2}$&$5.1e{-2}$ &$4.9e{-2}$&$1.62e{-1}$&$1.54e{-1}$  & $1.50e{-1}$ & \\
\hline
  parallel  &$1.4e{-2}$ &$1.4e{-2}$ &$1.3e{-2}$  &$9.19e{-2}$&$8.95e{-2}$ &$8.99e{-2}$&$1.80e{-1}$&$1.77e{-1}$  & $1.77e{-1}$ & \multirow{4}{*}{$\rho=0.5$}  \\
  STC       &$1.6e{-2}$ &$1.3e{-2}$ &$1.3e{-2}$  &$9.35e{-2}$&$8.97e{-2}$ &$8.89e{-2}$&$1.79e{-1}$&$1.71e{-1}$  & $1.71e{-1}$ & \\
  fusion    &$1.4e{-3}$ &$1.2e{-2}$ &$1.2e{-2}$  &$9.12e{-2}$&$8.79e{-2}$ &$8.55e{-2}$&$1.72e{-1}$&$1.71e{-1}$  & $1.71e{-1}$ & \\
  threshold &$1.4e{-2}$ &$1.3e{-2}$ &$1.0e{-2}$  &$9.19e{-2}$&$8.80e{-2}$ &$8.70e{-2}$&$1.80e{-1}$&$1.78e{-1}$  & $1.77e{-1}$ & \\
\hline

  parallel  &$3.1e{-2}$ &$3.0e{-2}$ &$2.9e{-2}$  &$1.25e{-1}$&$1.25e{-1}$ &$1.25e{-1}$&$2.10e{-1}$&$2.10e{-1}$  & $2.09e{-1}$ & \multirow{4}{*}{$\rho=0.8$}  \\
  STC       &$3.0e{-2}$ &$2.8e{-2}$ &$2.8e{-2}$  &$1.25e{-1}$&$1.25e{-1}$ &$1.25e{-1}$&$2.12e{-1}$&$2.09e{-1}$  & $2.09e{-1}$ &\\
  fusion    &$2.9e{-2}$ &$2.9e{-2}$ &$2.8e{-2}$  &$1.25e{-1}$&$1.25e{-1}$ &$1.25e{-1}$&$2.09e{-1}$&$2.09e{-1}$  & $2.09e{-1}$ & \\
  threshold &$3.0e{-2}$ &$3.0e{-2}$ &$3.0e{-2}$  &$1.29e{-1}$&$1.29e{-1}$ &$1.29e{-1}$&$2.17e{-1}$&$2.20e{-1}$  & $2.22e{-1}$ &\\
\hline

\hline
  \end{tabular}

\label{table:correlation}
\end{table}

%=================================================
%\begin{table}
%\footnotesize
%%\hrulefill
%% The spacer can be tweaked to stop underfull vboxes.
%\vspace*{4pt}
%\caption{all schemes, LRT rule, $\rho$ =0.3, $\pi_0=0.6$} % title of Table
%\centering % used for centering table
%\begin{tabular}{||l|c|c|l||}
%\hline
% & \multicolumn{3}{|c|}{SNR$_c=10$dB}\\
%\hline
%
%  SNR$_h$&$5$dB & $10$dB & $15$dB \\
%  parallel &$4.8e{-3}$ & $4.6e{-3}$ & $4.4e{-3}$\\
%  STC &$5.6e{-3}$ &$4.5e{-3}$ &$4.3e{-3}$\\
%  fusion &$4.8e{-3}$ &$4.4e{-3}$ &$4.3e{-3}$\\
%  threshold&$5.0e{-3}$ &$4.4e{-3}$ &$4.0e{-3}$\\
%
%
%\hline
%
%\end{tabular}
%  \label{table:7} % is used to refer this table in the text
%\end{table}
\begin{table}
\footnotesize

\vspace*{4pt}
\caption{all schemes, LRT rule, $\rho=0$, $\pi_0=0.6$, $K=20$ } % title of Table

\hspace{-1cm}
\begin{tabular}{||l|c|c|c|c|c|c|c|c|l||}
\hline
 & \multicolumn{3}{|c|}{SNR$_c=10$dB}&\multicolumn{3}{|c|}{SNR$_c=6$dB}&\multicolumn{3}{|c|}{SNR$_c=2$dB}\\
\hline

  SNR$_h$&$5$dB & $10$dB & $15$dB & $5$dB & $10$dB & $15$dB & $5$dB & $10$dB & $15$dB\\
  parallel  &$3.8e{-7}$ &$4.0e{-8}$ &$3.0e{-9}$  &$7.6e{-4}$&$3.3e{-4}$ &$2.3e{-4}$&$2.3e{-2}$&$1.6e{-2}$  & $1.5e{-2}$ \\
  STC       &$3.8e{-6}$ &$2.0e{-8}$ &$2.0e{-9}$  &$1.6e{-3}$&$2.6e{-4}$ &$1.8e{-4}$&$2.7e{-2}$&$1.5e{-2}$  & $1.3e{-2}$ \\
  fusion    &$3.6e{-7}$ &$5.7e{-9}$ &$2.0e{-9}$  &$6.0e{-4}$&$1.7e{-4}$ &$9.5e{-5}$&$2.1e{-2}$&$1.3e{-2}$  & $1.1e{-2}$ \\
  threshold &$8.5e{-7}$ &$6.0e{-9}$ &$1.0e{-10}$ &$7.3e{-4}$&$1.1e{-4}$ &$3.8e{-5}$&$2.1e{-2}$&$9.1e{-3}$  & $6.0e{-3}$ \\
\hline

  \end{tabular}

\label{table:incresing-number-sensors}
\end{table}

\begin{table}
\footnotesize
%\hrulefill
% The spacer can be tweaked to stop underfull vboxes.
\vspace*{4pt}
\caption{all schemes for a group of four sensors, LRT rule, $\rho=0$, $K=4$} % title of Table
%\centering % used for centering table
\hspace{-1cm}
\begin{tabular}{||l|c|c|c|c|c|c|c|c|c|l||}
\hline
 & \multicolumn{3}{|c|}{SNR$_c=10$dB}&\multicolumn{3}{|c|}{SNR$_c=6$dB}&\multicolumn{3}{|c|}{SNR$_c=2$dB}&\multicolumn{1}{|l|}{}\\
\hline

  SNR$_h$&$5$dB & $10$dB & $15$dB & $5$dB & $10$dB & $15$dB & $5$dB & $10$dB & $15$dB&\\
  parallel&$1.3e{-2}$ & $5.9e{-3}$ & $5.3e{-3}$&$7.3e{-2}$ & $5.8e{-2}$ & $5.6e{-2}$&$1.8e{-1}$ & $1.5e{-1}$ & $1.5e{-1}$&\multirow {4}{*}{$\pi_0=0.6$}\\
  STC4&$6.1e{-2}$ &$6.7e{-3}$ &$5.1e{-3}$&$1.3e{-1}$ &$6.0e{-2}$ &$5.4e{-2}$&$1.5e{-1}$ &$1.5e{-1}$ &$1.5e{-1}$&\\
  fusion4&$1.3e{-2}$ &$2.8e{-3}$ &$1.6e{-3}$&$7.8e{-2}$ &$4.5e{-2}$ &$3.7e{-2}$&$1.8e{-1}$ &$1.4e{-1}$ &$1.3e{-1}$&\\
  threshold4&$2.8e{-1}$ &$2.4e{-1}$ &$2.0e{-1}$&$3.0e{-1}$ &$2.7e{-1}$ &$1.9e{-1}$&$3.9e{-1}$ &$3.0e{-1}$ &$2.2e{-1}$&\\

  STC&$2.1e{-2}$ &$5.8e{-3}$ &$5.1e{-3}$&$8.8e{-2}$ &$5.7e{-2}$ &$5.4e{-2}$&$1.9e{-1}$ &$1.5e{-1}$ &$1.5e{-1}$&\\
  fusion&$1.3e{-2}$ &$3.8e{-3}$ &$3.6e{-3}$&$7.4e{-2}$ &$5.2e{-2}$ &$4.9e{-2}$&$1.8e{-1}$ &$1.5e{-1}$ &$1.4e{-1}$&\\
  threshold&$1.5e{-2}$ &$2.8e{-3}$ &$1.5e{-3}$&$7.8e{-2}$ &$4.4e{-2}$ &$3.5e{-2}$&$1.7e{-1}$ &$1.3e{-1}$ &$1.2e{-1}$&\\
\hline

  parallel&$1.0e{-2}$ & $5.5e{-3}$ & $4.8e{-3}$&$6.7e{-2}$ & $5.3e{-2}$ & $5.0e{-2}$&$1.6e{-1}$ & $1.5e{-1}$ & $1.4e{-1}$&\multirow {4}{*}{$\pi_0=0.7$}\\
  STC4&$5.7e{-2}$ &$8.3e{-3}$ &$3.5e{-3}$&$1.3e{-1}$ &$5.8e{-2}$ &$4.2e{-2}$&$2.0e{-1}$ &$1.5e{-1}$ &$1.4e{-1}$&\\
  fusion4&$1.5e{-2}$ &$4.5e{-3}$ &$2.6e{-3}$&$8.2e{-2}$ &$5.0e{-2}$ &$4.4e{-2}$&$1.7e{-1}$ &$1.4e{-1}$ &$1.3e{-1}$&\\
  threshold4&$2.3e{-1}$ &$1.7e{-1}$ &$2.0e{-1}$&$1.4e{-1}$ &$2.0e{-1}$ &$1.4e{-1}$&$2.9e{-1}$ &$2.7e{-1}$ &$1.2e{-1}$&\\
  STC&$1.8e{-2}$ &$5.0e{-3}$ &$4.1e{-3}$&$7.9e{-2}$ &$5.0e{-2}$ &$4.8e{-2}$&$1.7e{-1}$ &$1.4e{-1}$ &$1.4e{-1}$&\\
  fusion&$9.7e{-3}$ &$4.6e{-3}$ &$3.8e{-3}$&$6.7e{-2}$ &$5.3e{-2}$ &$4.7e{-2}$&$1.6e{-1}$ &$1.4e{-1}$ &$1.4e{-1}$&\\
  threshold&$1.2e{-2}$ &$3.1e{-3}$ &$2.2e{-3}$&$7.0e{-2}$ &$4.5e{-2}$ &$3.4e{-2}$&$1.6e{-1}$ &$1.3e{-1}$ &$1.2e{-1}$&\\

\hline
  \end{tabular}

\label{table:9} % is used to refer this table in the text
\end{table}

\begin{table}
\footnotesize

\vspace*{4pt}
\caption{``parallel'', ``fusion@sensors'' and ``fusion6@sensors'', LRT rule, $\rho=0$, $\pi_0=0.6$, $K=6$} % title of Table

\hspace{-1cm}
\begin{tabular}{||l|c|c|c|c|c|c|c|c|l||}
\hline
 & \multicolumn{3}{|c|}{SNR$_c=10$dB}&\multicolumn{3}{|c|}{SNR$_c=6$dB}&\multicolumn{3}{|c|}{SNR$_c=2$dB}\\
\hline

  SNR$_h$&$5$dB & $10$dB & $15$dB & $5$dB & $10$dB & $15$dB & $5$dB & $10$dB & $15$dB\\
  parallel&$3.0e{-3}$ &$1.3e{-3}$ &$9.0e{-4}$  &$3.7e{-2}$&$2.9e{-2}$ &$2.7e{-2}$&$1.3e{-1}$&$1.1e{-1}$  & $1.1e{-1}$  \\
  fusion  &$2.8e{-3}$ &$8.0e{-4}$ &$3.0e{-4}$  &$3.5e{-2}$&$2.1e{-2}$ &$1.7e{-2}$&$1.2e{-1}$&$9.5e{-2}$  & $9.0e{-2}$ \\
  fusion6 &$3.1e{-3}$ &$1.3e{-3}$ &$3.1e{-4}$  &$3.8e{-2}$&$2.9e{-2}$ &$2.6e{-2}$&$1.4e{-1}$&$1.1e{-1}$  & $9.9e{-2}$ \\
\hline
  \end{tabular}

\label{table:fusion6}
\end{table}
%============================================================

\begin{table}
\footnotesize
\vspace*{4pt}
\caption{all schemes, LRT rule, $\rho=0$, $\pi_0=0.6$, $K=4$} % title of Table

\hspace{-1cm}
\begin{tabular}{||l|c|c|c|c|c|c|c|c|c|l||}
\hline
 & \multicolumn{3}{|c|}{SNR$_c=10$dB}&\multicolumn{3}{|c|}{SNR$_c=6$dB}&\multicolumn{3}{|c|}{SNR$_c=2$dB}&\multicolumn{1}{|l|}{}\\
\hline

  SNR$_h$&$5$dB & $10$dB & $15$dB & $5$dB & $10$dB & $15$dB & $5$dB & $10$dB & $15$dB&\\
  parallel  &$1.3e{-2}$ &$5.9e{-3}$ &$5.3e{-3}$&$7.3e{-2}$ &$5.8e{-2}$ &$5.6e{-2}$&$1.8e{-1}$ & $1.5e{-1}$&$1.5e{-1}$&\multirow {4}{*}{homogenous}\\
  STC       &$2.1e{-2}$ &$5.8e{-3}$ &$5.1e{-3}$&$8.8e{-2}$ &$5.7e{-2}$ &$5.4e{-2}$&$1.9e{-1}$ &$1.5e{-1}$ &$1.5e{-1}$&\\
  fusion    &$1.3e{-2}$ &$3.8e{-3}$ &$3.6e{-3}$&$7.4e{-2}$ &$5.2e{-2}$ &$4.9e{-2}$&$1.8e{-1}$ &$1.5e{-1}$ &$1.4e{-1}$&\\
  threshold &$1.5e{-2}$ &$2.8e{-3}$ &$1.5e{-3}$&$7.8e{-2}$ &$4.4e{-2}$ &$3.5e{-2}$&$1.7e{-1}$ &$1.3e{-1}$ &$1.2e{-1}$&\\

\hline

   ${\cal P}$&$3.2$mW & $10$mW & $32$mW & $3.2$mW & $10$mW & $32$mW & $3.2$mW & $10$mW & $32$mW&\\
  parallel  &$1.1e{-2}$ &$6.4e{-3}$ &$5.2e{-3}$  &$7.5e{-2}$&$6.0e{-2}$ &$5.8e{-2}$&$1.7e{-1}$&$1.5e{-1}$  & $1.5e{-1}$&\multirow {4}{*}{inhomogenous} \\
  STC       &$1.4e{-2}$ &$6.9e{-3}$ &$5.1e{-3}$  &$8.4e{-2}$&$5.8e{-2}$ &$5.6e{-2}$&$1.8e{-1}$&$1.5e{-1}$  & $1.5e{-1}$&\\
  fusion    &$1.1e{-2}$ &$4.0e{-3}$ &$3.6e{-3}$  &$7.3e{-2}$&$5.4e{-2}$ &$4.9e{-2}$&$1.7e{-1}$&$1.5e{-1}$  & $1.4e{-1}$&\\
  threshold &$1.0e{-2}$ &$3.6e{-3}$ &$1.4e{-3}$  &$6.3e{-2}$&$4.5e{-2}$ &$3.3e{-2}$&$1.6e{-1}$&$1.4e{-1}$  & $1.2e{-1}$&\\
\hline

  \end{tabular}

\label{table:heterogenous}
\end{table}
\end{document}